\documentclass[letterpaper,12pt]{report}

\usepackage[top=1in, bottom=1in, left=1.5in, right=1in]{geometry}
\usepackage{setspace}
\usepackage{graphicx}
\usepackage{subfigure}
\usepackage{amssymb}
\usepackage{amsmath}
\usepackage{listings}
\usepackage{verbatim}
\usepackage{multirow}
\usepackage{docmute}
\usepackage{color}
\usepackage{url}
\usepackage{natbib}
\usepackage{cancel}
\usepackage{threeparttable}
\usepackage{textcomp}
\usepackage[T1]{fontenc}
\usepackage{lmodern}
\usepackage{mathrsfs}

\DeclareGraphicsExtensions{.pdf,.eps,.jpg,.png,.ps}

\newcommand{\simgt}{\,\hbox{\lower0.6ex\hbox{$\sim$}\llap{\raise0.6ex\hbox{$>$}}}\,}
\newcommand{\simlt}{\,\hbox{\lower0.6ex\hbox{$\sim$}\llap{\raise0.6ex\hbox{$<$}}}\,}

\newcommand{\Hi}{\mbox{$H_\mathrm{i}$}}
\newcommand{\dhor}{\mbox{$d_\mathrm{hor}$}}
\newcommand{\dprop}{\mbox{$d_\mathrm{p}$}}
\newcommand{\Nmp}{\mbox{$N_\mathrm{mp}$}}
\newcommand{\ti}{\mbox{$t_\mathrm{i}$}}
\newcommand{\tf}{\mbox{$t_\mathrm{f}$}}
\newcommand{\tls}{\mbox{$t_\mathrm{ls}$}}
\newcommand{\trm}{\mbox{$t_\mathrm{rm}$}}
\newcommand{\tml}{\mbox{$t_\mathrm{m\Lambda}$}}

\newcommand{\Tcmb}{T_\mathrm{CMB}}
\newcommand{\celcius}{\ensuremath{^\circ \mathrm{C}}}
\newcommand{\fahrenheit}{\ensuremath{^\circ \mathrm{F}}}
\newcommand{\NETP}{\ensuremath{\mathrm{NET}_\mathrm{P}}}
\newcommand{\NETT}{\ensuremath{\mathrm{NET}_\mathrm{T}}}

\newcommand{\Ptot}{P_\mathrm{tot}}
\newcommand{\Pe}{P_\mathrm{e}}
\newcommand{\Popt}{P_\mathrm{opt}}
\newcommand{\Psat}{P_\mathrm{sat}}
\newcommand{\RL}{R_\mathrm{L}}
\newcommand{\Rn}{R_\mathrm{n}}
\newcommand{\Rtes}{R_\mathrm{TES}}
\newcommand{\sI}{s_\mathrm{I}(\omega)}
\newcommand{\sT}{s_\mathrm{T}(\omega)}
\newcommand{\Tc}{T_\mathrm{c}}
\newcommand{\Tb}{T_\mathrm{b}}
\newcommand{\taue}{\tau_\mathrm{e}}
\newcommand{\taueff}{\tau_\mathrm{eff}}
\newcommand{\Ttes}{T_\mathrm{TES}}
\newcommand{\Vbias}{V_\mathrm{bias}}   

\newcommand{\asz}{\mbox{$A_\mathrm{SZ}$}}
\newcommand{\aprx}{\mbox{$\ensuremath{\sim}$}}
\newcommand{\ax}{\mbox{$A_\mathrm{X}$}}
\newcommand{\bsz}{\mbox{$B_\mathrm{SZ}$}}
\newcommand{\bx}{\mbox{$B_\mathrm{X}$}}
\newcommand{\csz}{\mbox{$C_\mathrm{SZ}$}}
\newcommand{\cx}{\mbox{$C_\mathrm{X}$}}
\newcommand{\da}{\mbox{$D_\mathrm{A}(z)$}}
\newcommand{\dainv}{\mbox{$D_\mathrm{A}^{-1}(z)$}}

\newcommand{\dx}{\mbox{$D_\mathrm{X}$}}
\newcommand{\fx}{\mbox{$f_x$}}
\newcommand{\onefifty}{$150\,$GHz}
\newcommand{\LCDM}{\mbox{$\mathrm{\Lambda}$CDM}}
\newcommand{\LX}{\mbox{$L_\mathrm{X}$}}
\newcommand{\Mfh}{\mbox{$M_{500}$}}

\newcommand{\MSZ}{\mbox{$M_{\mathrm{SZ}}^{500}$}}
\newcommand{\Mvir}{\mbox{$M_{\mathrm{vir}}$}}
\newcommand{\MX}{\mbox{$M_{\mathrm{X}}^{500}$}}
\newcommand{\msun}{\ensuremath{M_\odot}}

\newcommand{\rfh}{\mbox{$r_{500}$}}
\newcommand{\sqdeg}{\ensuremath{\mathrm{deg}^2}}
\newcommand{\tcmb}{\mbox{$T_\mathrm{CMB}$}}
\newcommand{\thcore}{\mbox{$\theta_\mathrm{c}$}}
\newcommand{\thcoresq}{\mbox{$\theta^2_\mathrm{c}$}}
\newcommand{\thint}{\mbox{$\theta_\mathrm{int}$}}

\newcommand{\uk}{\mbox{$\mu \mbox{K}$}}
\newcommand{\tzero}{\mbox{$\Delta T_0$}}
\newcommand{\Yfh}{\mbox{$Y_{\mathrm{SZ}}^{500}$}}

\newcommand{\Yrho}{\mbox{$Y_{\mathrm{SZ}}^{\rho}$}}
\newcommand{\Ysf}{\mbox{$Y_{\mathrm{SZ}}^{0.75'}$}}
\newcommand{\YSZ}{\mbox{$Y_{\mathrm{SZ}}$}}
\newcommand{\Yvir}{\mbox{$Y_{\mathrm{vir}}$}}
\newcommand{\Ytheta}{\mbox{$Y_{\mathrm{SZ}}^{\theta}$}}
\newcommand{\YthM}{\mbox{$Y_{\mathrm{SZ}}^{0.3\mathrm{Mpc}}$}}
\newcommand{\Ythresh}{\mbox{$Y_{\mathrm{SZ}}^{\mathrm{\phi}}$}(z)}
\newcommand{\YX}{\mbox{$Y_{\mathrm{X}}$}}
\newcommand{\yzero}{\mbox{$y_{\mathrm{0}}$}}

\begin{document}

%Title page, table of contents, abstract, etc.
\pagestyle{empty} 

%%%%%%%%%%%%%%
%Titlepage
{
\centering 
{\Large\bfseries MULTICHROIC TES BOLOMETERS AND GALAXY CLUSTER MASS SCALING RELATIONS WITH THE SOUTH POLE TELESCOPE}
\par
\vspace{7\baselineskip}
{by}\\[0.1\baselineskip]
{BENJAMIN ROMAN BERNARD SALIWANCHIK\\[1.9\baselineskip]
Submitted in partial fulfillment of the requirements\\[0.5\baselineskip]
For the degree of Doctor of Philosophy\\[0.5\baselineskip]
}\par

\centering
\vspace{7\baselineskip}
Dissertation Adviser: Dr. John Ruhl \\[\baselineskip]

\centering
\vspace{7\baselineskip}
Department of Physics \\
CASE WESTERN RESERVE UNIVERSITY \\[\baselineskip]
January, 2016\par
\vfill
}

\clearpage

{
\centering
{\bfseries CASE WESTERN RESERVE UNIVERSITY \\ SCHOOL OF GRADUATE STUDIES}\par
\vspace{3\baselineskip}
\flushleft
{We hereby approve the thesis/dissertation of}\par
\vspace{1.5\baselineskip}
\underline{Benjamin Roman Bernard Saliwanchik \hspace{7.6em}} \\
%\rule{25em}{0.4pt}\\
[1\baselineskip]
candidate for the \underline{Doctor of Philosophy \hspace{3em}} degree *.\\ %\rule{15em}{0.4pt} degree *.\\
[4\baselineskip]
(signed) \underline{John Ruhl \hspace{16.3em}} \\ %\rule{21.2em}{0.4pt}\\
~~~~~~~~~~~~(chair of the committee) \\[1.5\baselineskip]
\textcolor{white}{(signed) } \underline{Corbin Covault \hspace{13.9em}} \\ %\rule{21.2em}{0.4pt}\\
~~~~~~~~~~~~\textcolor{white}{(chair of the committee)} \\[1.5\baselineskip]
\textcolor{white}{(signed) } \underline{Glenn Starkman \hspace{13.5em}} \\ %\rule{21.2em}{0.4pt}\\
~~~~~~~~~~~~\textcolor{white}{(chair of the committee)} \\[1.5\baselineskip]
\textcolor{white}{(signed) } \underline{Paul Harding \hspace{15em}} \\ %\rule{21.2em}{0.4pt}\\
~~~~~~~~~~~~\textcolor{white}{(chair of the committee)} \\[3\baselineskip]
(date) \underline{July $30^\mathrm{th}$, 2015 \hspace{3em}} \\

\vspace{1\baselineskip}
*We also certify that written approval has been obtained for any proprietary material contained therein.
\vfill
}
%%%%%%%%%%%%%%

\clearpage

\vspace*{7in}

\begin{centering}

\textcopyright \ 2015, Benjamin Saliwanchik

\end{centering}

\clearpage

%\tengwarannataritalic[2.5]
%\begin{centering}

%\ \\
%\tengwa{254}
%\Ttelco \TTthreedots \Tnuumen \Ts \Ttelco \TTdot \Ts \Ttinco \TTdoubleleftcurl \Troomen \TTthreedots \Talda \TTthreedots \Ts \Ttelco \TTthreedots \Tlambe \Tcalma \TTthreedots \Toore \Ts \Ttelco \TTacute \Troomen \TTleftcurl \Ttelco \TTrightcurl
%\tengwa{255}
%\ \\
%\Ttelco \TTdot \Textendedtinco \TTthreedots \tengwa{67} \Toore \Ts \Tlambe \TTacute \Tando \TTnasalizer \TTacute \Tlambe \TTdot \Ts\Ttelco \TTthreedots \Tanto \TTthreedots \Ts \Tmalta \Tyanta \TTthreedots \Troomen \TTdoubleacute \Ts \Tsilmenuquerna \TTdot \Tnuumen \TTthreedots \tengwarannatar[2.5] \Tcolon \Tcentereddot \tengwarannataritalic[2.5]
%\ \\
%\end{centering}

\ \\
\ \\

\begin{figure}
\begin{center}
\includegraphics[width=4.75in]{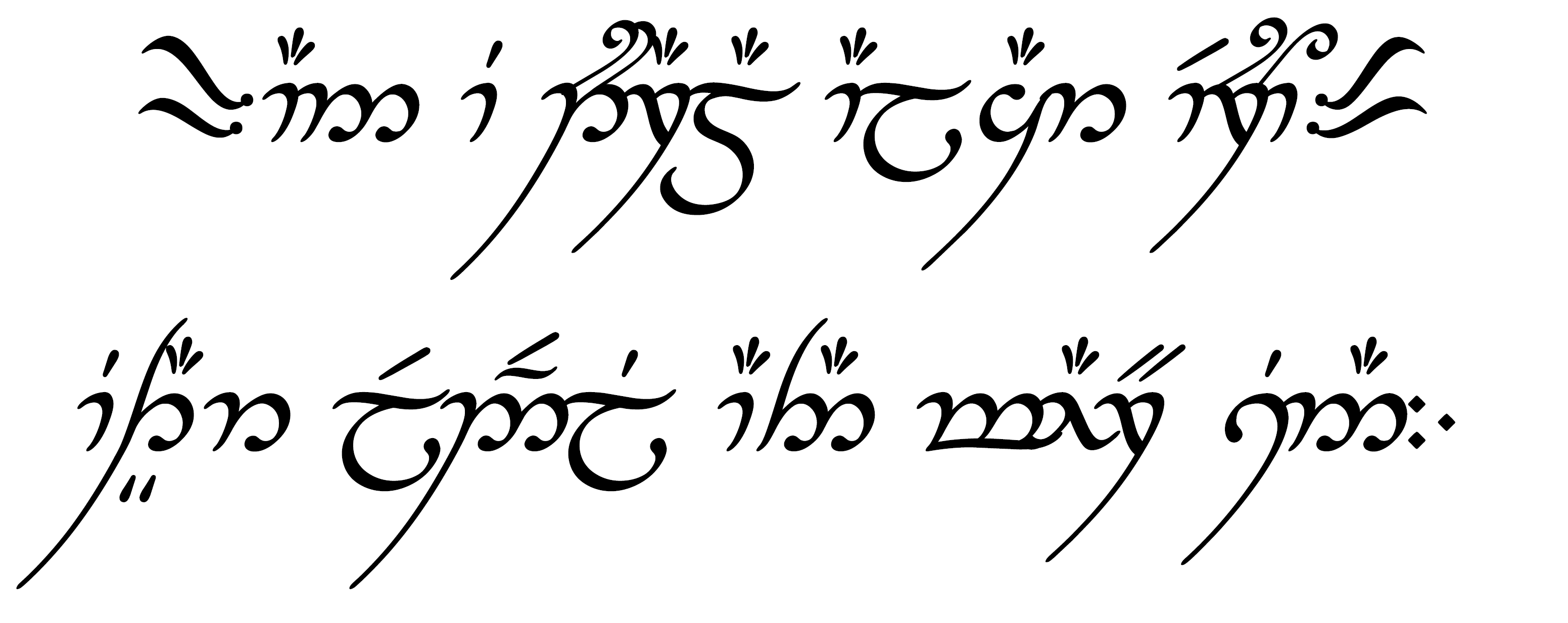}
%An i turalda alcar Eruo, istyar Elendil anta maire sina.
%For the greater glory of God, the scholar Elendil gives this work.
\end{center}
\end{figure}

\ \\ 
\ \\
\ \\

\clearpage

\setcounter{page}{1}
\pagenumbering{roman}
\pagestyle{plain}

\tableofcontents

\listoftables

\listoffigures

\chapter*{Acknowledgments}

I would first like to thank John Ruhl, and everyone in Ruhl Lab: Tom, for all his help with galaxy clusters, Sean and JT for many years of interesting lab discussions, Johanna for all her unsung labors for SPT, Scott for taking up the mantle of the Blue Dewar, and all the undergrads along the way for their hard work.

Working on the SPT collaboration has been extremely rewarding, and I am honored to have worked with such excellent scientists. I have enjoyed working with and spending time with all of you. I would especially like to thank Liz and Nick for their tireless efforts during our long days of wirebonding at Berkeley. Traveling to the South Pole was an incredible experience: from exploring Christchurch with Abby and Martin, to rock climbing with Wendy and Kyle, to making maps and greasing gears with Tom, and playing Durak with the whole crew.

Irina, thank you for your patience with my endless physics lectures, and your help debugging Fortran. Thank you to everyone in my family for keeping me on my toes with random astrophysics questions, and for always being there for me. Thank you Dad for wandering science discussions on the drive to Kalamazoo, and Mom for your critique of my first paper, and my overuse of the word ``cluster''.

\clearpage

\thispagestyle{empty}
\begin{centering}
{\Large \textbf{Multichroic TES Bolometers for the South Pole Telescope and Galaxy Cluster Mass Scaling Relations} }\\ 
\ \\
by \\
\ \\
Benjamin Roman Bernard Saliwanchik \\
\ \\
\ \\
\ \\
\ \\
\textbf{Abstract} \\
\end{centering}

\ \\
The South Pole Telescope (SPT) is a high-resolution microwave-frequency telescope designed to observe the Cosmic Microwave Background (CMB). To date, two cameras have been installed on the SPT to conduct two surveys of the CMB, the first in intensity only (SPT-SZ) and the second in intensity and polarization (SPTpol). A third-generation polarization-sensitive camera is currently in development (SPT-3G). This thesis describes work spanning all three instruments on the SPT. I present my work in time-reversed order, to follow the canonical narrative of instrument development, deployment, and analysis. First, the development and testing of novel 3-band multichroic Transition Edge Sensor (TES) bolometers for the SPT-3G experiment is detailed, followed by the development and deployment of the frequency multiplexed cryogenic readout electronics for the SPTpol experiment, and concluding with the analysis of data taken by the SPT-SZ instrument. I describe the development of a Bayesian likelihood based method I developed for measuring the integrated Comptonization (\YSZ) of galaxy clusters from the Sunyaev-Zel'dovich (SZ) effect, and constraining galaxy cluster \YSZ-mass scaling relations. 

\clearpage
 
\pagenumbering{arabic}
\setcounter{page}{1}

\doublespacing

%Introduction

\doublespacing

\chapter{Introduction}

\section{History of Modern Cosmology}

The era of modern cosmology began in 1927 when Georges Lema\^itre discovered that space was expanding \cite{lemaitre27}. The existence of galaxies as entities independent of the Milky Way had only recently been discovered, and it had been known for several years that most of the galaxies observed were redshifted, indicating motion away from the Milky Way. Lema\^itre calculated that more distant galaxies were receding more quickly, and inferred that the Universe was expanding. He also derived an equation predicting the apparent recessional velocity of galaxies based on their distance. For a variety of reasons his discovery was not well recognized. Notably, his initial publication was in a relatively minor French-language journal in Belgium, and it was not until 1931 that a more widely circulated English translation was published in the Monthly Notices of the Royal Astronomical Society \cite{lemaitre31a}, concurrent with an article further developing his expanding-Universe solution to the equations of General Relativity \cite{lemaitre31b}. In accordance with Stigler's Law of Eponymy, Lema\^itre's equation became known as Hubble's Law, after Edwin Hubble published a derivation of the same equation from similar data in 1929 \cite{hubble29}.

The Lema\^itre-Hubble law can be written as:
\begin{equation}
z = \frac{H_0}{c}r,
\end{equation}
where the redshift $z$ is the fractional difference between the emitted and observed wavelengths of a source:
\begin{equation}
z \equiv \frac{\lambda_\mathrm{ob}-\lambda_\mathrm{em}}{\lambda_\mathrm{em}},
\end{equation}
$c$ is the speed of light, $r$ is the distance to the source, and $H_0$ is the current expansion rate of the Universe, usually given in km s$^{-1}$ Mpc$^{-1}$.  The current best fit value of this expansion rate is $H_0 = 67.8\pm0.8$ km s$^{-1}$ Mpc$^{-1}$ \cite{olive14} \footnote{This is the 2014 best-fit value given by the Particle Data Group at Lawrence Berkeley National Laboratory. It is derived using the combination of the Planck 2013 temperature data, the WMAP 9-year polarization data, ACT and SPT high $\ell$ data, and BAO data from the SDSS, BOSS, 6dF and WiggleZ surveys.}. Lema\^itre and Hubble initially estimated the expansion rate as 625 km s$^{-1}$ Mpc$^{-1}$ and 530 km s$^{-1}$ Mpc$^{-1}$ respectively, due to vast underestimates of the distances to the galaxies in their data.

Hubble interpreted the redshifts of galaxies as Doppler shifts from a physical velocity but Lema\^itre saw them as the result of an isotropic expansion of space itself. Under this interpretation, the redshift of light emitted by a source can be used to determine the size of the observable Universe at the time that light was emitted relative to its present size. If $a(t_0)$ is the size of the observable Universe, or scale factor, at this moment, then the scale factor at the time the light was emitted, $a(t_\mathrm{em})$, is given by
\begin{equation}
a(t_\mathrm{em}) = \frac{a(t_0)}{1+z}.
\end{equation}

Tracing this expansion back in time naturally led to the inference of a high density, high temperature initial state of the Universe, which Lema\^itre referred to as the ``Cosmic Atom.'' This model of the Universe would eventually come to be known as the ``Hot Big Bang'' theory.

In 1948 Alpher, Gamow, and Herman used General Relativity and the Hot Big Bang model of the Universe to predict the primordial relative abundances of the elements \cite{alpher48}. As a side result, they also predicted the existence of relic radiation from the early Universe, and calculated a present-day temperature of 5K. It was not until nearly two decades later, in 1965, that Penzias and Wilson, working on the Holmdel Horn Antenna at the Bell Telephone Laboratories, detected excess noise in their antenna, with an effective temperature of 3K at 4.08GHz \cite{penzias65}. After accounting for or removing numerous sources of noise, including the presence of pigeon droppings in the antenna horn, the signal remained. They determined that it was an isotropic signal, not associated with the Sun, Galaxy, or any other localized source. Dicke, Peebles, Roll, and Wilkinson, who were building a microwave telescope at the time to search for the cosmic black-body radiation, interpreted Penzias and Wilson's discovery as a signature of the Big Bang \cite{dicke65}. In 1978, Penzias and Wilson received the Nobel Prize in Physics for the discovery of the Cosmic Microwave Background (CMB), which was seen as crucial evidence in favor of the Big Bang cosmological model.

In 1989 the COBE satellite was launched, marking the beginning of the precision cosmology era \cite{mather94}. The FIRAS instrument on the COBE satellite confirmed that the CMB followed a nearly perfect black-body spectrum, and measured its effective black-body temperature to be $2.728\pm0.004$K \cite{fixsen96b}. The COBE experiment was also the first to publish a measurement of the anisotropies in the CMB, which are a factor of $\aprx10^{-5}$ below the DC level of the CMB in amplitude \cite{smoot92}. 

The dipole moment of the CMB, caused by the velocity of our local reference frame with respect to the CMB rest frame, is comparatively large, at $3.355$mK \cite{olive14} ($\aprx10^{-3}$ times the DC level of the CMB). This motion is the sum of the velocities of the Earth around the Sun ($\aprx 30$~km/s), the Sun around the Milky Way ($\aprx 220$~km/s), and the Milky Way with respect to the CMB ($\aprx 550$~km/s). 

The anisotropies of the CMB are due to density variations in the primordial plasma at the time of recombination, $\aprx 380,000$ years after the Big Bang, and \aprx13.7 Gyrs before the present. These anisotropies are sourced both by temperature variations in the plasma due to the density variations, and by gravitational redshifting as radiation propagated out of overdense or underdense regions. This second source is called the Sachs-Wolfe effect. 

There is also a third component to the anisotropies introduced at later times. If a gravitational potential is evolving as photons pass through it, a gravitational redshift will be imparted. For example, if a galaxy cluster is collapsing and growing more dense, then photons will lose more energy exiting the potential than they gained entering it, resulting in a redshift. Likewise, if a region is becoming less dense, like the voids between clusters, photons will gain energy while passing through them, resulting in a blue shift. This effect depends on the integrated change in the potential along the path of the photon. The first-order linear portion of this effect is called the Integrated Sachs-Wolfe effect \cite{sachs67}, while the second-order nonlinear portion is called the Rees-Sciama effect \cite{rees68b}.

Measurements of the power spectrum of the CMB anisotropies provide crucial information about the constituents of the primordial plasma, and the physics at work in the early Universe.

Since COBE, the WMAP \cite{spergel03} and Planck \cite{planck06} satellite experiments have continued the characterization of the CMB, as have numerous ground based experiments (including, for example, ACBAR \cite{runyan03a}, QUaD \cite{bowden04}, BICEP \cite{takahashi08}, ACT \cite{fowler07}, Keck \cite{sheehy10}, and SPT \cite{carlstrom11}) and balloon-borne experiments ( including BOOMERANG \cite{montroy03b}, EBEX \cite{reichborn10}, and SPIDER \cite{filippini10}).

The picture of the early Universe that has emerged from measurements of the CMB has been in amazing agreement with other cosmological data, including measurements of the Hubble Law expansion of the Universe (e.g. supernovae redshifts), the primordial deuterium abundance, and large scale structure formation. 
These data all support the cosmological model developed in the decades before the COBE experiment, called the Inflationary \LCDM\ model. In this model, an early period of cosmic inflation solves some problems with the original Big Bang cosmological model (including the ``Horizon Problem'', the ``Flatness Problem'' and the ``Magnetic-Monopole Problem''), while a cosmological constant, $\Lambda$, causes the observed accelerating expansion of the Universe. The evolution of the early Universe as it is currently understood is described in the following sections. A more thorough treatment of these topics can be found in Ryden \cite{ryden03} or Mukhanov \cite{mukhanov05}.

\section{Inflation}

We will speak of the Universe as beginning at some initial time, $t=0$. What the Universe looks like at precisely this moment cannot be said with confidence, as the energy density of the Universe asymptotically approaches infinity at this point.
Fortunately, the state of the system at later times should not be sensitive to the initial conditions at $t=0$, because of the process of cosmic inflation.

There are three main inconsistencies with the picture of the early Universe we get from the CMB. First, the CMB is very uniform in temperature, with fluctuations on the order of one part in $10^5$. This implies that antipodal points on the CMB must have been in causal contact for some time to reach thermal equilibrium. In the case of a radiation dominated Universe, we can calculate the scale factor of the Universe, $a(t)$, at a time $t$ as
\begin{equation}
a(t) = \left( \frac{t}{t_0} \right) ^{1/2},
\label{eq:scale_rad}
\end{equation}
and the age of the Universe as
\begin{equation}
t_0 = \frac{1}{2H_0}.
\end{equation}
In the case of a spatially flat, radiation-only Universe the horizon distance is
\begin{equation}
\dhor(t_0)=2ct_0=\frac{c}{H_0},
\end{equation}
or the Hubble distance. Plugging in the age of the Universe at the time of last scattering during recombination, $t_\mathrm{ls}$, we find that the causal horizon was far smaller than the observed Hubble distance: $\dhor(\tls) \approx 0.4$Mpc. Taking in to account the expansion of the Universe since then, the size of causally related patches on the sky now should be only $\aprx2^\circ$ in diameter. This is the ``Horizon Problem'': how is the observed Universe essentially homogenous and isotropic on large scales?

The second problem is that the Universe appears to be very close to spatially flat. That is, to within the errors on our best measurements, the Universe appears to be Euclidean 3-space, not spherical or hyperbolic. Moreover, flatness is inherently unstable. That is, a curved Universe will tend to move away from flatness over time. Therefore, the observed spatial flatness of the universe is not necessary, and appears to be fine tuned. So why is the Universe spatially flat?

Historically, it was expected that at sufficiently high temperatures (possibly at the Grand Unification Theory (GUT) scale, at which the strong nuclear force merges with the electro-weak force) there should be thermal processes that produce magnetic monopoles. However, there is no experimental evidence for the existence of magnetic monopoles in the modern Universe. The lack of monopoles in the observed universe would also be explained by an inflationary period, however it is also possible that the theories predicting the existence of monopoles are incorrect, and no monopoles were ever produced. The lack of observed monopoles was historically one of the reasons inflationary theories were first explored, but is not currently seen as a necessary justification for the theory.

These problems are all potentially solved by a period of exponential inflation in the earliest moments of the Universe. This inflation is driven by one or more fields which are refered to as inflaton fields. One possible inflationary theory is that inflation is driven by a massless scalar field, $\phi(\vec{r},t)$, which varies as a function of position and time. We will explore this model as an example of the dynamics of inflation.

The energy density of this field, over a region where $\phi$ is homogenous, will be:
\begin{equation}
\varepsilon_{\phi} = \frac{1}{2} \frac{1}{\hbar c^3} \dot{\phi}^2 + V(\phi),
\end{equation} 
where $\hbar$ is the reduced Planck constant. The inflaton field must be homogenous over the space that would expand into our Hubble volume, in order to explain the observed homogeneity of the Universe. 

The pressure of the inflaton field will be:
\begin{equation}
P_{\phi} = \frac{1}{2} \frac{1}{\hbar c^3} \dot{\phi}^2 - V(\phi).
\end{equation}
So, if the inflation field varies slowly compared to the amplitude of the potential:
\begin{equation}
{\dot{\phi}}^2 \ll \hbar c^3 V(\phi),
\end{equation}
then it will behave like a cosmological constant, with energy density: 
\begin{equation}
\varepsilon_{\phi} \approx - P_{\phi} \approx V(\phi). 
\end{equation}

To determine how the Universe will respond to this field we use the Friedmann equation, which is an analytic solution to the Einstein field equation, assuming that the stress-energy tensor is isotropic and homogenous:
\begin{equation}
\left( \frac{\dot{a}}{a} \right)^2 = \frac{8 \pi G}{3c^2} \varepsilon - \frac{\kappa c^2}{R^2_0 a^2},
\end{equation}
where $G$ is the gravitational constant, $\varepsilon$ is the energy density, $\kappa$ specifies the spatial curvature (-1 for hyperbolic, 0 for Euclidean, and +1 for spherical), and $R_0$ gives the curvature of the space, in the case where $\kappa \neq 0$.
In addition, we need the acceleration equation, which can be derived from the fluid equation and the Friedmann equation:
\begin{equation}
\frac{\ddot{a}}{a} = - \frac{4 \pi G}{3 c^2} (\varepsilon + 3 P),
\label{eq:acceleration}
\end{equation}
where $P$ is pressure.

From the acceleration equation, we can see that iff $V(\phi) > 0$ then our slow-roll inflaton field ($\varepsilon_{\phi} = -P_{\phi}$) gives us accelerating expansion:
\begin{equation}
\frac{\ddot{a}}{a} = \frac{8 \pi G}{3 c^2}\varepsilon > 0,
\end{equation}
and the Friedmann equation reduces to:
\begin{equation}
\left( \frac{\dot{a}}{a} \right)^2 = \frac{8 \pi G}{3 c^2} \varepsilon = H_i^2.
\end{equation}
Solving for $a(t)$ we find:
\begin{equation}
a(t) = e^{H_i t}.
\end{equation}
The ratio of the scale factor after inflation, at time $\tf$, to the scale factor before inflation, at time $\ti$, is then $e^N$, where $N = \Hi (\tf - \ti)$. N is referred to as the number of e-folds of inflation.

Let us assume for the sake of example that inflation began at the GUT time, $\ti \approx 10^{-36}$s, with a Hubble parameter $H_i \approx 2 \times 10^{36} \mathrm{s}^{-1}$. 

The horizon distance at a time $t'$ is:
\begin{equation}
\dhor(t') = a(t') c \int_{0}^{t'} \frac{dt}{a(t)} 
\end{equation}
Since the Universe is radiation dominated at this point, the radius of the causal horizon at $\ti$ would be: 
\begin{equation}
\dhor(\ti) = a(\ti) c \int_{0}^{\ti} \left( \frac{t}{t_i} \right)^{-1/2} dt = 2 c \ti \approx 6 \times 10^{-28}\mathrm{m}.
\end{equation}
If inflation lasts for 62 e-folds, then after inflation the causal horizon is:
\begin{equation}
\dhor(\tf) = a(\ti)e^N c \left[ \int_{0}^{\ti} \left( \frac{t}{t_i} \right)^{-1/2} dt + \int_{\ti}^{\tf} \mathrm{exp}[-H_i(t-\ti)] dt \right] = 3 c \ti e^N \approx 0.76\mathrm{m}.
\end{equation}
At the time inflation ended, the current Hubble volume had a proper radius $\dprop(\tf) \aprx 0.6$m, calculating back from the present through the $\Lambda$, matter, and radiation dominated phases described in the following section.\footnote{The current proper radius of the Universe is $\dprop(t_0) = 14.5$ Gpc or $4.4\times10^{26}$ m. Using Equations \ref{eq:scale_rad}, \ref{eq:scale_matter}, and \ref{eq:scale_lambda}, this means at $\tf$ the proper radius was approximately:
\begin{equation}
\dprop(\tf) = \dprop(t_0) \ e^{H_0(t_0-\tml)} \left( \frac{\trm}{\tml} \right)^{2/3} \left( \frac{\ti}{\trm} \right)^{1/2} \approx 0.6\mathrm{m}.
\nonumber
\end{equation}} 
This implies that a minimum of 62 e-folds of inflation are needed to expand the pre-inflation causal horizon to above the observed Hubble volume.
It is possible that inflation lasted longer, but this is the minimum number of e-folds required to allow the current Hubble volume to come from a causally connected volume before inflation and solve the horizon problem.

It is further easy to see that any deviation from flatness in the pre-inflation Universe would have been exponentially suppressed. The curvature of the Universe is related to the density, $\Omega(t)$, by:
\begin{equation}
1-\Omega(t) = \frac{c^2}{R^2_0 \ a(t)^2 \ H(t)^2}.
\end{equation}
Then during the inflationary epoch $\Omega(t)$ goes as:
\begin{equation}
1-\Omega(t) \propto e^{-2N},
\end{equation}
and after 62 e-folds of inflation, $\Omega(t)$ is reduced to $\aprx 10^{-54}$ times its initial value.

As a side note, in inflationary theories, the density variations that are observed in the early Universe, and which grew into the present day large scale structure of galaxy clusters, were sourced by quantum fluctuations during inflation, and enlarged to macroscopic scales by inflation.

Similarly, the density of magnetic-monopoles would have been suppressed by a factor of the volume after inflation over the volume before inflation: $e^{3N}$. If inflation lasted for 62 e-folds, and the density of magnetic-monopoles before inflation was approximately one per causal volume \cite{ryden03}, $\Nmp \approx 10^{82}$~m$^{-3}$ , then today their density would be $\Nmp \approx 1.3 \times 10^{-3}$~Gpc$^{-3}$, or $\aprx 4$ monopoles in our entire Hubble volume ($2.9 \times 10^{3}$~Gpc$^3$).

After inflation, the inflaton field decays into radiation and matter through couplings to other fields. This process is known as ``reheating.'' All the matter in the current Universe is composed of the products of this inflaton decay. As the argument about magnetic-monopoles shows, essentially no matter from the pre-inflationary period persists after inflation.

\section{The Evolving Composition of the Universe}

\subsection{From Radiation Dominated to Matter Dominated}

After inflation ends, and the inflaton field decays, the Universe transitions to a radiation dominated state. In this state, the scale factor grows as the square root of time, as in Equation \ref{eq:scale_rad}. Similarly, the temperature will fall as the square root of time:
\begin{equation}
T(t) = \left( \frac{45}{32 \pi^2} \right)^{1/4} T_P \left( \frac{t}{t_P} \right)^{-1/2},
\end{equation}
where $t_P$ is the Planck time, and $T_P$ is the Planck temperature. 

However, as the Universe expands, the energy density of radiation falls as:
\begin{equation}
\varepsilon_{r}(a) = \frac{\varepsilon_{r,0}}{a^4},
\end{equation}
while the energy density of matter falls as:
\begin{equation}
\varepsilon_{m}(a) = \frac{\varepsilon_{m,0}}{a^3}.
\end{equation}
Therefore, eventually the energy density of radiation will fall below that of matter, and the Universe will become matter dominated. This point occurs at approximately $\trm \approx 50,000$~yrs (Table \ref{tab:thermal_history}).

At this point the scale factor goes as
\begin{equation}
a(t) \simeq \left( \frac{t}{t_0} \right) ^{2/3}.
\label{eq:scale_matter}
\end{equation}

After this, the matter density continues to fall, while the energy density in dark energy, $\Lambda$, is constant (assuming that $w = -1$, and dark energy is a cosmological constant):
\begin{equation}
\varepsilon_\Lambda(a) = \varepsilon_{\Lambda,0}.
\end{equation}
Eventually the Universe becomes $\Lambda$ dominated, at $\aprx 9$~Gyrs, after which the scale factor goes as:
\begin{equation}
a(t) \simeq e^{H_0(t-t_0)},
\label{eq:scale_lambda}
\end{equation}
where $H_0$ is the current Hubble parameter. The dynamics of dark matter will be described further in Section \ref{sec:dark_energy}.

\subsection{Phase Transitions and Decoupling Events}

A number of important events take place during the radiation dominated era. Many of these key events are listed in Table \ref{tab:thermal_history}, and the few most relevant will be described here. 

\begin{table*}
\begin{center}
{\caption{Events in the Thermal History of the Universe}
\label{tab:thermal_history}
\begin{tabular}{l|c|c|c|c}
\hline \hline
 Event & Time & Redshift & Temperature & Temperature \\
       &      &          & (eV)        & (K) \\
\hline
 Planck Era & $10^{-43}$s? & $10^{60}$? & $10^{28}$ eV? & $10^{32}$ K? \\
 GUT Era & $10^{-36}$s? & $10^{54}$? & $10^{25}$ eV? & $10^{29}$ K? \\  
 Inflation & $10^{-36}$-$10^{-35}$s? & $10^{54}$-$10^{27}$? & $10^{25}$-$10^{23}$ eV? & $10^{29}$-$10^{27}$ K? \\
 Baryogenesis & $<20$ ps & $>10^{15}$ & $>100$ GeV & $>10^{15}$ K \\
 EW Phase Transition & 20 ps & $10^{15}$ & 100 GeV & $1.16 \times 10^{15}$ K \\
 QCD Phase Transition & $20~\mu$s & $10^{12}$ & 150 MeV & $1.74 \times 10^{12}$ K \\
 DM Decoupling & 1 s? & $6 \times 10^{9}$? & 1 MeV? & $1.16 \times 10^{10}$ K? \\
 Neutrino Decoupling & 1 s & $6 \times 10^{9}$ & 1 MeV & $1.16 \times 10^{10}$ K \\
 e$^{-}$+e$^{+}$ Annihilation & 6 s & $2 \times 10^{9}$ & 500 keV & $5.80 \times 10^{9}$ K \\
 Nucleosynthsis & 3 min & $4 \times 10^{8}$ & 100 keV & $1.16 \times 10^{9}$ K \\
 Radiation-Matter Equality & 50 kyr & 3400 & 0.75 eV & 8,700 K \\
 Recombination & 260-380 kyr & 1400-1100 & 0.33-0.26 eV & 3,800-3000 K \\
 Photon Decoupling & 380 kyr & 1100 & 0.26 eV & 3000 K \\
 Reionization & 100-400 Myr & 30-11 & 7.0-2.6 meV & 80-30 K \\
 Matter-$\Lambda$ Equality & 9 Gyr & 0.4 & 0.33 meV & 3.8 K \\
 Present & 13.7 Gyr & 0 & 0.24 meV & 2.7 K \\
 \hline
\end{tabular}}
\begin{tablenotes}
Note -- As the Universe expands and cools it undergoes a number of phase transitions. Particle species fall out of thermal equilibrium (decouple) when their mean free path exceeds the horizon size. 
Values up to inflation calculated by this author based on a lower limit of 62 e-folds of inflation. Later data from Mukhanov \cite{mukhanov05} and Baumann \cite{baumann14}.
\end{tablenotes}
\end{center}
\end{table*}

At some unknown point, CP-symmetry violating processes produce a baryon anti-baryon imbalance, leading (much later) to the anti-baryonic content of the Universe being annihilated, and the Universe containing essentially only baryons.

At $ T \aprx 150$ GeV, approximately 20ps after reheating, the electroweak (EW) field separates into the electromagnetic field and the weak nuclear field.

At some unknown point dark matter decouples from the radiation and baryonic content of the Universe. This thermal decoupling happens when the rate of dark matter-baryon interactions becomes smaller than the expansion rate, $H$. If dark matter couples to baryonic matter via the weak force, dark matter decoupling would occur at approximately the same time as the neutrino decoupling. This occurs at energies of 1 MeV, approximately 1s after reheating. The fact that dark matter decouples before photon decoupling is important, since it allows density perturbations in the dark matter to begin to grow at an early time. Since the energy density in dark matter is $\aprx 10 \times$ the energy density in baryonic matter, the baryons will be drawn into the dark matter overdensities, seeding the density perturbations  at the time of photon decoupling which are seen in the CMB.

Throughout this period, nuclei heavier than hydrogen are forming through nuclear fusion. This is referred to as Big Bang Nucleosynthesis (BBN), and it continues until the temperature falls below 100 keV ($\aprx 10^9$ K), approximately 3 minutes after reheating. After BBN, the dominant nuclei are isotopes of hydrogen and helium, with a smattering of heavier elements. 

The density perturbations in this baryonic plasma cannot gravitationally collapse until after recombination. The length scale above which a fluid can collapse gravitationally is given by the Jeans length:
\begin{equation}
\lambda_\mathrm{J} = v_s \sqrt{\frac{\pi}{G \rho}},
\label{eq:jeans_length}
\end{equation}
where $v_s$ is the speed of sound in the fluid, and $\rho$ is its density. Here $v_s$ is $\aprx 2/3 c$, since the energy density of the plasma is still dominated by radiation at this point. It can be shown \cite{ryden03} that before recombination $\lambda_\mathrm{J}$ is greater than the Hubble scale, meaning that no density perturbations can collapse. They contract for a time under gravity, but when the pressure becomes too great they bounce back and expand again. This is the source of the acoustic oscillations in the CMB power spectrum.

The combination of atomic nuclei with electrons into neutral atoms, at 260-380 kyr after the Big Bang, is the next major phase transition in the Universe, and the photon decoupling that follows produces the CMB. Despite this being the first time neutral hydrogen had formed in the history of the Universe, this process is known as ``recombination.''

\section{The Cosmic Microwave Background}

\subsection{Origin of the CMB}

If we consider the baryonic content of the Universe before recombination to be exclusively hydrogen, then until the temperature falls significantly below the ionization energy of hydrogen, $Q = 13.6$eV, the following reaction will be in statistical equilibrium:
\begin{equation}
\mathrm{H} + \gamma \rightleftharpoons p + e^-.
\label{eq:ionization_reaction}
\end{equation}
Hydrogen can either be ionized by a photon with an energy above $13.6$eV, or a free proton and electron can recombine into hydrogen, emitting a photon in the process to carry away the excess energy. Equation \ref{eq:ionization_reaction} determines the atomic fraction of the Universe above the ionization temperature of hydrogen.

The ionization energy of helium is higher than that of hydrogen (24.6eV), so helium will recombine slightly earlier than hydrogen. However, since it is a small fraction of the total baryonic content, the total ionization fraction remains high, and photons are still well coupled to the baryons. For these reasons it is reasonable to treat the recombination phase transition as occurring solely in hydrogen. The addition of helium to the model will only introduce a small correction to the temperature at recombination. 

The degree to which the Universe is ionized is given by the ratio of free electrons to the total number of baryons:
\begin{equation}
X_e = \frac{n_e}{n_e+n_\mathrm{H}}
\end{equation}
where $n_e$ is the number density of electrons, and $n_\mathrm{H}$ is the number density of neutral hydrogen.

If we assume that the reaction in Equation \ref{eq:ionization_reaction} is in statistical equilibrium, then the number density, $n_x$, of a particle species with mass $m_x$ is given by the Maxwell-Boltzmann equation:
\begin{equation}
n_x = g_x \left( \frac{m_x k T}{2 \pi \hbar^2} \right)^{3/2} \mathrm{exp} \left( \frac{-m_x c^2}{k T} \right),
\end{equation}
where the statistical weight $g_x$ is given by the number of spin states of the particle ($g_e~=~g_p~=~2$, and $g_\mathrm{H}~=~4$). Combining the number densities of the different species we get:
\begin{equation}
\frac{n_\mathrm{H}}{n_p n_e} = \frac{g_\mathrm{H}}{g_p g_e} \left( \frac{m_\mathrm{H}}{m_p m_e} \frac{2 \pi \hbar}{k T} \right)^{3/2} \mathrm{exp} \left( \frac{[m_p + m_e - m_\mathrm{H}]c^2}{kT} \right).
\end{equation}
We may assume that the particles are not relativistic near the recombination temperature, so $kT \ll m_x c^2$. We also assume that the mass of the electron is small, so $m_\mathrm{H} \approx m_p$, and note that $(m_p + m_e - m_H)c^2 = Q$. This results in the Saha equation:
\begin{equation}
\frac{n_\mathrm{H}}{n_p n_e} = \left( \frac{2 \pi \hbar^2}{m_e k T} \right)^{3/2} \mathrm{exp} \left( \frac{Q}{kT} \right).
\end{equation}
Or, in terms of the ionization fraction:
\begin{equation}
\frac{1-X}{X} = n_p \left( \frac{2 \pi \hbar^2}{m_e k T} \right)^{3/2} \mathrm{exp} \left( \frac{Q}{kT} \right).
\end{equation}

Since the photons are in thermal equilibrium they will have a blackbody spectrum, and the number density of photons, $n_\gamma$, as a function of temperature will be:
\begin{equation}
n_\gamma = \frac{2.404}{\pi^2} \left( \frac{kT}{\hbar c} \right)^3.
\end{equation}
This allows us to express the Saha equation in terms of the photon to baryon ratio:
\begin{equation}
\eta = \frac{n_\gamma}{n_p+n_\mathrm{H}} = \frac{X n_\gamma}{n_p}.
\end{equation}
We will assume $\eta \approx 2 \times 10^9$, though given the exponential dependence on temperature, the value of $\eta$ will not strongly affect the recombination temperature. If for simplicity we define $t_\mathrm{rec}$ as being the moment when $X=1/2$, we can use the Saha equation to calculate that $T_\mathrm{rec} = 3650$.

Since the number of free electrons which can scatter photons falls rapidly during recombination, photons decouple from baryons shortly after $t_\mathrm{rec}$. This alters the precise course of recombination, because the photoionization reaction is no longer in equilibrium. A full non-equilibrium kinematic treatment of recombination results in a slightly lower recombination temperature, at a slightly later time \cite{mukhanov05}. The end result is that photon decoupling happens at approximately $T = 3000$K, at a redshift of $z = 1100$, when the Universe is $380,000$ years old. 

Since decoupling is defined as the time at which the mean free path of photons is longer than the Hubble distance (or equivalently when the Compton scattering rate falls below the expansion rate), the CMB photons will scatter off of electrons for the last time shortly after recombination. Taken collectively, these last scattering events form the surface which is seen by observing this radiation, and is known as the surface of last scattering. The Universe is then essentially transparent, and the majority of the photons emitted at this time travel to us without scattering again. A small number of photons will interact with intracluster gas through the SZ effect, and all will be slightly gravitationally lensed by intervening structures, but these are small corrections, which will be discussed further in later sections. 

Since the Universe has expanded by a factor of $1100$ since the time of recombination, and the energy density of radiation is proportional to $a^{-4}$, the blackbody radiation which was originally in the infrared portion of the spectrum has since been redshifted by a factor of $1100$, and now peaks in the microwave portion of the spectrum. It is therefore referred to as the Cosmic Microwave Background, or CMB.

\subsection{CMB Anisotropies}

The CMB is largely uniform, but its anisotropies contain a wealth of information about the state of the Universe at the time of last scattering, and at earlier times. Measuring these anisotropies is thus of great interest. The CMB scalar anisotropies are typically parameterized with the spherical harmonic functions, $Y_{\ell m}(\theta,\phi)$:
\begin{equation}
T(\theta,\phi) = \sum_\ell \sum^{\ell}_{m=-\ell} a_{\ell m} Y_{\ell m}(\theta,\phi).
\end{equation}

Since the density perturbations in the primordial plasma are expected to be Gaussian random, the average amplitude of the $a_{\ell m}$ coefficients at a given $\ell$ is zero, and the statistics of the scalar anisotropies can be described by a single dimensional parameter, $C_\ell$, the variance of the $ a_{\ell m}$ coefficients:
\begin{equation}
C_\ell = <a_{\ell m}a_{\ell m}^*> = \frac{1}{2 \ell + 1} \sum^{\ell}_{m=-\ell} |a_{\ell m}|^2.
\end{equation}

This power spectrum has been extensively measured by numerous CMB experiments, and used to contrain cosmological parameters.

\subsection{CMB Polarization}

\begin{figure}
\begin{center}
\includegraphics[width=3.0in]{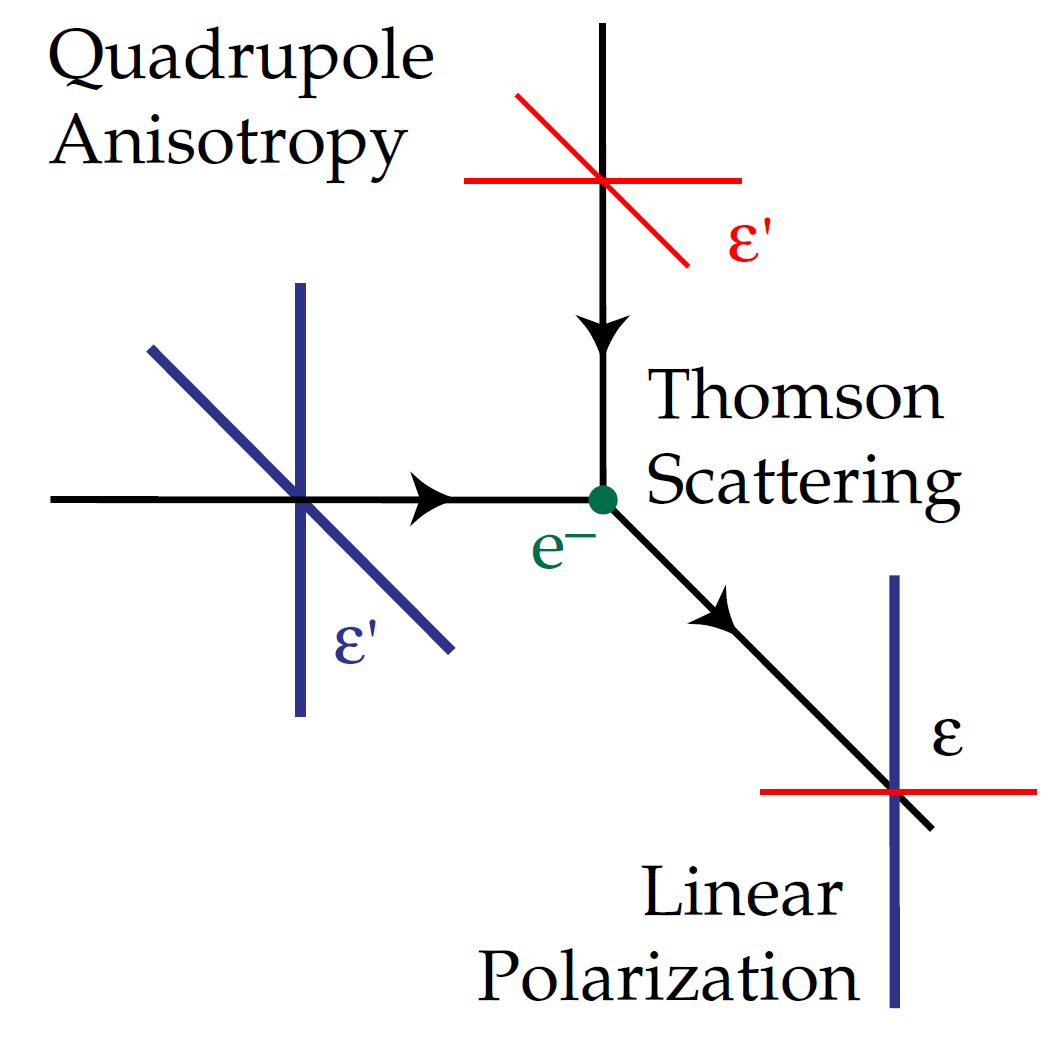}
\caption[Thompson Scattering]{Diagram of the production of linearly polarized radiation via Thompson scattering in a thermal quadrupole. The radiation incident on the electron, $\varepsilon'$, is hotter in the horizontal direction (thick blue lines) and colder in the vertical direction (thin red lines), resulting in partially polarized scattered radiation, $\varepsilon$, in the orthogonal direction. Figure from Hu and White \cite{hu97d}.}
\label{fig:thompson_scattering}
\end{center}
\end{figure}

Though of significantly smaller amplitude than the intensity fluctuations, there is also a polarized component to the CMB, sourced by the plasma density fluctuations. This polarized radiation is produced by Thompson scattering of photons in a thermal quadrupole. As shown in Figure \ref{fig:thompson_scattering}, if the radiation incident on an electron is hotter in one direction than in another direction, the scattered radiation in the direction orthogonal to the quadrupole will be partially polarized.

The polarization state of any light is commonly described by the Stokes parameters. For light traveling in the $z$ direction, they are defined as:
\begin{equation}
I = <a_x^2> + <a_y^2>,
\end{equation}
\begin{equation}
Q = <a_x^2> - <a_y^2>,
\end{equation}
\begin{equation}
U = 2a_x a_y \cos(\theta_x - \theta_y),
\end{equation}
\begin{equation}
V = 2a_x a_y \sin(\theta_x - \theta_y).
\end{equation}

Here $a_x$ and $a_y$ are the amplitudes of the electric field components, and $\theta_x$ and $\theta_y$ are the phase angles:
\begin{equation}
E_x = a_x(t) \cos(kz -\omega t - \theta_x),
\end{equation}
\begin{equation}
E_y = a_y(t) \sin(kz -\omega t - \theta_y).
\end{equation}

In this parameterization, $I$ is the total intensity, $Q$ and $U$ are the two linear polarizations, and $V$ is the circular polarization. Only linear polarizations are produced by Thompson scattering, so most experiments do not measure the $V$ component of the polarization state of the CMB.

The polarization fraction is given by:
\begin{equation}
P = \frac{\sqrt{Q^2+U^2}}{I},
\end{equation}
and the polarization angle is:
\begin{equation}
\psi = \frac{1}{2} \arctan{2}(U,Q),
\end{equation}
where $\arctan{2}(U,Q)$ is the two component signed arctangent:
\begin{equation}
\arctan{2}(U,Q) = 2 \tan^{-1} \left( \frac{Q}{\sqrt{U^2+Q^2}+U} \right).
\end{equation}

\subsection{Scalar and Tensor Perturbations}
\label{sec:e_and_b_modes}

\begin{figure}
\begin{center}
\includegraphics[width=3in]{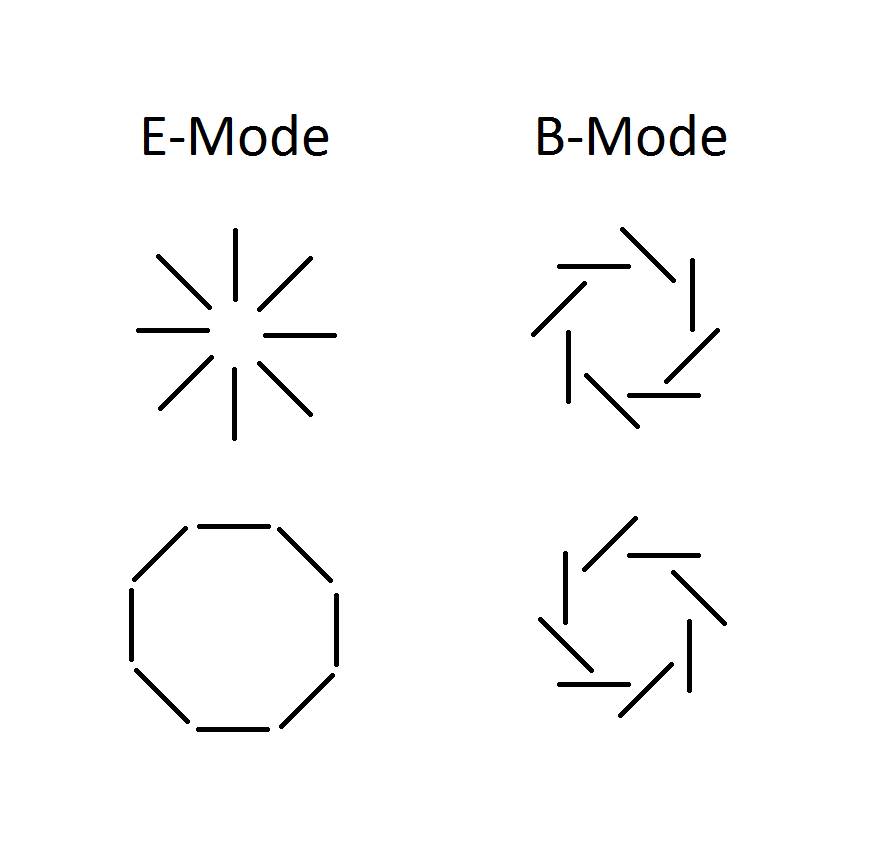}
\caption[E and B Polarization Modes]{Gradient only (E) and curl only (B) polarization modes. Scalar (density) perturbations in the primordial plasma can only produce E-mode polarization in the CMB. Tensor perturbations, produced by gravitational waves from the inflationary epoch, produce E and B-mode polarization. Gravitational lensing of the CMB by large scale structure also converts E-modes into B-modes, and is a foreground which must be accounted for in a detection of inflationary B-modes.}
\label{fig:e_and_b_modes}
\end{center}
\end{figure}

Since density perturbations are a scalar quantity, they have spin 0, and are invariant under a parity flip. Intuitively, this means that if they are viewed in a mirror, they remain unchanged. Therefore the polarization modes produced by the density perturbations in the plasma at the surface of last scattering are also invariant under a parity flip. This means these polarization modes have no curl component, only divergence. We therefore refer to them as E-modes, by analogy with electromagnetism (Figure \ref{fig:e_and_b_modes}). 

\begin{figure}
\begin{center}
\includegraphics[width=4.0in]{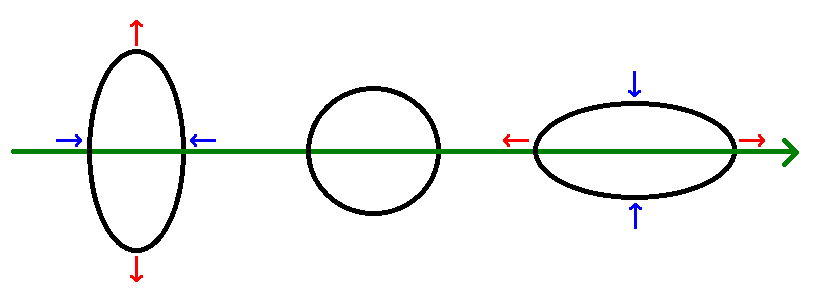}
\caption[Gravitational Wave]{Diagram of the metric and thermal distortions produced by a gravitational wave. A gravitational wave alternately compresses and rarefies the plasma along the dimensions orthogonal to its direction of travel, with the oscillations in the two axes 180 out of phase. This produces quadrupolar thermal perturbations, which result in partially polarized scattered emissions. Here the blue arrows represent compression, resulting in increased temperatures along the corresponding axis, while the red arrows represent rarefaction, resulting in decreased temperatures.}
\label{fig:gravity_wave}
\end{center}
\end{figure}

Perturbations to the metric during inflation are expected to produce gravitational waves, which will propagate through the primordial plasma. A gravitational wave produces time varying metric distortions as they propagate through the primordial plasma (Figure \ref{fig:gravity_wave}). In the plane orthogonal to the direction of travel, and at a given point in time, these distortions compress the plasma along one axis, and rarefy it along the orthogonal axis. As the wave propagates, these distortions oscillate, alternately compressing and rarefying the plasma along each axis, with the compression and rarefaction cycles in the orthogonal axes 180 degrees out of phase. This process produces quadapolar thermal perturbations at various points along the path of the wave, and thus produces polarized scattered emissions in the CMB. 

Since these gravity waves are tensor perturbations (spin 2), they are not generally invariant under a parity flip, and neither are the polarization modes produced by them. In general, the polarization modes produced by gravitational waves will have both gradient and curl components, in roughly equal amounts. By analogy with electromagnetism, these curl polarization modes are referred to as B-modes (Figure \ref{fig:e_and_b_modes}).  The only source of B-modes at the surface of last scattering is gravitational waves, and the only source capable of producing gravitational waves in the early universe is believed to be inflation, making primordial B-modes strong evidence of inflation. Their detection is therefore of great interest, and is currently one of the chief goals of CMB cosmology.

B-mode polarization is also produced at later times, when E-mode polarized light is gravitationally lensed by large scale structure in the Universe. This happens when CMB photons pass by or through galaxies or larger structures, and are gravitationally lensed around them. This gravitational process can distort the light in such a way that the divergence-only E-modes gain a curl component. These B-modes are called lensing B-modes, to distinguish them from the primordial B-modes produced by gravitational waves. The lensing B-modes constitute a foreground of higher amplitude than the primordial B-modes above $\ell \aprx 100$, which must be taken into account in searches for primordial B-modes. The first detection of lensing B-modes was made recently by the SPT \cite{hanson13}. SPT excels at mapping the lensing B-modes, due to its high angular resolution.  These lensing B-mode maps will be critical for delensing in future searches for primordial B-modes.

The E and B modes form a complete orthonormal basis for the second rank tensor space of polarization modes on a sphere, so the polarization state of a map can be written in terms of the E and B-modes in a similar fashion to the temperature anisotropies: 
\begin{equation}
Q(\hat{n})+i U(\hat{n}) = \sum_{\ell>0} \sum_{m=-\ell}^{\ell} (a_{\ell m}^{E} + i a_{\ell m}^{B}) \mathcal{Y}_\ell^m(\hat{n})
\end{equation}
where $\mathcal{Y}_\ell^m(\hat{n})$ are the spin-2 spherical harmonic functions, and $a_{\ell m}^{E}$ and $a_{\ell m}^{B}$ are the amplitudes of the E and B-modes.

We can also generalize the definition of the temperature $C_\ell$ parameter to:
\begin{equation}
C_{\ell}^{XY} = <a_{\ell m}^{X} a_{\ell m}^{Y*}>
\end{equation}
where $X,Y \in \{T,E,B\}$. Of chief importance are the autospectra ($C_\ell^{TT}$, $C_\ell^{EE}$, and $C_\ell^{BB}$), and the $C_\ell^{TE}$ cross-spectrum. Hu and White \cite{hu97d} show that in the standard cosmology, because of the parity transformations of E and B, the TB and EB cross-spectra are zero. There are possible non-standard cosmological effects (such as cosmic birefringence) which could induce non-zero TB and EB cross-correlations, and so measuring that these cross spectra are zero, to within measurement error, is an interesting cosmological constraint. 

Additionally, power leaking from temperature into polarization, or from E-modes into B-modes, due to imperfectly calibrated detector angles, or cross polar detector sensitivity, can lead to non-zero TB or EB correlations. The power in E-modes is substantially less than the power in T, and the power in B-modes is smaller still, making it important to control power leakage into E and B. 

\subsection{Constraining Inflationary Parameters with the CMB}

Gravity waves from inflation produce tensor perturbations in the primordial plasma, as discussed in the previous section, but inflation also produces scalar density perturbations in the primordial plasma. Since the inflaton field decays at the end of inflation, producing the matter and radiation density of the Universe, the shape of the potential influences the spectrum of density perturbations as well. This means both the tensor and scalar perturbations can be used to constrain inflationary models.

The spectrum of the primordial B-modes can in principle be used to infer the spectrum of tensor modes which produced them, which we parameterize as:
\begin{equation}
\Delta_t(k) = \Delta_t(k_0) \left( \frac{k}{k_0} \right)^{n_t}.
\end{equation}
Here $\Delta_t(k_0)$ is the amplitude of the tensor perturbations at the pivot scale $k=k_0$, and $n_t$ is the tensor spectral index, which is used to parameterize departures from scale invariance ($n_t = 0$ in the case of a scale invariant tensor spectrum), and $k_0$ is the pivot scale. 
The pivot scale is usually taken to be $k_0 = 0.002$Mpc$^{-1}$, which is the horizon scale at the time of last scattering. 
In effect however, given the small expected value of $r$, and the limited $\ell$ range over which the BB spectrum can be measured, constraints on $n_t$ are likely to be cosmic variance limited. 
It will therefore most likely not ever be possible to measure a non-zero $n_t$, or place more than loose constraints on the scale invariance of the tensor spectrum, if single-field slow-roll inflation is correct.

Similarly, the measured TT spectrum can be used to calculate the spectrum of the scalar modes:
\begin{equation}
\Delta_s(k) = \Delta_s(k_0) \left( \frac{k}{k_0} \right)^{n_s(k_0)-1},
\end{equation}
where $\Delta_s(k_0)$ is the amplitude of the scalar perturbations at the pivot scale $k=k_0$, and $n_s$ is the scalar index ($n_s = 1$ in the case of scale invariance). Currently the best-fit value, as given by the Particle Data Group, is $n_s = 0.958 \pm 0.007$ \cite{olive14}.

The tensor-to-scalar ratio is defined as:
\begin{equation}
r \equiv \frac{\Delta_t(k_0)}{\Delta_s(k_0)},
\end{equation}
and provides a convenient way of discriminating among different inflationary theories, which may predict different tensor-to-scalar ratios. The value of $r$ predicted by a theory will depend on both the amplitude of the inflationary potential, $V_\phi$, and the change in the amplitude of the inflaton field during inflation, $\Delta \phi$.

A measurement of non-zero $r$ would be strong evidence for inflation, because inflationary theories generically predict tensor fluctuations, while without inflation it is difficult to produce the metric perturbations that would be needed to account for tensor fluctuations. Lyth \cite{lyth97} shows that a measurement of $r$ would constitute a measurement of the energy scale of the inflaton potential
\begin{equation}
V_{\phi}^{1/4} \approx 10^{16}\mathrm{GeV} \left( \frac{r}{0.01} \right)^{1/4}.
\end{equation}

Furthermore, if we assume slow-roll inflation, the value of $r$ determines the change in the amplitude of the inflaton field during inflation
\begin{equation}
\frac{\Delta \phi}{M_\mathrm{P}} = \sqrt{\frac{r}{0.01}}.
\end{equation}
Here $M_\mathrm{P}$ is the Planck mass, and $\Delta \phi = \phi_\mathrm{\textsc{cmb}} - \phi_\mathrm{end}$, or the difference in the amplitude of the inflaton field between the time the CMB scale fluctuations passed out of the horizon during inflation, $\phi_\mathrm{\textsc{cmb}}$, and the time inflation ended, $\phi_\mathrm{end}$. 

The slow-roll parameters $\epsilon$ and $\eta$ are used to describe the shape of the inflaton potential \cite{peacock99}:
\begin{equation}
\epsilon = \frac{M_\mathrm{P}^2}{16\pi} \left( \frac{V'}{V} \right)^2,
\end{equation}
\begin{equation}
\eta = \frac{M_\mathrm{P}^2}{8\pi} \left( \frac{V''}{V} \right).
\end{equation}

\begin{figure}
\begin{center}
\includegraphics[width=\columnwidth]{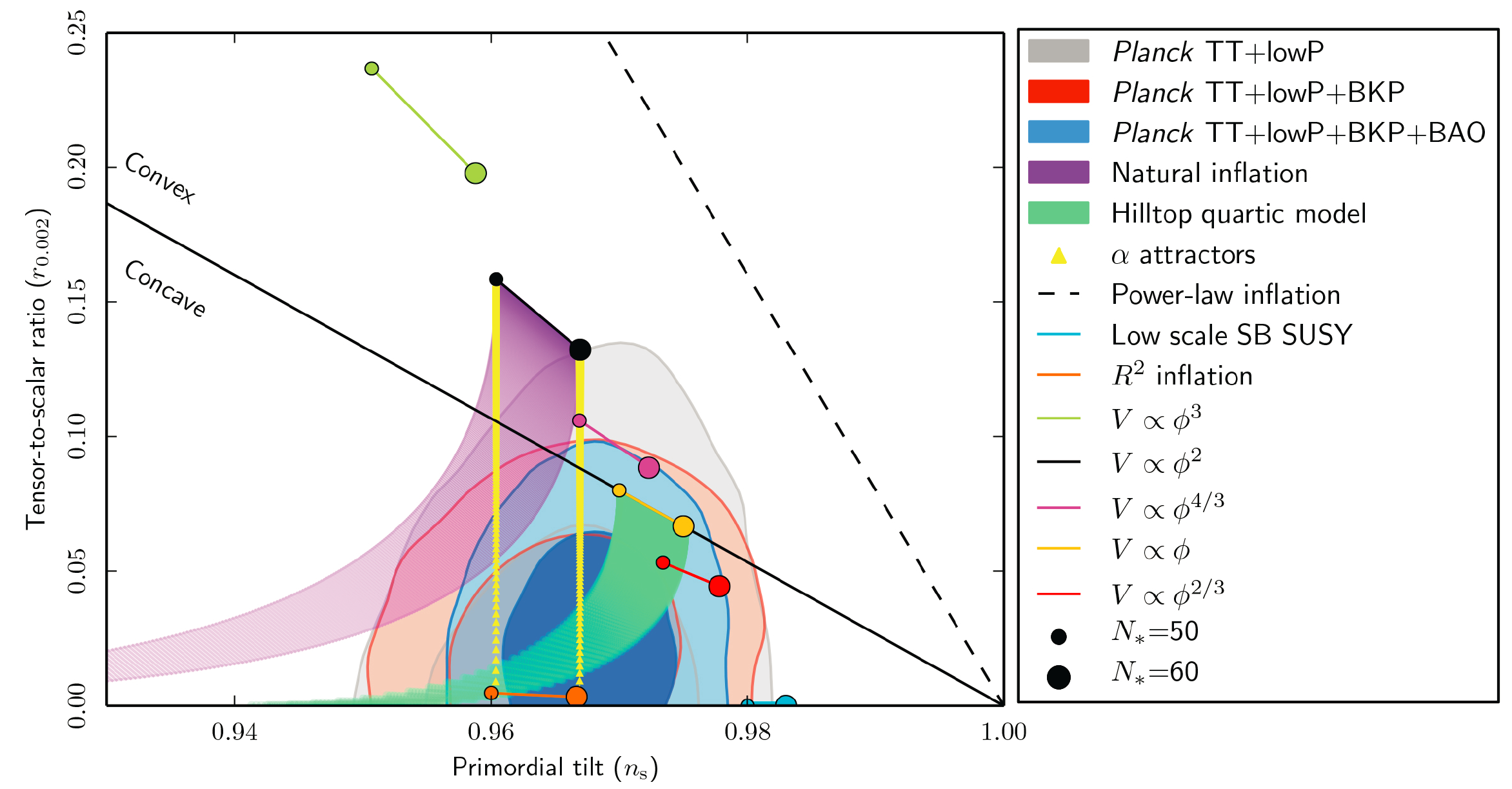}
\caption[Constraints on $r$ and $n_s$]{Current constraints on the inflationary parameters $r$ and $n_s$. Lighter shading represents $2\sigma$ constraints, darker shading $1\sigma$ constraints. Overplotted are the values of $r$ and $n_s$ predicted by various inflationary theories. We are beginning to rule out some models (e.g.\ $V \propto \phi^2$). 
The generation of experiments currently being developed expect to constrain $r \simlt 0.01$, or if $r > 0.01$ measure it with $\sigma(r) \approx 0.01$, assuming simple foreground models. 
Figure from Planck Collaboration \cite{planck15-20}.}
\label{fig:planck_r_vs_ns}
\end{center}
\end{figure}

The shape of the inflaton potential determines how the inflaton field decays during reheating, and governs the spectrum of density modes produced. It can be shown \cite{peacock99} that the $\epsilon$ and $\eta$ are related to the scalar spectral index by
\begin{equation}
1-n_s = 6\epsilon-2\eta.
\end{equation}

For most polynomial inflaton potentials, $\epsilon \approx \eta$, and \cite{peacock99} demonstrates that in this case $r$ and $n_s$ are related by
\begin{equation}
r \approx 3(1-n_s).
\end{equation}

The best current constraints on $r$ and $n_s$ are shown in Figure \ref{fig:planck_r_vs_ns}. The next generation of experiments, including SPT-3G, typically expect to constrain $r \leq 0.01$ at the $\aprx1\sigma$ level, for simple foreground models. Equivalently, if $r > 0.01$, they expect to measure $r$ with an error of $\sigma(r) \approx 0.01$. Constraining inflationary parameters, in addition to the standard \LCDM \ parameters is one of the major drivers behind the development of the SPT-3G experiment, which will be discussed in Chapters \ref{ch:spt_overview} and \ref{ch:tes_bolometers}.

\section{Cosmology in the Late Universe}

\subsection{Galaxy Clusters}

After photon decoupling, a long period followed in which very little optical light was produced anywhere in the Universe. These ``Dark Ages'' lasted several hundred million years, from the time of photon decoupling, $t=380,000$ years ($z = 1100$), to the formation of the first stars around 100-400 Myrs ($z$ \aprx 30-11). This period has not been studied directly yet, though future radio astronomy experiments hope to probe this era, notably the Square Kilometer Array (SKA) in South Africa \cite{taylor12, klockner12}.

During this time the neutral baryonic gas (chiefly hydrogen and helium) continued to gravitationally collapse into the dark matter potentials that began to form after the earlier dark matter decoupling. The length scale above which the baryonic gas can collapse is again given by the Jeans length (Equation \ref{eq:jeans_length}):
\begin{equation}
\lambda_\mathrm{J} = v_s \sqrt{\frac{\pi}{G \rho}}.
\nonumber
\end{equation}
Here $\rho$ is the density of the baryonic gas, and $v_s$ is the speed of sound in the gas. 

There are two competing models of how galaxy and star formation progressed. In the first model, the Jeans length falls rapidly after recombination to the order of tens of parsecs. 
Since the time it takes the gas to fall into a gravitational potential is smaller at lower length scales, the first structures which will form are ones at the Jeans length. Therefore in this model the first structures to form would be globular clusters, on the tens of parsecs scale. As the gas becomes denser, the Jeans length shrinks further and stars can form. Galaxies and eventually galaxy clusters are formed by mergers between smaller structures. This is called the bottom-up structure formation model.

In the alternate model, which is referred to as the top-down structure formation model, galaxy scale objects are the first to form, followed by smaller star cluster scale objects and eventually stars.
In principle these models could be distinguished by finding the locations of the first stars to form. In the bottom-up scenario, they would be located in globular clusters, while in the top down scenario they would be located in galactic cores. However, due to the expected extremely large size and resulting short lifespans of the first stars, they have yet to be observed. One recent study claims to have detected the first galaxy consisting largely of Pop-III stars, at a redshift of $z = 7$ \cite{sobral15}. In either scenario, galaxy clusters form through galaxy mergers.

Galaxy clusters are complex objects, and rich sources of information about cosmological parameters.
In particular, their number density and mass as a function of redshift provides information about the rate at which they collapse, and therefore about the competing forces of gravity (sourced principally by the dark matter density) and cosmic expansion.

Accurate and precise measurements of cluster masses are powerful tools for constraining dark matter density and the dark energy equation of state, and other cosmological parameters. Numerous methods of measuring galaxy cluster masses have been developed for this purpose, including gravitational lensing, measurements of the velocity dispersion of optically observed member galaxies, X-ray measurements of the temperature and luminosity of the intracluster gas that makes up the majority of the baryonic mass of clusters, and microwave measurements of the intracluster gas using interactions with CMB photons. 

In the Sunyaev-Zeldovich effect \cite{sunyaev72}, CMB photons inverse Compton scatter off of hot intracluster gas, resulting in a distortion in the observed CMB thermal spectrum. This is essentially a measurement of the line-of-sight integral of the pressure of the intracluster gas, and is expected to be a low scatter proxy for the cluster mass \cite{carlstrom02, kravtsov06a}. Chapter \ref{ch:ysz-m} describes a method I developed to infer the masses of galaxy clusters using the SZ effect.

\subsection{Reionization}

Once stars and other energetic objects such as quasars form, the Universe becomes luminous again, and the interstellar medium begins to be reionized. The process of the reionization of the Universe is still poorly understood, as it has not been thoroughly mapped, and depends on the details of early star formation, and complicated fluid physics in the interstellar, and intracluster media. In very rough terms it is believed to span from approximately 100-400 Myrs (30 \simlt z \simlt 11) \cite{baumann14}. After that reionization, there is only one more phase transition in the history of the Universe: the emergence of dark energy.

\subsection{Dark Energy}
\label{sec:dark_energy}

As the Universe expands, the energy density of matter falls as $a^{-3}$. Dark energy is a component of the Universe which seems to be a cosmological constant \cite{benson13,hou13b}. That is, the energy density of dark energy is constant in any fixed volume of space. At approximately $9$~Gyrs after the Big Bang, the energy density of dark energy exceeded the energy density of matter, and dark energy became the dominant form of energy density in the Universe. After this point, the dynamics of the Universe are chiefly determined by the dark energy. 

The pressure of a cosmological constant dark energy ($w = -1$) is:
\begin{equation}
\varepsilon_\Lambda = -P_\Lambda,
\end{equation}
and since the dark energy has negative pressure, this results in an accelerating expansion of space, as we saw previously in Equation \ref{eq:acceleration}:
\begin{equation}
\frac{\ddot{a}}{a} = \frac{8\pi G}{3 c^2}\varepsilon_\Lambda >0,
\end{equation}
and the Friedmann equation is the same as we found for the inflationary epoch:
\begin{equation}
\left( \frac{\dot{a}}{a} \right)^2 = \frac{8 \pi G}{3 c^2} \varepsilon_\Lambda = H_0^2.
\end{equation}
Solving for the scale factor we find:
\begin{equation}
a(t) = e^{H_0 t}.
\end{equation}
This equation implies that the Universe will continue expanding at an increasing rate indefinitely.

However, this is only true globally. On smaller scales, where matter had the opportunity to gravitationally collapse, and increase in density, matter remains the dominant form of energy density. This situation pertains on the scale of galaxy clusters and smaller. On these scales, since matter is the dominant form of energy density, and space is no longer expanding, dark energy will never become dominant. This means gravitationally bound systems will remain gravitationally bound, but the space between them will expand at an exponential rate. 

If we extrapolate this into the distant future, the space between galaxy clusters will eventually be expanding faster than the speed of light, resulting in islands whose futures are causally separated. That is, if the space between two points is expanding faster than the speed of light, they will no longer be able to communicate, and events at one point will not be able to affect future events at the other point.

This transition to dark energy dominance is the final phase transition in the Universe thus far, though it is not impossible that the Universe will undergo further phase transitions in the distant future.  One possibility is that the dark energy may not be a true cosmological constant, but instead a dynamic scalar field called quintessence \cite{tsujikawa13}. The energy density of the quintessence field results in an expansion of spacetime, similar to the inflaton field. There are many possible dynamics for such a field, but in some versions the quintessence field would eventually decay, resulting in another phase transition, and an end to the accelerating expansion phase of the Universe \cite{kamionkowski14}. Barring further events, the expansion of the Universe would then slow and eventually it would collapse. 

%However, in the opinion of this author, it is dangerous to extrapolate any theory to infinity. The only things we can speak of with certainty are the past, and the near future. (Though for a cosmologist the near future may be billions or trillions of years.) Beyond that, only time will tell.

%General overview of telescope and experiment generations

\doublespacing

\chapter{The South Pole Telescope}
\label{ch:spt_overview}

\subsubsection{South Pole Site}

The South Pole Telescope (SPT) is a 10-meter diameter off-axis Gregorian telescope with a 1 \sqdeg \ field of view, designed to operate at millimeter and submillimeter wavelengths \cite{carlstrom11}. The SPT is located at the Amundsen-Scott South Pole station, within approximately a mile of the geographic south pole. The SPT is situated in the Dark Sector of the Amundsen-Scott station, an optically and radio dark sector which has housed several other millimeter wave experiments including the BICEP \cite{keating03a,keating03b} and Keck experiments \cite{sheehy10}. 

The South Pole is an attractive site for a millimeter wave observatory because of the high altitude, low temperature, and extremely low precipitable water vapor. The South Pole is 2835 m (9301') above sea level, though the pressure elevation can exceed $3350$ m ($11,000'$) due to atmospheric thinning from the low temperatures, and atmospheric bulging at the equator. The temperature in the winter is approximately $-60\celcius$ ($-80\fahrenheit$), with record low temperatures of $< -80\celcius$ ($-110\fahrenheit$).  The elevation and temperature provide an exceptionally clear and stable atmosphere. The lack of solar perturbations during the six months of austral winter further reduces atmospheric activity. 

Winds are low, a $\aprx 5 \ \mathrm{m \ s}^{-1}$ katabaric wind typically blows from the East Antarctic Plateau \cite{schwerdtfeger84}. The highest recorded wind speed is only $24 \ \mathrm{m \ s}^{-1}$. Snow accumulation is only $\aprx 150 \ \mathrm{mm \ yr}^{-1}$, but snow drift formation around surface structures is considerable ($\aprx 3 \ \mathrm{m \ yr}^{-1}$). To reduce the drift accumulation, most buildings are elevated, including the Dark Sector Laboratory housing the SPT.

The precipitable water vapor is $< 0.32$  mm $75\%$ of the time \cite{lane98}, with a median precipitable water vapor of $0.25$ mm \cite{chamberlin01}. The low precipitable water vapor is essential for microwave observations, as water is the dominant source of atmospheric emissions in the microwave spectrum. The median brightness fluctuation power at \onefifty\ is $\aprx 31 \ \mathrm{mK}^2 \ \mathrm{rad}^{-5/3}$ in $\Tcmb$ \ \cite{bussmann05}. This is at least an order of magnitude better than other developed millimeter wave sites \cite{bussmann05,sayers10}.

The South Pole station is accessible principally by air, and only in the Austral summer, between approximately November 1 and February 14. Personnel and cargo are typically transported to the South Pole by LC-130 aircraft with 11,500 kg of cargo capacity, operated by the United States Air National Guard 109th Airlift Wing. The logistics chain for Antarctic operations runs out of Christchurch, New Zealand. From there C-17 Globemaster and LC-130 Hercules aircraft transport materials and personnel to the McMurdo base on Ross Island. Air transport to the South Pole is via LC-130, or the lighter DHC-6 Twin Otter which can land on unprepared snow for very early season transport. 

The winter fuel supply for the station is transported in LC-130s and on the ground in fuel bladders on sleds drawn by tracked vehicles. This supply train is referred to as ``the traverse''. It can take up to a month to travel the 1000 miles between McMurdo and the Amundsen-Scott Station, across the Transantarctic Mountains and the central Antarctic Plateau. Each year over 100,000 gallons of fuel are transported in two traverses.

\subsubsection{Instrument}

\begin{figure}
\includegraphics[width=\columnwidth]{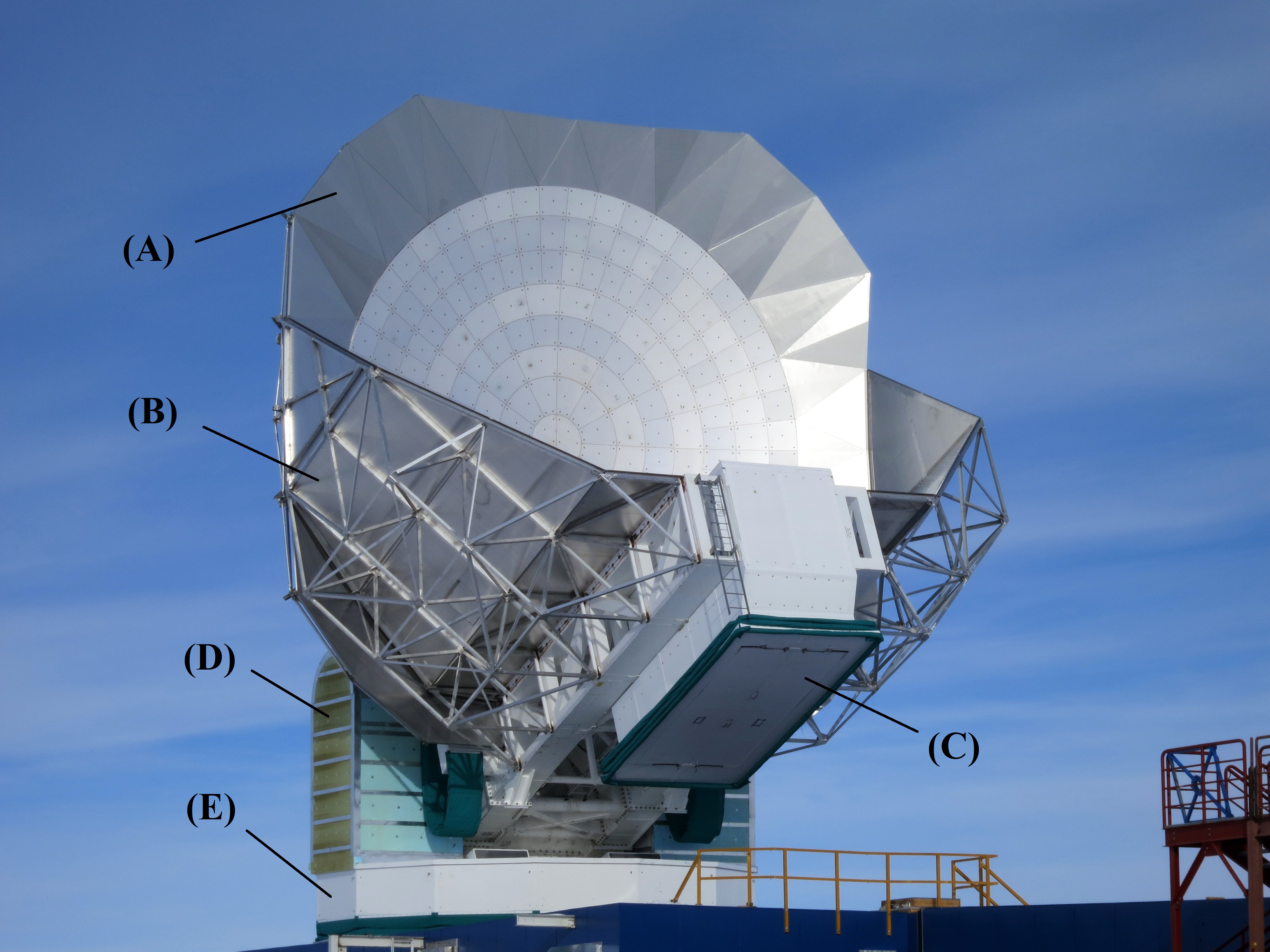}
\caption[The South Pole Telescope]{The South Pole Telescope in the Austral summer of 2013-2014. Labeled are (A) the primary mirror shield, (B) comoving side shields, (C) the receiver cabin, (D) the elevation yoke, and (E) the azimuth bearing. The receiver cabin houses the secondary mirror, optics and receiver cryostats, and readout electronics.  It can be docked against the roof of the Dark Sector Laboratory and accessed via the door panels on the bottom of the cabin. This provides a shirt-sleeves environment for \textit{in situ} work, and allows the cryostats to be lowered directly into the laboratory space when necessary.}
\label{fig:spt_primary}
\end{figure}

The SPT was constructed in 2006 and 2007, and saw first light in the Austral winter of 2007. It is described in detail in Carlstrom et al. \cite{carlstrom11}. Figure \ref{fig:spt_primary} shows the telescope as it was in the Austral summer of 2013-2014. The off axis Gregorian design of the SPT allows for unobstructed illumination of the primary mirror, minimizing noise from object in the beam path, and ground pick up. Comoving side shields further reduce ground pick up. The original optics included only two elements, a primary and secondary mirror. Having fewer optical elements has the advantage of reducing loss, scatter, and instrument polarization. 

The primary mirror is a 10 m diameter mirror with a 7 m focal length. It is constructed from 218 machined aluminum panels. 2 mm gaps between the panels (at $-80\celcius$) are covered with 5 mm wide BeCu strips with spring fingers to fix and center them in the panel gaps. After alignment of the panels, and with the BeCu strips in place, the surface smoothness is $23 \mu$m rms. To prevent snow and ice accumulation on the primary mirror, each panel is equipped with a 50 W m$^2$ electric heating pad on the back. The heating system collectively uses 4 kW of power, and raises the temperature $1-2\celcius$ above the ambient air temperature. This is enough to prevent ice deposition on the primary, and sublimate snow that blows onto it, without deforming the mirror. 

The secondary is a 1 m diameter aluminum 7075-T6 mirror, with $50 \mu$m rms. The secondary is contained in the optics cryostat, and cooled to $\aprx 10$K. The optics cryostat also contains a cold stop, made of microwave absorbing HR-10 foam, and cooled to 10 K. The optics cryostat is cooled by a model PT-410 pulse tube, with 80 W of cooling power at 70 K and 10 W of power at 10 K. 

\begin{figure}
\begin{center}
\includegraphics[width=\columnwidth]{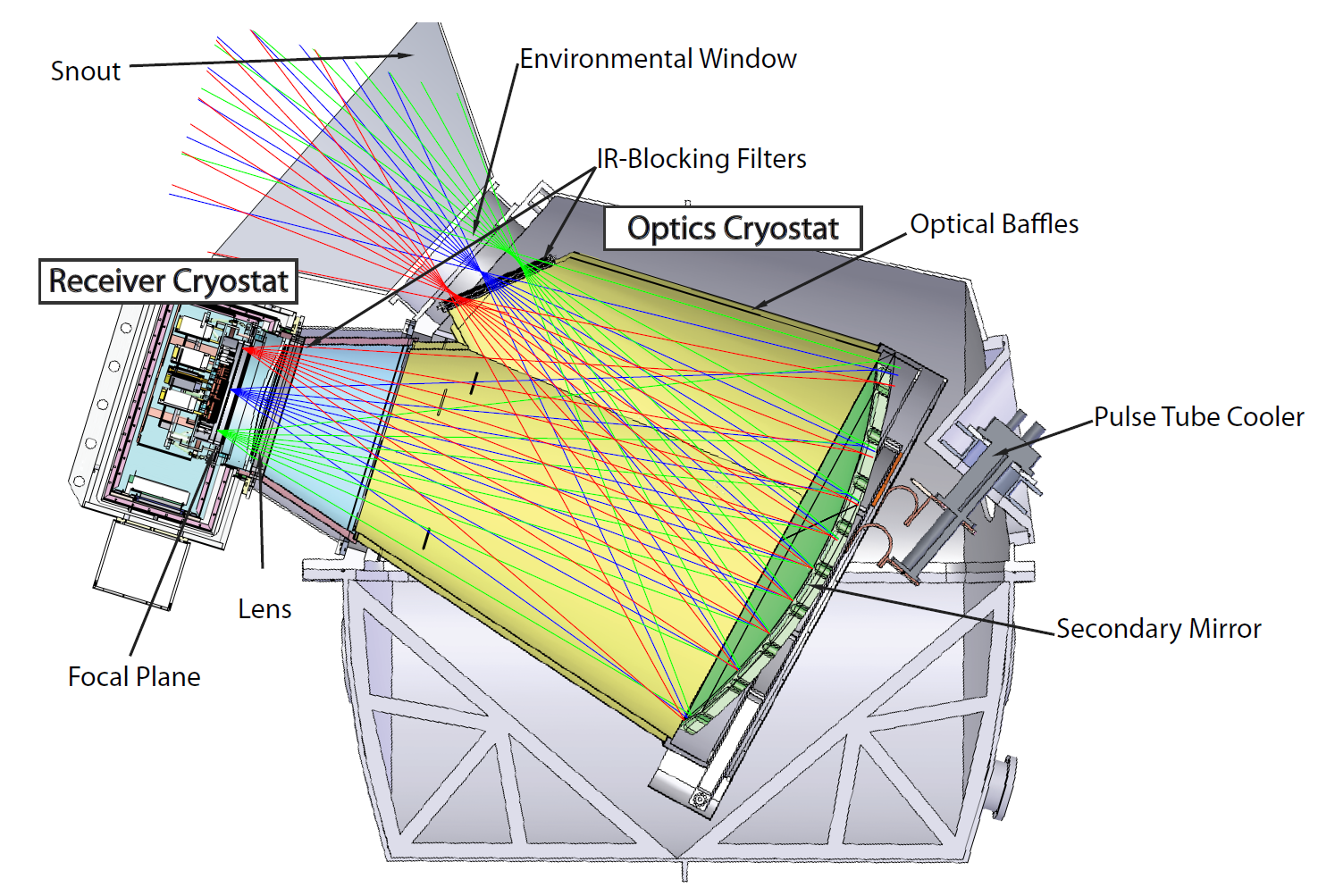}
\caption[SPT Optics and Receiver Cryostats]{The SPT optics and receiver cryostats. The two cryostats share a common vacuum space. A Zotefoam environmental window seals the vacuum of the cryostats, while resulting in minimal loss. IR blocking filters reduce loading on the secondary mirror and focal plane. A forebaffle, referred to as the snout, reduces the spillover to the primary shield. Figure from Sayre, 2014 \cite{sayre14}.}
\label{fig:spt_optics}
\end{center}
\end{figure}

The receiver cryostat is cooled by a PT-415 pulse tube, which provides approximately 40 W of cooling power at 45 K, and 1.5 W at 4.2 K. The focal plane is cooled by a $^4$He$^3$He$^3$He sorption refrigerator (model CRC10), which provides $80 \mu$W of cooling power at $380$mK, and $4 \mu$W of cooling power at 250mK. The helium sorption fridge takes $\aprx 3$ hours to cycle, and has a hold time of $\aprx 36$ hours. Figure \ref{fig:spt_optics} shows the optics and receiver cryostats.

The telescope mount rotates the instrument in azimuth and elevation, with pointing accuracy of a few arcseconds. The telescope can safely be driven at speeds up to $2^\circ$ s$^{-1}$, with a typical observing scan speed of $0.25^\circ$ s$^{-1}$ in azimuth. The total mass of the SPT is $\aprx 300,000$kg, with $20\%$ of this being the elevation drive counterweight.

\section{The SPT-SZ Experiment}

In 2007-2011, the SPT surveyed 2,500 \sqdeg \ in three frequency bands centered at 95, 150, and $220\,$ GHz. This survey is referred to as the SPT-SZ survey. 

\begin{figure}
\begin{center}
\includegraphics[width=\columnwidth]{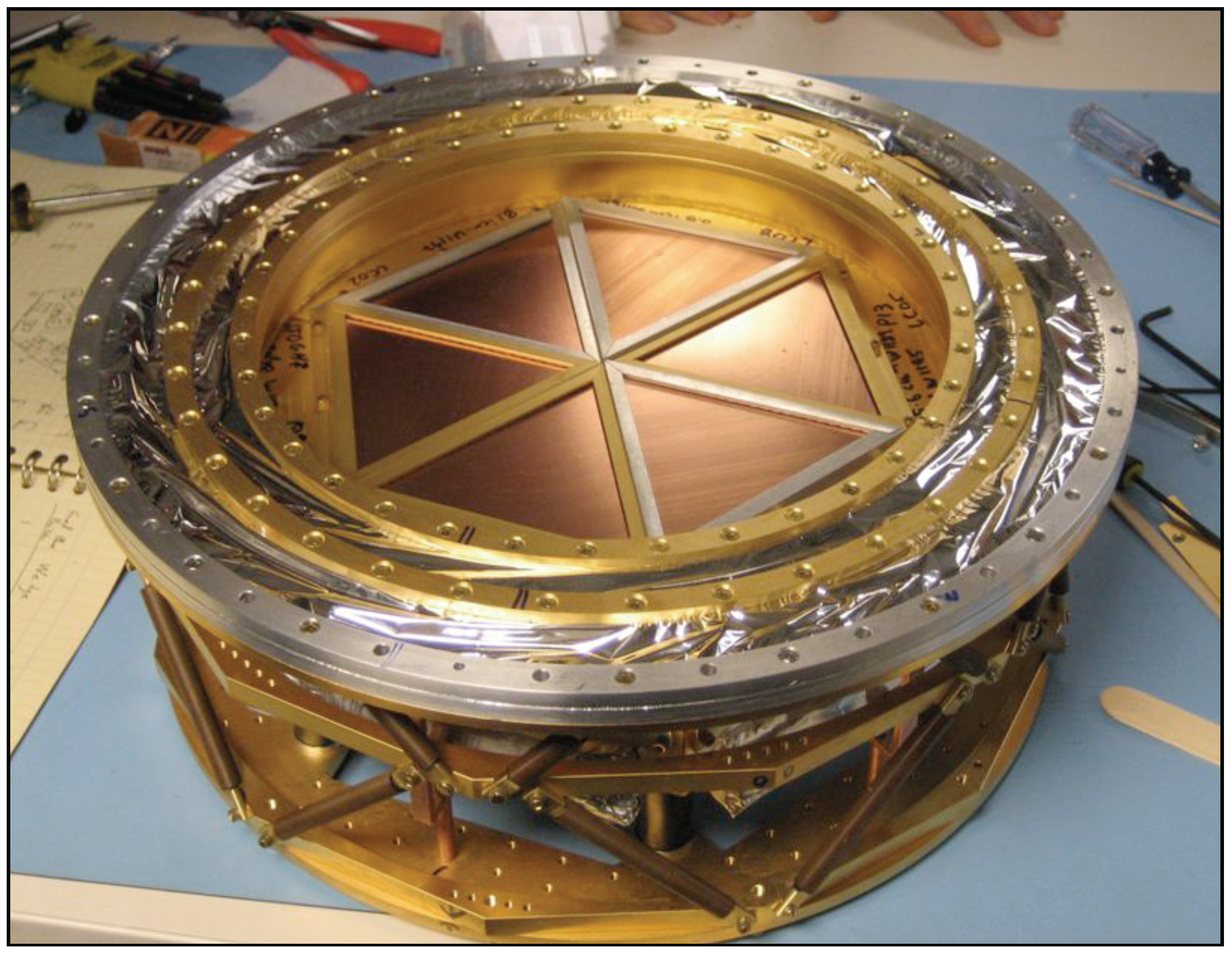}
\caption[SPT-SZ Focal Plane]{The fully integrated SPT-SZ focal plane. Each 100mm triangular wafer (located under the copper colored filters) contains 161 feedhorn coupled spider web bolometers. Circular waveguides set the low end of the bolometer frequency bands, while the metal-mesh filters visible over each wafer define the high end. The separately fabricated wafers allow different frequency pixels to be fielded in different wafers. Four of the wafers contain 150 GHz pixels, and the remaining two are 90 GHz and 220 GHz. The detectors and 300mK ultracold stage are thermally isolated with brown Vespel legs from an intermediate 500mK stage, which is similarly isolated from the main 4K stage in the cryostat. The foil surrounding the wafers is part of the RF shielding, protecting the detectors from radio frequency interference. Image from Benson, 2014 \cite{benson14}.}
\label{fig:spt-sz_focal_plane}
\end{center}
\end{figure}

SPT-SZ used a focal plane consisting of six triangular 100mm diameter wafers, each containing 161 total-power sensitive bolometers (Figure \ref{fig:spt-sz_focal_plane}). One triangular wafer contained 95 GHz detectors, one contained 220 GHz detectors, and four contained 150 GHz detectors. The back surface of the wafers were metalized to provide a $\lambda/4$ backshort. A smooth-wall conical feedhorn array coupled free space radiation into each detector array. The low ends of the detector frequency bands were defined by a circular single moded waveguide between the detectors and the feedhorns. The upper ends were defined by metal-mesh low-pass filters \cite{ade06}, mounted in front of the feedhorn arrays.

The pixels were spiderweb absorber devices, with AlTi Transition Edge Sensor (TES) bolometers \cite{gildemeister99,gildemeister00}. The antenna is a $\aprx 1\mu$m thick silicon nitride (SiN) mesh 3mm in diameter, coated with gold, and suspended by six 0.5mm long SiN legs. A gold layer is applied to the TES to increase the thermal time constant and prevent instability in the electrothermal feedback loop (see Section \ref{sec:responsivity}).

\begin{figure}
\includegraphics[width=\columnwidth]{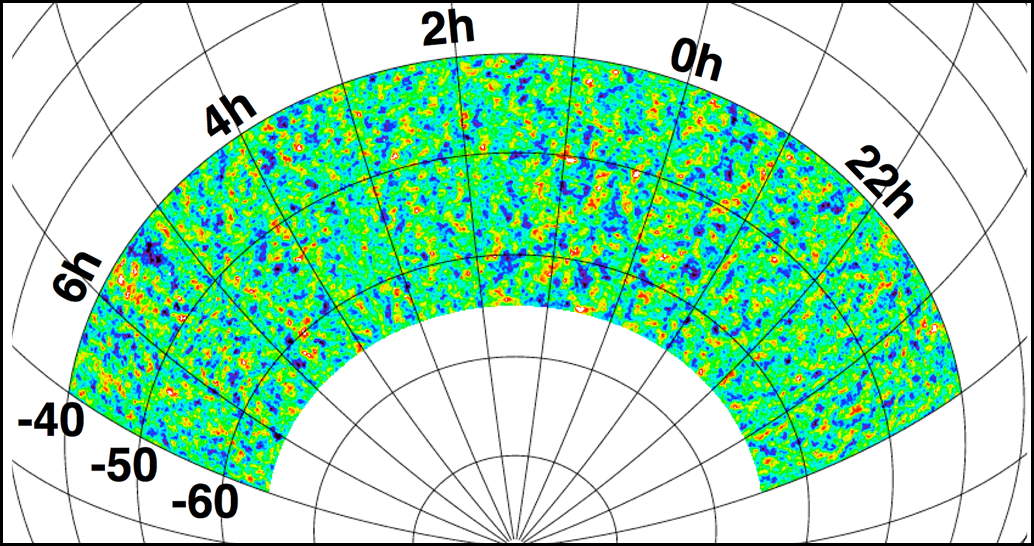}
\caption[SPT-SZ Survey Full Map]{The SPT-SZ survey 150 GHz full map. The SPT-SZ survey covers 2,500 \sqdeg \ of sky between 20h and 7h right ascension, and between $-40^\circ$ and $-65^\circ$ declination, to a survey depth of $\leq 18 \mu$K-arcmin.  Visible are the degree scale anisotropies of the CMB. Figure from Benson, 2015 \cite{benson15}.}
\label{fig:spt-sz_survey}
\end{figure}

\begin{figure}
\includegraphics[width=7.0in]{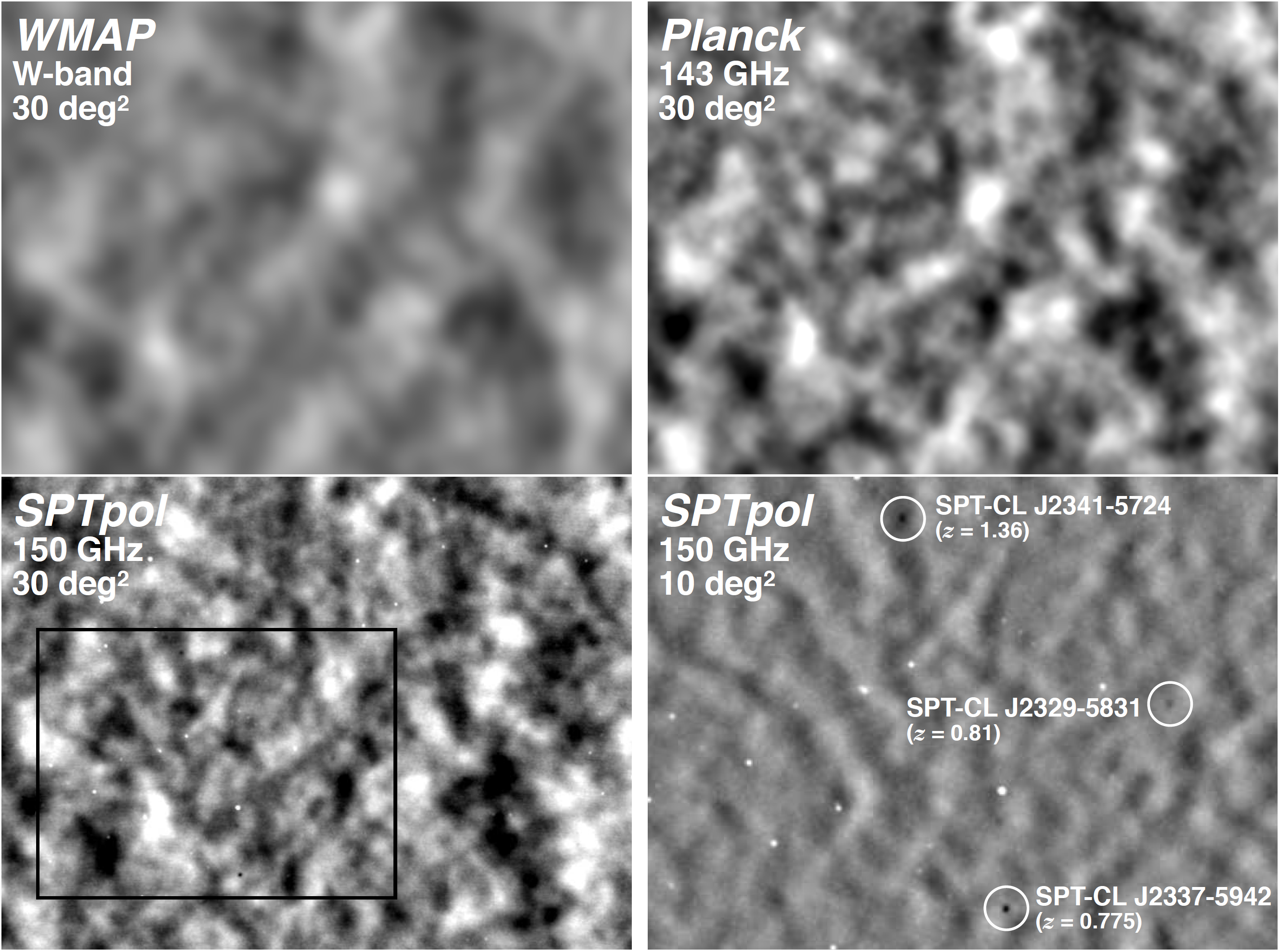}
\caption[Comparison of CMB Maps]{A comparison of CMB maps from several experiments: WMAP (top left), Planck (top right), and SPTpol (bottom). The WMAP and Planck maps are filtered at $\ell \aprx 50$ for comparison with the SPT maps. The bottom left map is a lightly filtered SPT map. The bottom right map shows a zoomed-in portion of the bottom left map (indicated with the black rectangle), and is spatially filtered to remove degree scale anisotropies, making sub-degree scale CMB anisotropies and SZ power more visible. Galaxy cluster are visible through the SZ effect in the filtered SPT map as arcminute scale flux decrements. The white points visible in both SPT maps are radio point sources such as Active Galactic Nuclei (AGN). Images from Benson et al., 2014 \cite{benson14}.}
\label{fig:cmb_map_comparison}
\end{figure}

The SPT-SZ instrument was deployed in 2007, and conducted a five-year survey, ending in 2011. In that time, SPT-SZ surveyed 2,500 \sqdeg \ of sky to a depth of $ \leq 18 \mu$K-arcmin at 150 GHz (Figure \ref{fig:spt-sz_survey}). The high angular resolution of the SPT optics, and the sensitivity of the instrument resulted in deeper and more precise maps of CMB anisotropies than previously had been obtained. The angular resolution of the SPT can be seen in Figure \ref{fig:cmb_map_comparison}, where SPTpol maps are compared to WMAP and Planck maps. The angular resolution of the SPT in particular was crucial for studies of galaxy clusters through the Sunyaev-Zel'dovich (SZ) effect (see Chapter \ref{ch:ysz-m}). The first galaxy clusters to be detected through the SZ effect were seen by the SPT in 2007 \cite{staniszewski09}. This has since become a valuable method of detecting galaxy clusters, particularly high redshift clusters. Traditional optical or X-ray surveys are limited by the $1/r^2$ fall off of the intensity of luminous objects (plus the Hubble expansion of the universe). The SZ effect is a scattering effect, its intensity depends not on the surface brightness of a galaxy cluster, but its gas mass and angular size. The integrated SZ flux, \YSZ, is conserved regardless of the distance to the galaxy cluster, allowing for detection of galaxy clusters out to high redshift. A cluster catalog selected by SZ flux will be highly complete above limiting mass and redshift thresholds, with relatively low contamination. 

This survey resulted in the detection of 516 optically confirmed galaxy cluster detections \cite{bleem15}, and in numerous improved constraints on cosmological parameters from power spectrum \cite{story12b,hou13b} and galaxy cluster \cite{reichardt13,benson13} measurements. It also led to new discoveries relating to submillimeter galaxies (SMGs) \cite{vieira10} and cool-core galaxy clusters \cite{semler12,mcdonald13}. A subset of the galaxy clusters detected in this survey will be analyzed in Chapter \ref{ch:ysz-m}, with the goal of measuring galaxy cluster integrated Comptonization, \YSZ, and constraining \YSZ-M scaling relations.

\newgeometry{top=0.25in, bottom=1.0in, left=0.75in, right=0.75in}
\newgeometry{top=1in, bottom=1in, left=1.5in, right=1in}
\doublespacing

\section{The SPTpol Experiment}

\begin{figure}
\begin{center}
\includegraphics[width=\columnwidth]{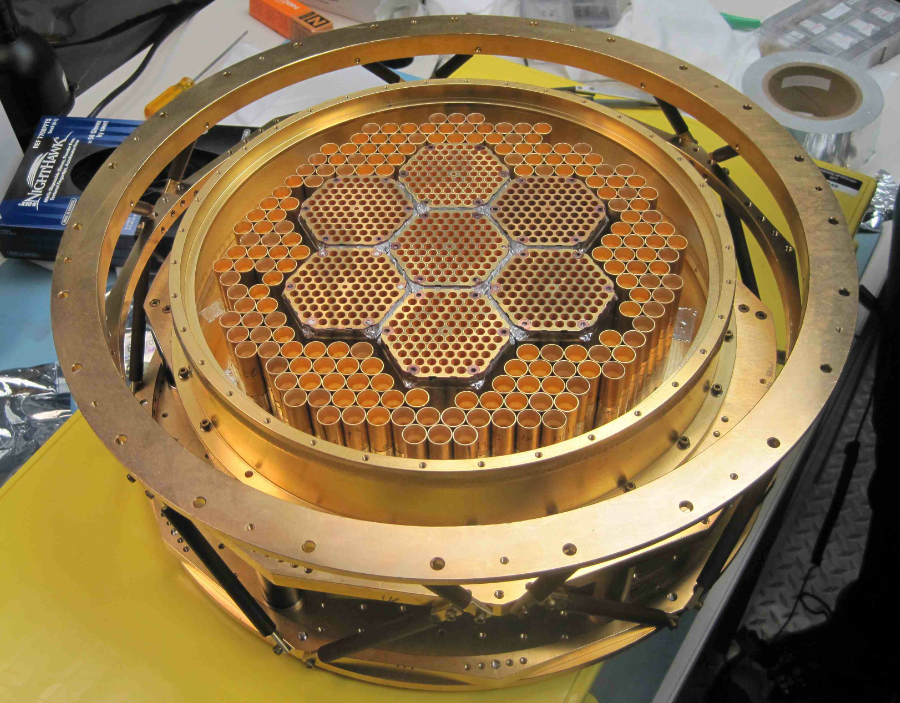}
\caption[SPTpol Focal Plane]{The SPTpol focal plane consists of seven \onefifty \ wafers, each containing 84 pixels, and 180 individually assembled 95 GHz pixels. Each pixel contains two bolometers for the two orthogonal linear polarization modes, for a total of 1,536 bolometers. The \onefifty \ pixels have corrugated silicon feedhorn arrays constructed in layers and gold plated to ensure a continuous high conductivity surface. The 95 GHz pixels have individually machined smooth wall feedhorns. Image from Benson, 2015 \cite{benson15}.}
\label{fig:sptpol_focal_plane}
\end{center}
\end{figure}

The second instrument deployed on the SPT was the SPTpol camera. This instrument consisted of seven wafers of \onefifty \ pixels, with 84 pixels per module, and 180 individually assembled 95 GHz pixels (Figure \ref{fig:sptpol_focal_plane}). The main improvement of this instrument over the previous SPT-SZ instrument was the addition of polarization sensitive bolometers. The SPT-SZ camera was equipped with total power sensitive absorbers, whereas in the detectors developed for the SPTpol instrument power is coupled into Orthomode Transducers (OMTs) which split the light into two orthogonal linear polarizations. Each polarization couples into a separate bolometer. With 768 pixels in the focal plane, there are a total of 1536 TES bolometers. The development of the \onefifty \ detectors is detailed in Henning et al. \cite{henning12}, and 95 GHz detectors in Sayre et al. \cite{sayre12}.

Austermann et al. \cite{austermann12} and George et al. \cite{george12} describe the overall SPTpol instrument development and characterization. Chapter \ref{ch:readout} describes the design and assembly of the cryogenic readout electronics for the SPTpol instrument. The increased number of bolometers for SPTpol necessitated a redesign of the readout electronics, and the development of cryogenic electronics with a higher multiplexing factor.

\begin{figure}
\begin{center}
\includegraphics[width=\columnwidth]{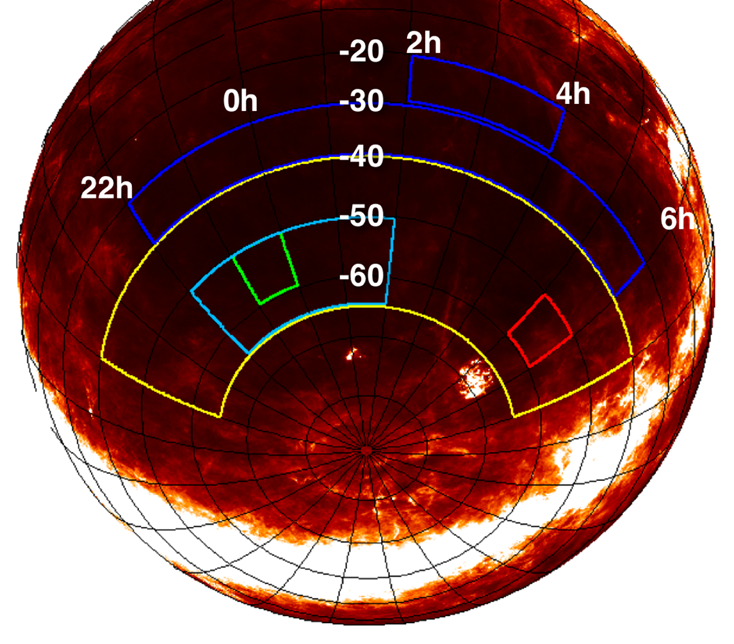}
\caption[SPT Survey Fields]{The SPT survey fields marked on an IRAS $100 \mu$m dust map. The South Celestial Pole is at the bottom of the image. Marked are the SPT-SZ survey field(yellow), the SPT-SZ east and west deep fields (red and green respectively), the SPTpol survey field (light blue), the SPTpol deep field (green), and summer survey fields (dark blue). Figure from Benson, 2015 \cite{benson15}.}
\label{fig:spt_survey_fields}
\end{center}
\end{figure}

The SPTpol instrument was deployed in 2011, and is scheduled to observe through the end of the 2015 season. The SPTpol survey field is a 500 \sqdeg \ subset of the SPT-SZ field (Figure \ref{fig:spt_survey_fields}). The main science goals of SPTpol focus on the polarization of the CMB, but it is also capable of continuing the broad science projects of SPT-SZ, enabled by its increased survey depth (Figure \ref{fig:cmb_map_comparison}). In particular, due to the rising cluster number density with decreasing mass, and the greater survey depth of the SPTpol survey, SPTpol is expected to detect a significant number of new galaxy clusters.

The linear polarization modes of the CMB can be decomposed into gradient modes and curl modes, as discussed in Section \ref{sec:e_and_b_modes}. These are commonly referred to as ``E'' and ``B'' modes in analogy with electromagnetism. Circularly polarized modes are not expected to be produced by any known physical processes in the early universe, and are not typically explored. The SPT and most other experiments are not equipped to measure circular polarization modes. 

As described in Section \ref{sec:e_and_b_modes}, E-modes are produced by Thompson scattering at the time of recombination, and also potentially by gravitational physics from the inflationary epoch. B-modes are only produced by gravity waves from the inflationary epoch, or by gravitational lensing.

The main science goal of the SPTpol experiment is the measurement of the EE and BB auto-correlation power spectra of the CMB. The first EE spectrum results were published in Crites et al. \cite{crites14}. Hanson et al. \cite{hanson13} show the first detection of B-modes in the CMB. These B-modes detected by SPTpol were shown to be correlated with large scale structure, and sourced primarily by gravitational lensing. 

\section{The SPT-3G Experiment}
\label{sec:3g_overview}

The third generation experiment on the SPT is the SPT-3G experiment. There are two primary innovations in this experiment. 

The first is the development of novel 3-band detectors. In the prior SPT experiments, each pixel was sensitive to only one frequency band. This was largely because the antennas had limited bandwidth, which was sufficient only for a single band. Chapter \ref{ch:tes_bolometers} describes the development and testing of broadband log-periodic sinuous planar antennas, which can couple to three frequency bands: 95 GHz, \onefifty, and 220 GHz.

The second improvement made for the SPT-3G experiment is the expansion of the focal plane. The $\aprx 4 \times$ increase in the area of the focal plane, coupled with the improvement in detector density, results in over an order of magnitude increase in the number of detectors fielded.

\begin{figure}
\includegraphics[width=\columnwidth]{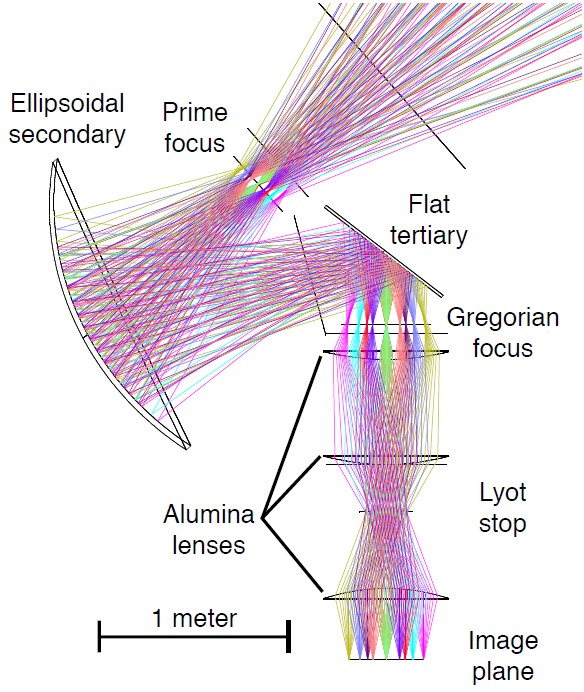}
\caption[SPT-3G Optics]{The reworked optics for the SPT-3G experiment. The secondary mirror is now warm (300K). Three cooled plano-convex alumina lenses have been added to illuminate a focal plane with $\aprx 4 \times$ the area of the SPT-SZ and SPTpol focal planes. Figure from Benson et al., 2014 \cite{benson14}.}
\label{fig:3g_optics}
\end{figure}

To accommodate the significantly larger focal plane of the SPT-3G instrument, the telescope optics are being completely reworked (Figure \ref{fig:3g_optics}). 
The new optics include a new warm aluminum elliptical secondary mirror and three cold alumina lenses. The lenses are contained in the optics cryostat which, as in previous generations, shares a vacuum space with the receiver cryostat. 

\begin{figure}
\includegraphics[width=\columnwidth]{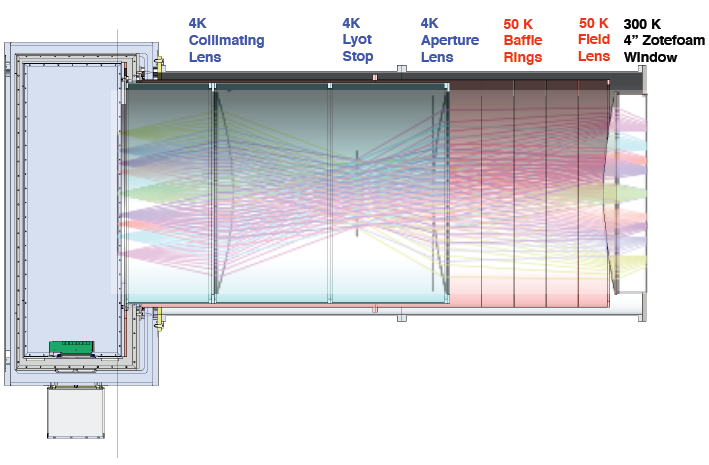}
\caption[SPT-3G Cryostat With Optical Elements]{Section view of the optics and receiver cryostats to show the optical elements. As in previous SPT instruments, the cryostat window is made of Zotefoam. The field lens and baffle rings are cooled to 50K, while the aperture lens, Lyot stop and collimating lens are cooled to 4K. Despite the warm secondary mirror, the thermal load from the telescope optics is projected to be reduced to 10K from 30K for the previous instruments. The new cryostat assembly is substantially longer (7.5') than the previous optics cryostats, and will be mounted vertically in the receiver cabin due to space constraints. Figure from Benson, 2015 \cite{benson15}.}
\label{fig:3g_cryostat_with_optics}
\end{figure}

Figure \ref{fig:3g_cryostat_with_optics} shows a section view of the optics and receiver cryostats, with the different elements labeled. The environment window for the optics cryostat is a 700mm diameter, 4'' thick Zotefoam window, supported from behind by a 50K alumina plate. The lenses are 720mm diameter alumina lenses, 50-65mm thick. There is also a 300mm diamter Lyot stop with 9 and 12 icm low pass filters. 

The field lens and baffle rings are cooled to 50K by the first stage of a PT-415 pulse tube. The aperture lens, Lyot stop, and collimating lenses are cooled to 4K by the second stage of the pulse tube. Despite the secondary mirror being warm in this version of the telescope, the thermal loading from the telescope optics is expected to be lower than in previous generations. The projected loading is 10K, as opposed to the 30K loading of the optics in SPT-SZ and SPTpol.

With the new optics in place, SPT-3G will have a $1.9^\circ$ diameter (2.8 \sqdeg) field of view. Strehl ratios will be $> 0.98$ in all three bands for the 430mm diameter focal plane.

\begin{figure}
\begin{center}
\includegraphics[width=4in]{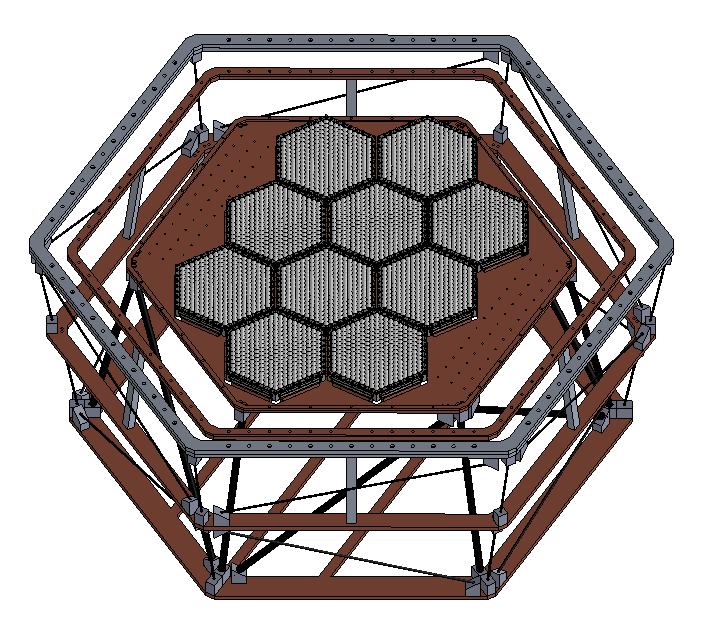}
\caption[SPT-3G Focal Plane]{The SPT-3G focal plane consists of ten 150mm (6'') hexagonal wafers, each containing 269 pixels. With TES bolometers in three frequency bands coupled to two orthogonal linear polarization modes in each pixel, the focal plane has a total of 16,140 detectors. Figure from Posada et al., 2015 \cite{posada15}.}
\label{fig:3g_focal_plane}
\end{center}
\end{figure}

The SPT-3G focal plane (Figure \ref{fig:3g_focal_plane}) consists of ten 150mm (6'') hexagonal wafers each containing 269 pixels. The wafers have hexagonal close packed pixels, ten on a side, minus two pixels per wafer for alignment marks which are necessary for processing. With two linear polarizations, and three frequency bands in each linear polarization, this results in 16,140 bolometers.

The expected Noise Equivalent Temperature (NET) in $\Tcmb$ of the SPT-3G focal plane is $\NETT = 3.4 \mu \mathrm{K} \sqrt{s}$ in temperature, and $\NETP = 4.8 \mu \mathrm{K} \sqrt{s}$ in polarization, including all frequency bands. See Table \ref{tab:net_and_speed} for a comparison of the NET values and mapping speeds of the SPT instruments. Due to the increased number of detectors deployed, and the decreased thermal loading, the expected mapping speed of SPT-3G is $17 \times$ higher than its predecessor SPTpol, in both temperature and polarization. 

\begin{table*}
\begin{center}
{
\caption{NETs and Mapping Speeds}
\begin{tabular}{l|c|cc|cc}
\hline \hline
 Experiment & N$_\mathrm{bolo}$ & \NETT & Speed$_\mathrm{T}$ & \NETP & Speed$_\mathrm{P}$ \\
          & & ($\mu \mathrm{K} \sqrt{s}$) & & ($\mu \mathrm{K} \sqrt{s}$) &  \\
\hline
 SPT-SZ & 960 & 22 & 1.0 & - & - \\
 SPTpol & 1,536 & 14 & 2.5 & 20 & 1.0 \\
 SPT-3G & 16,140 & 3.4 & 43 & 4.8 & 17 \\ \hline
\end{tabular}}
\begin{tablenotes}
Note -- Comparison of Noise Equivalent Temperature (NET) and mapping speed in temperature (and polarization where applicable) for the SPT-SZ, SPTpol, and SPT-3G experiments. Mapping speed in temperature is normalized to the SPT-SZ experiment, and mapping speed in polarization is normalized to SPTpol.
\end{tablenotes}
\label{tab:net_and_speed}
\end{center}
\end{table*}

This improved mapping speed will allow SPT-3G to reach a survey depth of $\aprx 3.5 \mu$K-arcmin in both $E$ and $B$ at \onefifty, and $\aprx 6 \mu$K-arcmin in $E$ and $B$ at 95 and 220 GHz. This is assuming a four-year survey over 2500 \sqdeg, with roughly the same footprint as the original SPT-SZ survey. The choice of survey area and location are motivated by maximizing the depth of the survey and improving constraints on $r$, which drive us toward a smaller area in a low foreground region, and improving constraints on $\Sigma m_\nu$, which drives us to larger survey area.

\begin{figure}
\begin{center}
\includegraphics[width=\columnwidth]{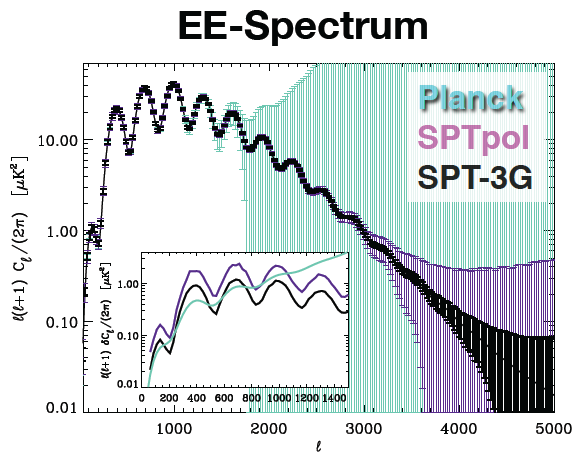}
\caption[SPT-3G Forecast EE Spectrum]{Forecast EE power spectrum for SPT-3G, for four years of observing, compared to the results from Planck and SPTpol. The constraints are from simulated observations including foregrounds, atmosphere, instrument $1/f$ noise, and E-B separation. The inset shows the low-$\ell$ EE uncertainty in the three experiments, showing SPT-3G will be competitive with \textit{Planck} down to $\ell \aprx 200$. SPT-3G will excel, however, at high multipoles ($\ell \simgt 3000$). Figure from Benson et al., 2014 \cite{benson14}.}
\label{fig:3g_ee_spectrum}
\end{center}
\end{figure}

\begin{figure}
\begin{center}
\includegraphics[width=\columnwidth]{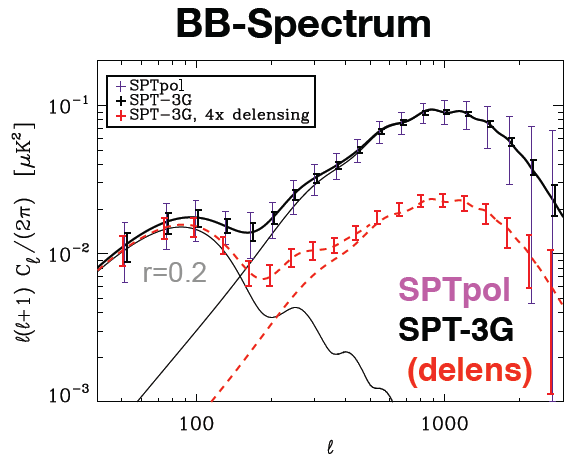}
\caption[SPT-3G Forecast BB Spectrum]{Projected BB spectrum constraints from the SPT-3G experiment. In purple are the expected SPTpol constraints, with SPT-3G in black. With a conservative estimate of $4\times$ delensing, SPT-3G expects to reach the constraints shown in red. SPT-3G expects to achieve a one sigma constraint on the tensor-to-scalar ratio, $r$, of $\sigma(r) = 0.01$. The constraints are from simulated observations including foregrounds, atmosphere, instrument $1/f$ noise, and E-B separation. Figure from Benson et al., 2014 \cite{benson14}, which also includes a more detailed description of the simulations.}
\label{fig:3g_bb_spectrum}
\end{center}
\end{figure}

SPT-3G will improve significantly on the EE and BB power spectra measured by SPTpol. Forecast EE and BB power spectra are shown in Figures \ref{fig:3g_ee_spectrum} and \ref{fig:3g_bb_spectrum} respectively. The E-mode polarization mapping will be exquisite. We expect to achieve a $150 \sigma$ detection of gravitational lensing of the CMB. This will also allow us to effectively delens the CMB, which is crucial for investigations of the tensor-to-scalar ratio. Additionally, the E-mode damping tail is expected to become foreground dominated at a significantly higher $\ell$ than the temperature damping tail, because of the low polarization of dusty point sources \cite{seiffert07}. Our high-resolution, low-noise measurements will allow us to use the E-mode damping tail to constrain the effective number of relativistic particle species, the primordial helium abundance, and the running of the scalar spectral index.

Including Planck priors in our analysis, we expect to place strong limits on the effective number of relativistic species and on the sum of the neutrino masses. We expect to achieve $\sigma(N_\mathrm{eff}) = 0.076$ and $\sigma(\Sigma m_\nu) = 0.06$eV. This constraint on $\Sigma m_\nu$ is approximately six times stronger than future beta decay experiments such as KATRIN \cite{wolf10}, and roughly the size of the largest neutrino mass splitting. SPT-3G will therefore either measure $\Sigma m_\nu$, and determine the mass scale for neutrinos, or put constraints on $\Sigma m_\nu$ that strongly disfavor the inverted neutrino hierarchy.

Using the potential of our E-mode maps for delensing, we anticipate being able to reduce the B-mode power from gravitational lensing by a factor of four. This will enable us to constrain $\sigma(r)=0.01$, or place a $95\%$-confidence upper limit of $r < 0.021$ in the case of no detection.  This estimate includes foregrounds, atmospheric and instrumental noise, E-B separation, and delensing \cite{benson15}.

\begin{figure}
\begin{center}
\includegraphics[width=\columnwidth]{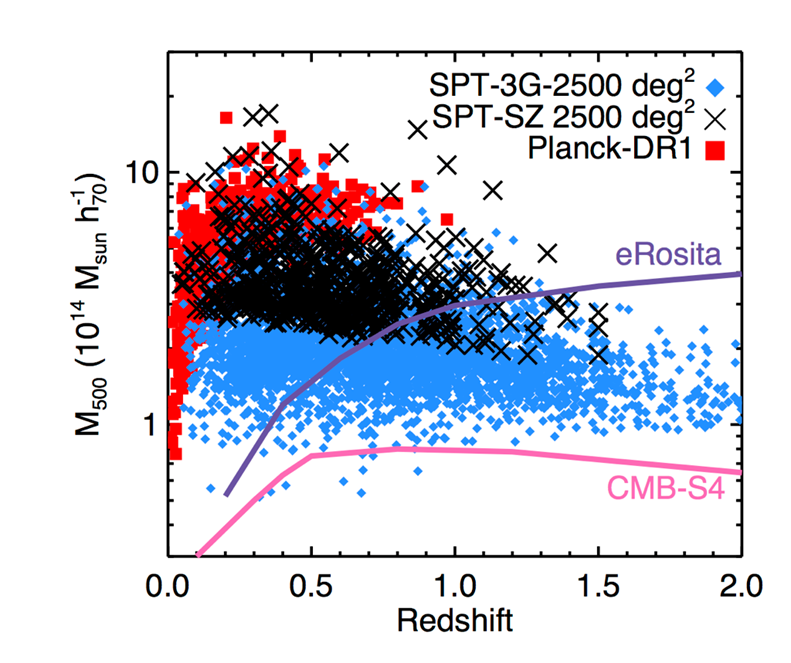}
\caption[SPT-3G Forecast Galaxy Cluster Detections]{Forecast of the expected number and mass-redshift distribution of galaxy clusters detected with the SPT-3G experiment. The clusters detected by the Planck experiment are shown in red. In black are the SPT-SZ 2,500 \sqdeg survey cluster detections. Approximate mass-redshift limits for eRosita and a CMB-S4 class experiment are shown in purple and pink respectively. SPT-3G is expected to discover an order of magnitude more clusters than SPTpol including an unprecedented number at high redshifts ($z > 1$). Figure from Benson et al., 2014 \cite{benson14}.}
\label{fig:3g_cluster_detections}
\end{center}
\end{figure}

Since the SZ effect allows us to detect essentially all clusters down to a mass limit, and given the rising number density of clusters at lower masses, the improved sensitivity of the SPT-3G survey will allow us to detect many more clusters than previous experiments (Figure \ref{fig:3g_cluster_detections}). The SPT-3G survey will achieve noise levels \aprx 12, 7, and 20 times lower than SPT-SZ in 95, 150 and 220 GHz, over the same survey area. This survey should detect \aprx10,000 galaxy clusters with signal-to-noise $> 4.5$, an order of magnitude improvement over the \aprx1,000 clusters detected in previous SPT surveys. Because these detections will reach to lower masses, they will also extend to higher redshifts than previous detections. The high median redshift of the SPT-3G cluster sample will make it an important complement to future optical surveys such as the Dark Energy Survey (DES) \cite{sanchez10}, which will partially overlap the SPT-3G survey area.

\section{Summary of Work on the SPT}

Unless otherwise specified, the work in the following chapters was performed by this author, including the detector testing in Chapter \ref{ch:tes_bolometers}, the design of the frequency multiplexing schedule and the assembly and testing of the cryogenic multiplexing electronics in Chapter \ref{ch:readout}, and the development of the \YSZ\ methods in Chapter \ref{ch:ysz-m}. I was involved with a number of smaller projects on the South Pole Telescope as well, which are not detailed here due to limitations of time and space, but which I will summarize briefly. 

I was involved with the design and assembly of a forebaffle for the SPT, located before the optics cryostat window and secondary mirror. This baffle, commonly referred to as the ``snout'' was deployed on the telescope in the Austral summer of 2011-2012.

In the Austral summers of 2012-2013 and 2013-2014 I was deployed to the South Pole for maintenance, calibration, and observations. In the 2012-2013 season I participated in the polarization calibration of the SPTpol instrument. In the 2013-2014 season I worked on improving the local mapmaking and analysis software which allows the winter season telescope operators to monitor the data quality on a daily basis, and quickly detect issues affecting the operation of the telescope and the final data products. In both seasons I performed regular maintenance on the telescope, and ran scheduled summer observations.

At Case Western, I worked to design a new pulse tube cryostat to replace the current wet cryostat as the primary testbed for SPT detector testing at Case Western, and possibly supplement our testing capabilities for the SPIDER experiment.

%TES Bolometer Design

\doublespacing

\chapter{Multichroic TES Bolometers}
\label{ch:tes_bolometers}

TES bolometers are photon shot-noise limited; one way to increase sensitivity is to increase the number of detectors deployed. The other two means of improving sensitivity are increasing optical efficiency, and to decrease non-sky loading. Work is also in progress to reduce non-sky loading for SPT-3G. For the SPT-3G experiment, improvements are being made in two avenues to increase the number of detectors in the instrument. The optics of the South Pole Telescope are being reworked, as discussed in Section \ref{sec:3g_overview}, to increase the area of the focal plane, and thus the real estate available for detectors. Additionally, novel 3-band detectors are being developed, which result in a factor of three increase in the density of detectors per unit area on the focal plane. See Benson et al. \cite{benson14} for a more thorough description of the SPT-3G experiment, and Posada et al. \cite{posada15} for details of the detector design and fabrication.

\section{Detector Structure}
\label{sec:detector_structure}

\begin{figure}
\begin{center}
\includegraphics[width=4.25in]{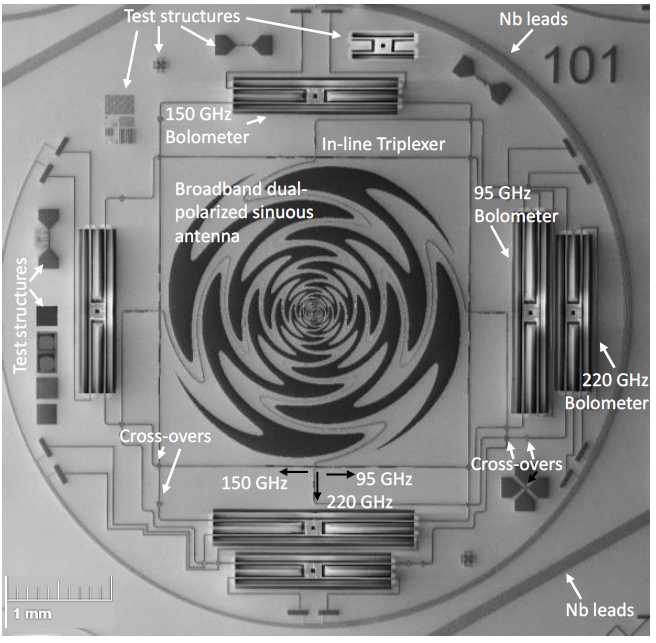}
\caption[Three-Band Multichroic Pixel]{Scanning Electron Microscope (SEM) image  of a 3-band multichroic detector for SPT-3G. The antenna is a two-octave bandwidth log-periodic sinuous planar antenna. The vertical and horizontal arms couple to orthogonal linear polarization modes. The antenna is connected to the detectors via microstrip lines which run along the back of the antenna arms, using them as a ground plane. The GHz signal is split into three bands with a lumped element triplexer, and deposited on three separate bolometers. The bolometer converts varying GHz power into variations in the temperature, and thus resistance, of the TES. This signal is then read out as variations in the current through the TES. Figure from Posada et al., 2015 \cite{posada15}.}
\label{fig:pixel_design}
\end{center}
\end{figure}

The multichroic detectors developed for SPT-3G use a two-octave bandwidth log-periodic sinuous planar antenna (See Figure \ref{fig:pixel_design}). This structure allows radiation from across our three frequency bands of 95, 150, and 220 GHz to couple to a single antenna. The two arms of the antenna couple to orthogonal linear polarization modes. The antenna also has a self-complementary structure, in which the metal and open areas have the same geometry and area. This produces an antenna impedance independent of frequency. 

The disadvantage to this design is that the plane of the polarization rotates periodically with frequency, with a $5^\circ$ amplitude. This effect is ameliorated in several ways. First, the variation in polarization angle is smaller over the $22.5\%$ fractional bandwidth of each frequency band. The average polarization rotation can also be smaller depending on the source spectrum, i.e. if there is not equal power at all frequencies. The expected polarization angle change between dust and CMB source spectra is $0.2^\circ$, which is small enough not to affect dust subtraction at nominal dust levels. Additionally, the focal plane will consist of equal numbers of left and right handed detectors, so that the array composite beam will have no net polarization angle rotation, assuming equal weights.

Each pair of antenna arms connects to the detectors via microstip lines which run along the back of the antenna arms, using them as a ground plane. A lumped element triplexer splits the GHz signal into three frequency bands, and the power from each is deposited on a separate TES bolometer. The bolometer converts varying GHz power into variations in the temperature of a resistive element (the TES) held in a superconducting transition. Small changes in the temperature of the TES induce strong variations in resistance. The TES is voltage biased, and the variation in resistance can then be read out as changes in the current through the TES.

Radiation is coupled into each antenna by a silicon lenslet 6mm in diameter, with a three layer broadband anti-reflection (AR) coating.
The beam of the antenna-lenslet system is determined by diffraction, with an aperture the size of the lenslet. The lenslets are registered to the antennas with circular depressions etched into a spacer wafer, which directly contacts the bolometer wafer. The spacer wafer is in turn optically aligned to the bolometer wafer with an accuracy of $< 0.01 \lambda$, and clamped into position with an invar clip.

\begin{figure}
\begin{center}
\includegraphics[width=5in]{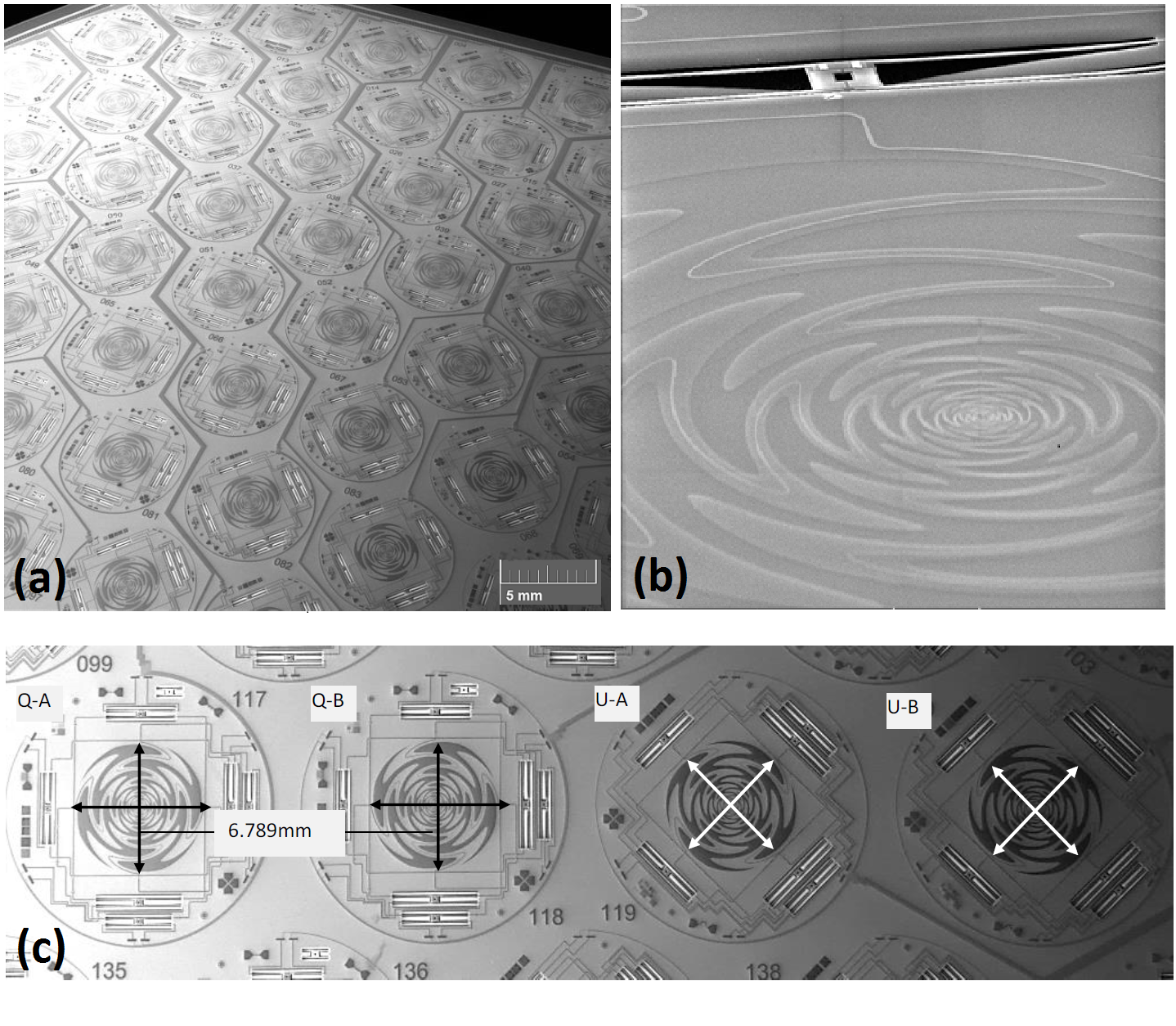}
\caption[Detector Wafer and Detail Views]{(a) A 6'' hexagonal wafer of 3-band multichroic sinuous antenna detectors. Surrounding each antenna are six TES bolometers, three for each polarization axis. (b) A close up of the sinuous antenna, showing the microstrip on the back of the antenna plane. (c) Q and U polarization pixels. For each polarization there is an A and B variant with opposite chirality of the antenna, for the suppression of systematic errors associated with the antenna chirality, such as polarization rotation. Images from Posada et al., 2015 \cite{posada15}.}
\label{fig:wafer_collage}
\end{center}
\end{figure}

Once detectors are fabricated (see Figure \ref{fig:wafer_collage}), there are a wide variety of thermal, electrical, and optical properties which must be measured to ensure that they are on target, and that variations in properties across detector wafers are acceptable. For SPT-3G this work is performed at three primary testing institutions, Case Western Reserve University, The University of Chicago, and The University of Colorado. At CWRU the testing is performed in a wet cryostat, with 77K and 4K stages cooled by liquid nitrogen and liquid helium respectively. A Simon-Chase He10 refrigerator cools an intermediate stage to \aprx 500mK, and an ultra-cold stage to \aprx 270mK. This cryostat is affectionately referred to as the Blue Dewar, to distinguish it from the other cryostats and dewars in the laboratory, all of which are blue. The results presented in this chapter are from data collected by me at Case Western, unless otherwise specified.

\section{Bolometer Thermal and Electrical Properties}
\label{sec:steady_state}

\begin{figure}
\begin{center}
\includegraphics[width=3in]{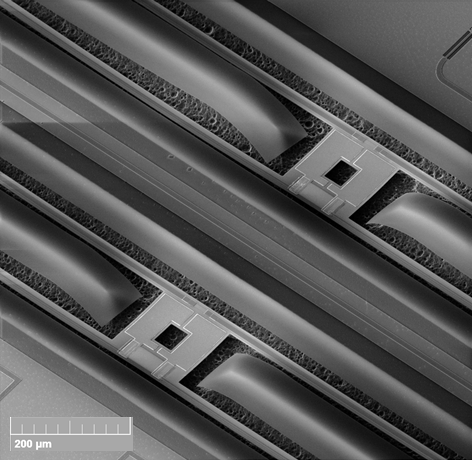}
\includegraphics[width=3in]{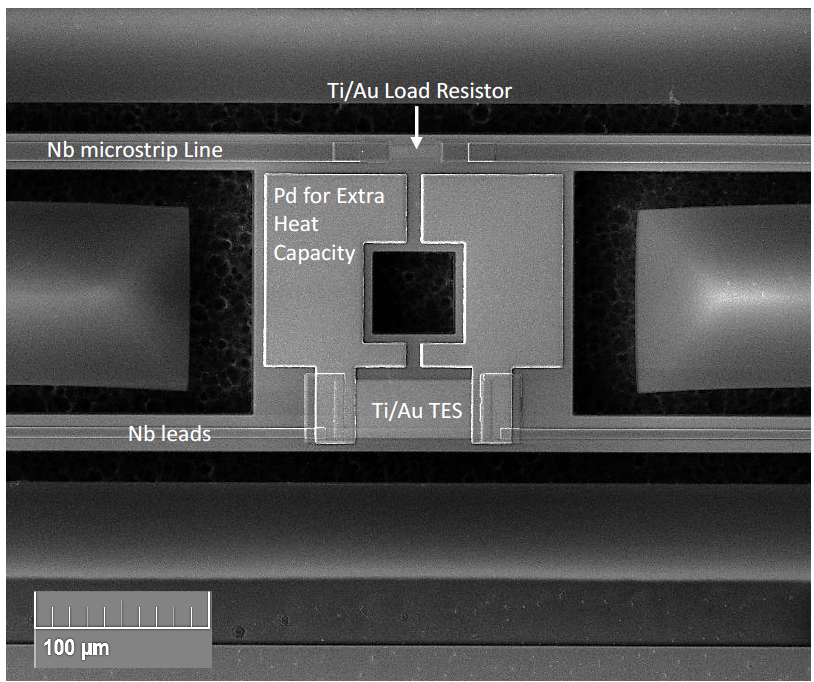}
\caption[Bolometer Structures]{(left) SEM image of two bolometers. The bolometer islands can clearly be seen suspended on SiN legs above the Si wafer. The island is released by XeF$_2$ chemical etching of the Si substrate beneath the island and legs. The hole in the center of the wafer is to improve the efficiency of the chemical etch under the island. (right) Close up SEM image of bolometer island. The TiAu load resistor is at the top of the island, and the TiAu TES is at the bottom. The Pd layer overlaps the TES to increase its heat capacity. The split shape of the Pd layer prevents it from increasing the TES resistance. The microstrips for GHz power, and the traces for the readout system are all Niobium, which is superconducting at the operational temperature of 500mK. Images from Posada et al., 2015 \cite{posada15}.}
\label{fig:TES_SEM}
\end{center}
\end{figure}
\doublespacing

Power from the antenna is coupled to the bolometer by means of a load resistor. 
The bolometer is suspended on long silicon nitride (SiN) legs, thermally isolating it from the rest of the detector wafer, which acts as a thermal bath (See \ref{fig:TES_SEM}). The temperature difference between the bolometer and bath is
\begin{equation}
\delta T = \frac{\Ptot}{\bar{G}},
\end{equation}
where $\Ptot$ is the total power deposited on the bolometer from all sources, and $\bar{G}$ is the mean thermal conductivity. The dynamic thermal conductivity, $G$, is defined as
\begin{equation}
G \equiv \frac{d\Ptot}{dT}.
\label{eq:dynamic_g}
\end{equation}
When the power on the bolometer is altered, it takes time for the system to thermalize. This process is characterized by a time constant defined by the heat capacity of the bolometer, and the dynamic thermal conductivity of the link to the cold bath:
\begin{equation}
\tau_0 = \frac{C}{G}.
\end{equation}
where $C$ is the heat capacity of the TES:
\begin{equation}
C \equiv \frac{dQ}{dT} \ .
\end{equation}
An exponential response with this time constant describes the detector response to a delta function input of power, in the limit that the bolometer resistance, $\Rtes$, is not a function of its temperature, and neglecting electrothermal feedback (see Section \ref{sec:responsivity}). 

For a simple bolometer, consisting of a resistive element to convert the GHz radiation coupled in from the microstrip to heat, and a TES with a superconducting transition temperature $\Tc$, connected to the readout circuitry, the heat capacity is very small, and thus the detector responds rapidly to changes in loading. There is also a time constant associated with the readout, which must be faster than to the detector time constant (see Section \ref{sec:responsivity}). In the development of the SPT-SZ instrument, it was found that the natural time constant of the detectors was too fast for the readout. This problem was solved by the addition of a relatively thick layer of high heat capacity metal over the TES which would remain normal down to below the transition temperature of the TES. This increases C, and lowers the time constant of the detector. This layer was originally gold, and unfortunately came to be known as bling. In the SPTpol instrument the bling consisted of palladium-gold, while for SPT-3G pure palladium is used (See \ref{fig:TES_SEM}).

The total power on the detector is the sum of the optical power from the antenna, and the electrical power dissipated in the TES
\begin{equation}
\Ptot = \Pe + \Popt.
\label{eq:ptot}
\end{equation}
The power on the detector as a result of Joule heating from the bias voltage, $\Vbias$, is
\begin{equation}
\Pe = \frac{\Vbias^2}{\Rtes}.
\label{eq:pelec}
\end{equation}
The total power deposited on the bolometer raises its temperature as
\begin{equation}
\Ptot = \kappa (\Ttes^n - \Tb^n),
\label{eq:power_dissipation}
\end{equation}
where $\Tb$ is the temperature of the cold bath, and $\kappa$ and $n$ are determined by the thermal link to the bath. For normal metals, $n=1$, while for insulators $n \aprx 3$. In SPT detectors with SiN legs, $n$ is usually measured to be $\aprx 2.7$.

If we then examine the case of small perturbations about the steady state, Equation \ref{eq:dynamic_g} says that the dynamic thermal conductivity can be found by taking the derivative of Equation \ref{eq:power_dissipation} with respect to $\Ttes$, yielding:
\begin{equation}
G(\Ttes) = nk \Ttes^{n-1}.
\end{equation}
The dynamic thermal conductivity, evaluated at $\Tc$, is important for understanding the behavior of the detector when it is in the operating mode.

The maximum power the bolometer can receive before it is driven normal is referred to as $P_\mathrm{turn}$. Once the detector is completely out of the transition, the resistance is nearly constant with varying temperature. This means that the bolometer will no longer be responsive to changes in optical loading. The resistance above the transition is called the normal resistance, or $\Rn$. Likewise, below a minimum power, the TES becomes superconducting, and unresponsive to changes in loading. This state is referred to as ``latched''.

\begin{figure}
\begin{center}
\includegraphics[width=5in]{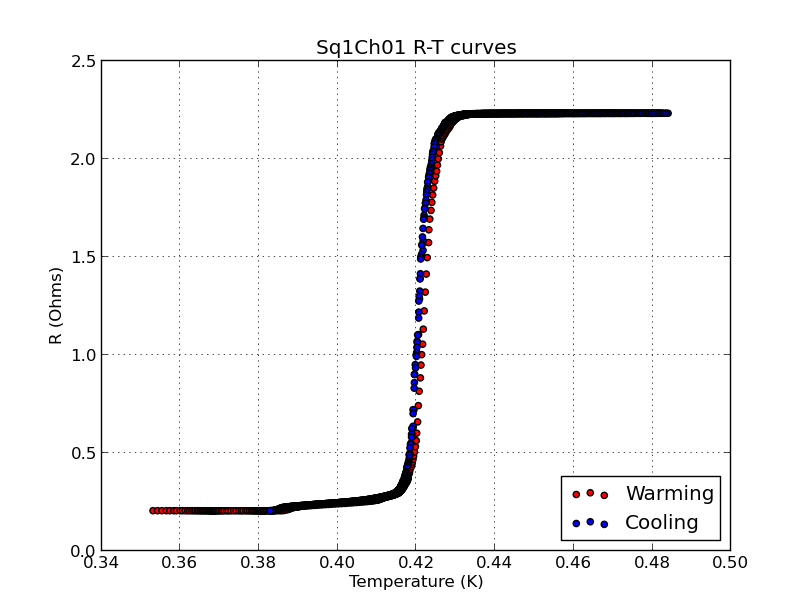}
\caption[Bolometer Superconducting Transition]{Measurements of the superconducting transition of a TES bolometer. The temperature of the stage to which the bolometer was mounted was varied by means of a heating resistor, using a proportional-integral-derivative (PID) controller. Current through the voltage biased TES was monitored, and used to calculate the resistance of the TES. Data was taken while passing through the transition in increasing and decreasing temperature directions to check for hysteresis, which was minimal. The ``foot'' feature at low temperature and resistance (\aprx 390mK - 410mK) is parasitic resistance due to normal metal on the detector. Data from testing at the University of Chicago.}
\label{fig:rt}
\end{center}
\end{figure}

When the detectors are being operated, they are kept in the superconducting transition (see Figure \ref{fig:rt}) with a negative feedback loop. When $\Popt$ increases, the TES heats up, and $\Rtes$ increases. Since $\Vbias$ is held constant the electrical power, $\Pe = \Vbias^2/\Rtes$, decreases, counteracting the increase in $\Popt$. This is referred to as an electrothermal feedback loop. Detectors are typically operated at a depth in the transition of $\aprx0.7\Rn$.  

Detectors that go normal or latch must be retuned, that is dropped into the transition to the operating depth, before they will be responsive to sky signals. If the detector is superconducting, it must be heated above $\Tc$ before the detector can be retuned, since in the superconducting state the current required to move the detector back into the transition is prohibitively large. This is accomplished either by heating the whole focal plane, or if there is a normal resistive element in series on the TES, by heating the latched bolometer individually. Therefore, if there is a small stray resistance on the TES this can actually be beneficial, as it facilitates retuning.

\begin{figure}
\begin{center}
\includegraphics[width=5in]{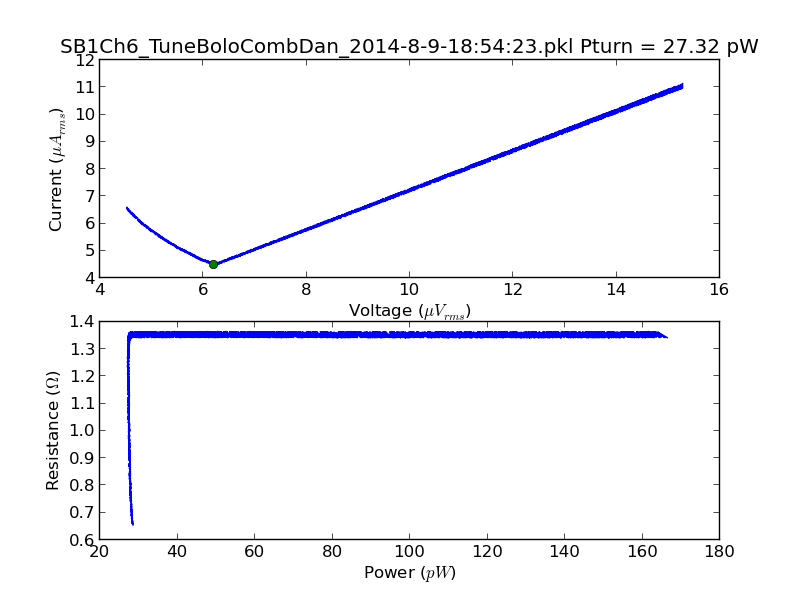}
\caption[Bolometer IV and RP Curves]{IV and RP curves for bolometers are obtained by varying the bias voltage, and measuring the current response. The portion of the IV curve with positive slope corresponds to the normal phase of the TES, while the portion with negative slope corresponds to the superconducting transition.  The minimum current in the IV curve occurs at the saturation point of the detector, $\Psat$, which is also the ``turning point'' in the RP curve. This is a high-G detector with a saturation power of 27pW. The ``bend back'' to higher power in the superconducting  portion of the RP curve (\aprx 30-27pW, $0.65 \Omega < R < 1.35 \Omega$) is due to the $\aprx 0.15 \Omega$ parasitic resistance in the circuit.}
\label{fig:iv}
\end{center}
\end{figure}

The saturation power of a bolometer can be measured by performing a load curve, or ``IV'' curve. For this measurement the detector is brought normal, and the voltage bias is dropped to lower the detector through the transition, while recording the current through the TES. Figure \ref{fig:iv} shows a load curve for a prototype SPT-3G detector, including the measured saturation power. Load curves are also useful diagnostics for several other detector properties. A simple transformation converts the data on TES current as a function of bias voltage into TES resistance as a function of power (an ``RP'' curve). From here it is trivial to calculate the normal resistance of the detector. It is also possible to calculate a lower limit on the parasitic resistance of the bolometer from the ``bend back'' feature in the RP curve. In an ideal superconductor, the resistance would be single valued and have positive slope everywhere in the RP curve. The slight curve toward higher power as the resistance falls toward zero in Figure \ref{fig:iv} is due to a small ($\aprx 0.15 \Omega$) parasitic resistance which maybe be due to normal metal on the TES. As stated above however, if it is located on the TES island this resistance can be useful for retuning individual latched bolometers. The parastic resistance does not interfere with the stable operation of the detector as long as it is significantly smaller than the normal resistance of the TES.

The characteristic V shape of the load curve is due to the unique behavior of the TES in the transition. For a normal resistor, we expect a positive slope in I as a function of V, as seen in the right hand side of the trace. When the TES passes through the transition, the resistance rapidly falls to zero, so the current increases despite the decreasing bias voltage. If the voltage bias is decreased further (not shown), the slope will return to positive after the TES goes superconducting, due to the parasitic resistance in the circuit.

The detector shown is a high-$G$ detector with a saturation power of 27pW. This high $G$, and thus high saturation power, is useful for testing in the laboratory, but is larger than the target $G$ value of final fieldable detectors. This is because the optical loading from the 300K laboratory environment is significantly larger than the $10-20K$ optical load from the sky at the South Pole. For laboratory testing, in addition to using detectors with higher saturation powers, it is also necessary to further reduce the loading on the detectors with low pass filters and possibly a neutral density filter (NDF). The target saturation powers are 12.5pW, 15pW, and 16.8pW for the 95 GHz, 150 GHz, and 220 GHz detectors respectively.

The thermal properties of the detector ($\kappa$, $G$, and $\Tc$) must be designed such that when the detector is held at $\Tc$ the ratio $\Pe/\Popt$ is within acceptable bounds. If the expected changes in optical loading are greater than the range of variation available in $\Pe$, then the feedback control will not be able to keep the detector in the transition, and it will either be driven normal, or into the superconducting state. However, the exact optical loading of an instrument is difficult to know before it is completed, as it depends on the final optical efficiency and internal thermal loading. The optical power from the sky also changes, for example because of differences in atmospheric loading between the highest and lowest elevation in an observing field. This change in atmospheric loading can be around $20\%$. Additionally, $P_e$ must be significant or $\mathcal{L}$ becomes too small.

The minimum condition for operation is that $\Pe/\Popt > 1$. In practice, a greater safety margin should be maintained. The ratio cannot be arbitrarily large however, as this would make thermal fluctuation noise from $G$ unnecessarily large. A target ratio of $\Pe/\Popt \ \aprx 1.5$ is usually used.

\begin{figure}
\begin{center}
\includegraphics[width=5in]{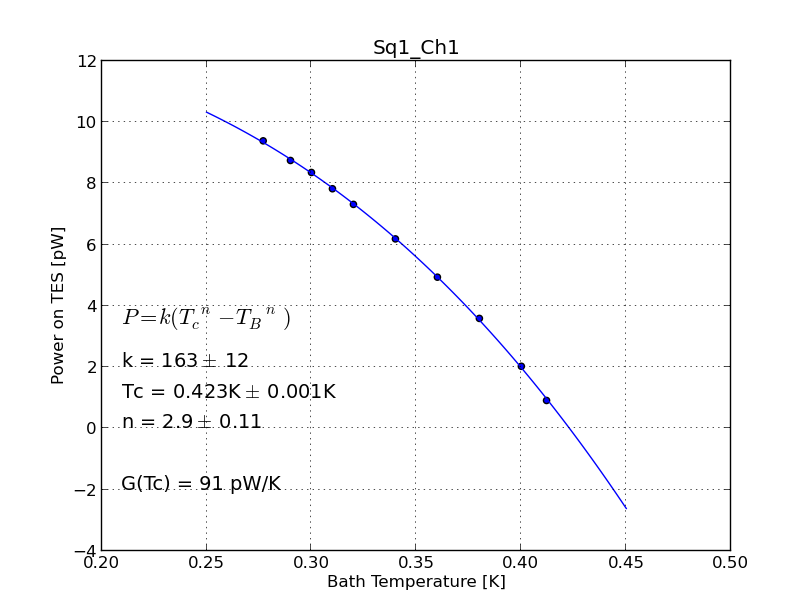}
\caption[Bolometer Dynamic Thermal Conductivity]{By varying the temperature of the cold bath, measuring the power dissipated in the bolometer, and fitting to the model $P = \kappa (\Ttes^n - \Tb^n)$, we can calculate several important properties of the bolometer. These include $G$, the thermal conductivity of the leg linking the TES to the thermal bath, $k$, which is determined by the geometry of the leg, and $n$, which contains information about the thermal transport mechanism in the thermal link.}
\label{fig:g}
\end{center}
\end{figure}

Figure \ref{fig:g} shows a measurement of the thermal conductivity of a prototype SPT-3G detector. The the temperature of the bolometer is varied, and a load curve is taken at each temperature. Equation \ref{eq:power_dissipation} can then be fit to the data, to determine the values of $G$, $k$, and $n$ for the detector.

\section{Responsivity}
\label{sec:responsivity}

Understanding the response of detectors to changes in optical power is important for characterizing detectors. Specifically, the variations in current through the TES and temperature of the TES as a function of power will be useful. We define the current and temperature responsivity of a detector as:
\begin{equation}
\sI \equiv \frac{\partial I}{\partial P},
\end{equation}
and
\begin{equation}
\sT \equiv \frac{\partial T}{\partial P}.
\end{equation}

The loopgain of the electrothermal negative feedback loop is an another important property of bolometers, and is defined as:
\begin{equation}
\mathcal{L} \equiv \frac{\Pe \alpha}{G \Tc}
\label{eq:loopgain}
\end{equation}
where $\alpha$ is defined as:
\begin{equation}
\alpha = \frac{d \ \mathrm{log}(R)}{d \ \mathrm{log}(T)} = \frac{T}{R}\frac{dR}{dT}
\label{eq:alpha}
\end{equation}
with $R$ the resistance of the TES and $T$ the temperature.\\

In Section \ref{sec:steady_state}, we discussed $\tau_0$, the response time of the detector to a delta-function input of heat. This picture of the thermal response neglected the electrothermal feedback loop, which tends to slow down the thermal response of the TES to power fluctuations. We may now give this response time more accurately as:
\begin{equation}
\taueff = \frac{\tau_0}{\mathcal{L}+1}.
\end{equation}
Additionally, we now have a second characteristic response time, the electrical response time $\taue$, determined by the high frequency response roll off of the LCR readout circuit: 
\begin{equation}
\taue = \frac{L}{\Rtes+\RL},
\end{equation}
where $L$ is the inductance of the resonant LCR readout circuit, and $\RL$ is the series resistance of the inductor. Under the assumption that $\RL \ll \Rtes$, this reduces to $\taue \approx L/\Rtes$. 

Irwin and Hilton \cite{irwin05} provide a detailed derivation of $\sI$ and $\sT$, demonstrating that:
\begin{equation}
\sI = -\frac{1}{V_0} \frac{\mathcal{L}}{\mathcal{L}+1} \left( \frac{1}{1+i\omega\taueff} \right) \left( \frac{1}{1+i\omega\taue} \right),
\end{equation}
and
\begin{equation}
\sT = \frac{1}{G} \left( \frac{1}{\mathcal{L}+1} \right) \left( \frac{1}{1+i\omega\taueff} \right).
\end{equation}
For high loopgains, $\frac{\mathcal{L}}{\mathcal{L}+1} \rightarrow 1$, and for $\omega < 1/\taueff$ and $\omega < 1/\taue$, $\sI \rightarrow -1/V_0$, demonstrating that for high $\mathcal{L}$ the detector current response is linear. Similarly, at high loopgain and $\omega < 1/\taueff$, $\sT \rightarrow 0$, and the TES temperature is insensitive to changes in optical power, because the electrothermal feedback corrects for changes in input power faster than the TES can thermalize.

However, loopgain cannot be arbitrarily large or power fluctuations induce either oscillations or exponential growth in the response current. The commonly used criteria for bolometer stability is:
\begin{equation}
\mathcal{L} < \frac{\tau_0}{5.8 \tau_e},
\end{equation}
or, if $\mathcal{L} >> 1$, $\taueff > 5.8 \tau_e$. Essentially, the electrical response time needs to be sufficiently small compared to the thermal response time that the electrical feedback can affect the temperature of the TES before it thermalizes. 

Lueker \cite{lueker10b,lueker10a} developed a method for testing the electrothermal feedback properties of detectors, including the effective time constant, $\taueff$, in frequency multiplexing systems. Consider a bolometer biased with a voltage $V_0$, at a frequency $\omega_0$. A perturbation with amplitude $\delta I$ is applied to this bolometer in a sideband of the bias voltage, that is, at a frequency offset from the bias frequency, $\omega_0 - \delta \omega$:
\begin{equation}
I_\mathrm{sb} = \delta I \mathrm{cos}\left( (\omega_0 - \delta \omega) t \right).
\end{equation}
This perturbation will result in modulated power at $\delta \omega$:
\begin{equation}
P_\mathrm{sb} = \delta P \mathrm{cos}(\delta \omega t),
\end{equation}
and amplitude modulation of the current at $\omega_0$:
\begin{equation}
I(t) = \left[ I_0 + 2 \delta P \lvert \sI \rvert \mathrm{cos} \left( (\omega_0 - \delta \omega) t \right) \right] e^{i\omega_0t}.
\end{equation}
The amplitude of the induced current in the sidebands, at $\omega_0 \pm \delta \omega$, is then
\begin{equation}
I_\mathrm{sb,\pm} = \lvert \sI \rvert \delta P.
\end{equation}
Measuring the current at the opposite sideband from where the power was applied avoids the injected current, and provides a clean measurement of the current responsivity, $\sI$, of the detector.

\begin{figure}
\begin{center}
\includegraphics[width=5in]{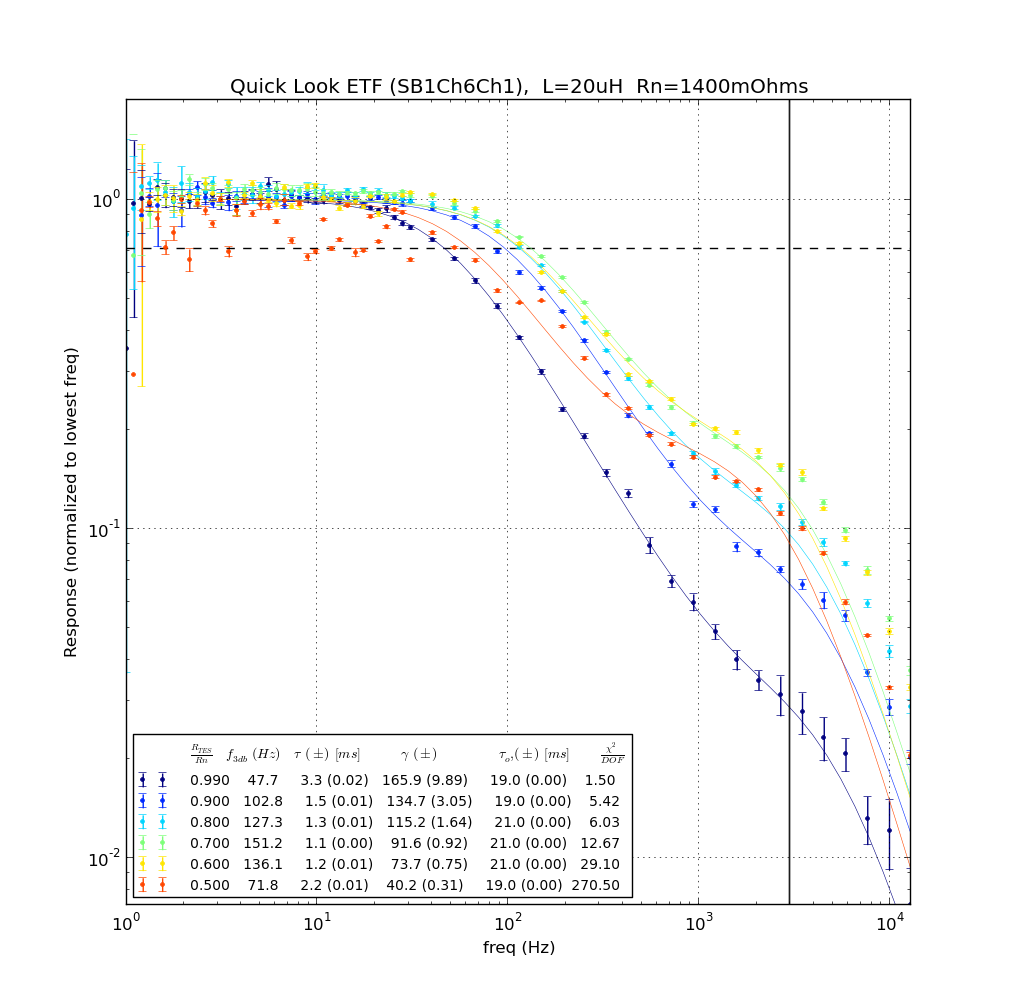}
\caption[Bolometer Electrothermal Feedback]{Sideband perturbation measurements of the current responsivity of an early monochroic SPT-3G detector. The different traces correspond to detector biasing at different points in the transition. The responsivity initially falls off due to $\taueff$. The fall off becomes shallower at \aprx1 kHz due to bling decoupling from the TES, but steepens again at \aprx5 kHz due to the LCR bandwidth.}
\label{fig:etf}
\end{center}
\end{figure}

Figure \ref{fig:etf} shows measurements of detector current responsivity as a function of $\delta \omega$ taken using this sideband perturbation method. As $\delta \omega$ increases the responsivity falls off, due to $\taueff$. At a $\delta \omega$ of approximately several kilohertz, the thermal conductivity between the TES and bling is insufficient for electrical power to drive both together, and the thermal response of the TES becomes that of the TES alone. The TES has a smaller thermal mass, and thus a lower $\taueff$, so the fall off becomes shallower.  It finally steepens again at $\aprx 5$kHz due to the LCR bandwidth. Testing whether the bling decoupling point is at a higher frequency than the electronics roll off is an important function of this sideband perturbation method.

\subsection{Understanding Loopgain}

The following is a simple, but novel, analysis I performed of how other physical properties affect detector loopgain. These results are useful primarily for detector testing, by giving insight into the connections between measurable detector properties. 

They can also be useful for optimizing loopgain in detector design. As discussed in Section \ref{sec:responsivity}, it is beneficial if loopgain is high, because it leads to linear detector current response, and makes the TES temperature insensitive to changes in optical loading. However, the loopgain cannot be arbitrarily high, as it leads to instability in the detector current response. We are therefore interested in physical means of optimizing loopgain.

The standard definition of of loopgain (Equation \ref{eq:loopgain}):
\begin{equation}
\nonumber
\mathcal{L} \equiv \frac{\Pe \alpha}{G \Tc},
\end{equation}
is problematic because it defines loopgain in terms of variable which are not independent. This makes it difficult to visualize how changes in one property will affect the others, or even the loopgain itself.

Using Equations \ref{eq:dynamic_g}, \ref{eq:ptot}, \ref{eq:power_dissipation}, and \ref{eq:loopgain} we can re-express loopgain in terms of independent variables:
\begin{equation}
\mathcal{L}  = \frac{\left[ k (\Tc^n - \Tb^n) - \Popt \right] \alpha}{nk\Tc^n}. \\
\label{eq:loopgain2}
\end{equation}

It is now apparent that there are only four independent variables that determine the loopgain of a bolometer: $k$, $\alpha$, $\Tc$, and $\Tb$, assuming that $n$ is essentially fixed. Of these properties, $k$, $\alpha$, and $\Tc$ are intrinsic properties of the detectors, while $\Tb$ is a property of the cryogenic system. 

We can now take the partial derivatives of Equation \ref{eq:loopgain2} with respect to each of these variables to find how each affects $\mathcal{L}$.

\begin{itemize}

\item Partial of $\mathcal{L}$ with respect to $\alpha$:

\begin{equation}
\frac{\partial \mathcal{L}}{\partial \alpha} =  \frac{ k (\Tc^n - \Tb^n) - \Popt}{nk\Tc^n}
\label{eq:partial_alpha}
\end{equation}

\item Partial of $\mathcal{L}$ with respect to $k$

\begin{equation}
\frac{\partial \mathcal{L}}{\partial k} =  \frac{\partial}{\partial k} \left[ \left( \cancel{\frac{k \Tc^n}{n k \Tc^n}} - \cancel{\frac{k \Tb^n}{n k \Tc^n}} - \frac{\Popt}{n k \Tc^n} \right) \alpha \right]
\label{eq:partial_k}
\end{equation}

\begin{equation}
\Rightarrow \frac{\partial \mathcal{L}}{\partial k} =  \frac{\Popt \alpha}{n k^2 \Tc^n} 
\label{eq:partial_k2}
\end{equation}

\item Partial of $\mathcal{L}$ with respect to $\Tc$

\begin{equation}
\frac{\partial \mathcal{L}}{\partial \Tc} =  \frac{\partial}{\partial \Tc} \left[ \left( \cancel{\frac{k \Tc^n}{n k \Tc^n}} - \frac{k \Tb^n}{n k \Tc^n} - \frac{\Popt}{n k \Tc^n} \right) \alpha \right]
\label{eq:partial_tc}
\end{equation}

\begin{equation}
\Rightarrow \frac{\partial \mathcal{L}}{\partial \Tc} =  \left( \frac{\Tb^n}{\Tc^{n+1}} + \frac{\Popt}{k \Tc^{n+1}} \right) \alpha
\label{eq:partial_tc2}
\end{equation}

\begin{equation}
\Rightarrow \frac{\partial \mathcal{L}}{\partial \Tc} =  \frac{(\Tb^n + k^{-1} \Popt) \alpha}{\Tc^{n+1}}
\label{eq:partial_tc3}
\end{equation}

\item Partial of $\mathcal{L}$ with respect to $\Tb$

\begin{equation}
\frac{\partial \mathcal{L}}{\partial \Tb} =  \frac{\partial}{\partial \Tb} \left[ \left( \cancel{\frac{k \Tc^n}{n k \Tc^n}} - \frac{k \Tb^n}{n k \Tc^n} - \cancel{\frac{\Popt}{n k \Tc^n}} \right) \alpha \right]
\label{eq:partial_tb}
\end{equation}

\begin{equation}
\Rightarrow \frac{\partial \mathcal{L}}{\partial \Tb} =  \frac{-\Tb^{n-1} \alpha}{\Tc^{n}}
\label{eq:partial_tb2}
\end{equation}

\end{itemize}

This reveals several interesting points which can be applied to bolometer testing and design.  

First, though it seems counter intuitive from the definition, $\mathcal{L} \equiv \Pe \alpha / G \Tc$, Equation \ref{eq:partial_tc3} shows that increasing $\Tc$ increases $\mathcal{L}$. However, this is effectively only a second order effect; it is only significant if $\Tb$ is comparable to $n \Tc$, or $\Popt$ is comparable to $nk \Tc^n$. 

In detector design, if you want to change $\mathcal{L}$, $\alpha$ provides the best lever arm. Using typical SPT values for all variables, and the partial derivatives of $\mathcal{L}$ above, it works out that $\alpha$ has a stronger effect on $\mathcal{L}$ than $k$, $\Tc$, or $\Tb$. Also, changing $k$, $\Tc$, or $\Tb$ changes the saturation power of the detector, making these less useful design parameters for independently adjusting loopgain.

These equations are especially useful for detector testing. Knowing how loopgain changes as a result of changes in other properties can be useful for predicting loopgain under different conditions. For example, in the iterative fabrication and testing process, there are often unexpected changes in $\Tc$ between fabrication runs. Using Equation \ref{eq:partial_tc3}, and a measurement of the loopgain in a detector with an undesirable $\Tc$, one can predict the loopgain in a detector with the desired $\Tc$. Similarly, Equation \ref{eq:partial_tb2} can be used to calculate $\alpha$ if loopgain is measured at two different base temperatures.

Lastly, and possibly most interesting, for dark tests ($\Popt = 0$), there is no change in loopgain for varying $k$. Practically, this means that measurements of high $G$ detectors can be used to accurately determine what the loopgain will be in the final low $G$ detectors. It was formerly believed that the loopgain of the detectors could not be determined until detectors with low $G$ were produced. This discovery allows us to measure loopgain earlier in the production cycle of an experiment, and begin fine tuning detector loopgain earlier, improving the efficiency of the detector design and testing process.

These results depend only on the thermal equations which describe voltage biased bolometers operating with negative electrothermal feedback, and are therefore applicable not only to SPT, but to the development and testing of such bolometric detectors generally.

\section{FTS Measurements of Filter Bands}

\begin{figure}
\begin{center}
\includegraphics[width=5in]{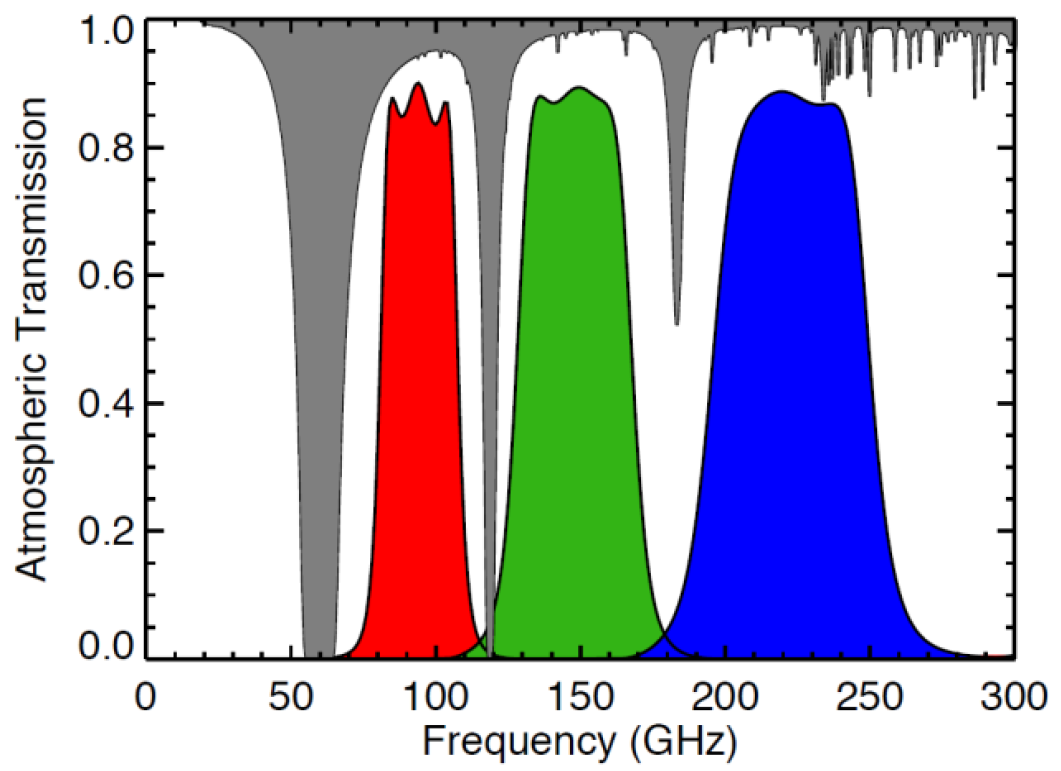}
\caption[Target Detector Frequency Bands]{Target filter frequency bands for the multichroic sinuous antenna detectors. The nominal band centers are 95, 150, and 220GHz, with $22.5\%$ fractional bandwidth. In grey is the atmospheric transmission spectrum with 0.25mm of precipitable water vapor (PWV), the median value at the South Pole \cite{chamberlin01}. Figure from Posada et al., 2015 \cite{posada15}.}
\label{fig:filter_bands}
\end{center}
\end{figure}

Atmospheric loading is an important consideration for microwave frequency experiments, and is one of the primary drivers for locating an instrument at the South Pole. The observing bands in a ground based instrument must be carefully designed to avoid portions of the spectrum where atmospheric transmission is low. Figure \ref{fig:filter_bands} shows the designed filter band-passes compared to the atmospheric transmission. Having a broader detector band is desirable since it allows you to integrate more sky signal. However, extending a band into a region with high atmospheric absorption is counter productive, as it will decrease the net signal to noise since less sky signal will be transmitted, and noise from atmospheric emissions will increase.

Given the importance of avoiding low atmospheric transmission spectral regions, it is important to verify that the pass bands of actual detectors match the designed bands. This is accomplished with Fourier Transform Spectroscopy (FTS) measurements. FTS measurements are performed at Case Western Reserve University using a Martin-Puplett interferometer \cite{martin70}, a polarized variant of the classic Michelson interferometer \cite{michelson87}. A wire grid beam splitter separates light from the input port into two polarizations, and sends each polarized beam down an independent arm of the interferometer. Rooftop mirrors flip the polarization of the beams at the end of each arm, and the beams are recombined at the wire grid. The recombined beam is fed out of the output port to the detectors. One of the arms is stationary, while the rooftop mirror of the second arm is translated with a stepper motor at a specified velocity. 

As the mirror translates, the recombined beam is modulated at a different audio frequency for each photon wavelength. The Fourier transform of the response of a detector to the white light fringes from the interferometer encodes the response of that detector to power modulations at varying photon wavelengths.

\begin{figure}
\begin{center}
\includegraphics[width=5in]{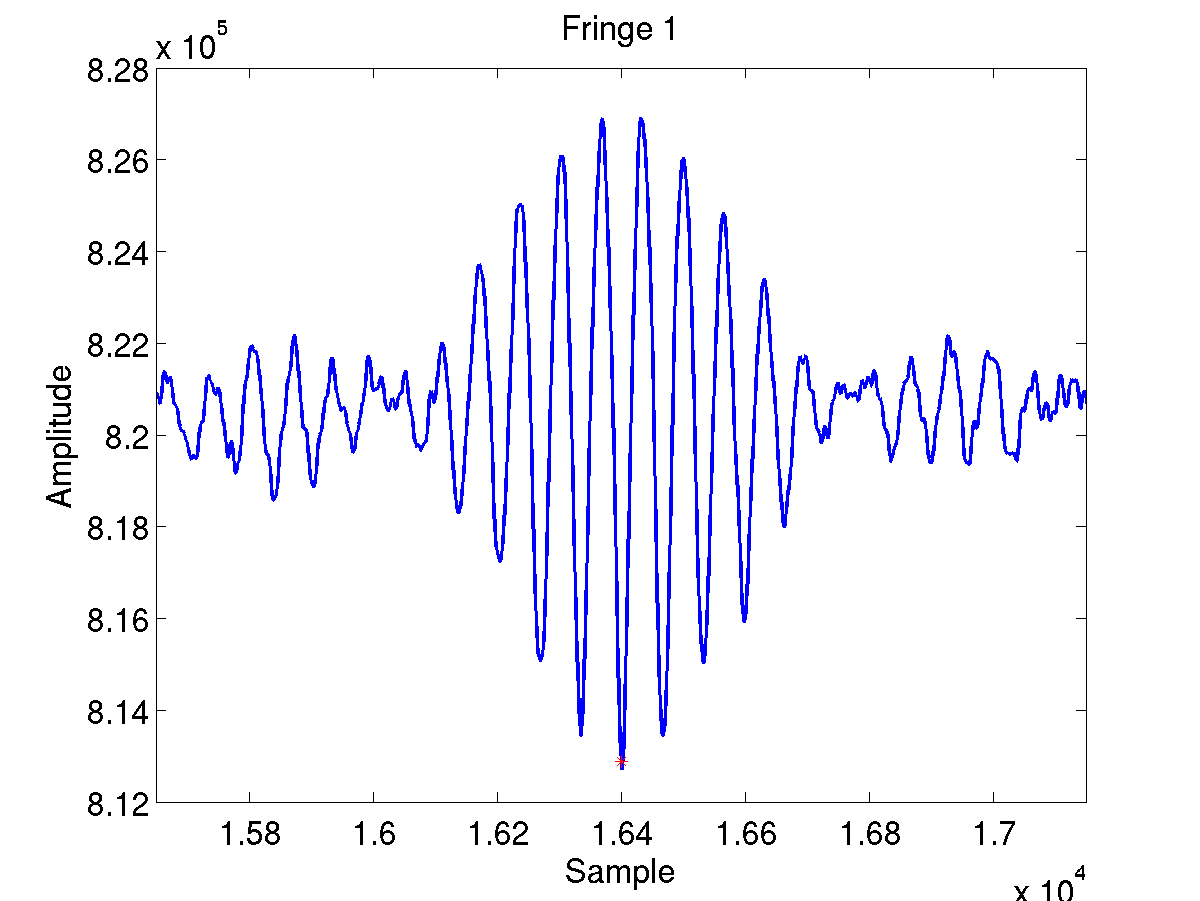}
\caption[FTS White Light Fringe]{An example white light fringe from an SPT-3G detector, taken with a Martin-Puplett interferometer. The Fourier transform of this fringe produces the spectrum of the detector filter, shown in Figure \ref{fig:meas_detector_band}. In practice, many fringes are measured and summed to produce a higher signal-to-noise measurement of the detector band.}
\label{fig:fringe}
\end{center}
\end{figure}

Figure \ref{fig:fringe} shows an example white light fringe from an SPT-3G detector. To produce high signal-to-noise detector band measurements, many fringes are obtained and summed, and the Fourier transform of the summed waveform is taken.

\begin{figure}
\begin{center}
\includegraphics[width=5in]{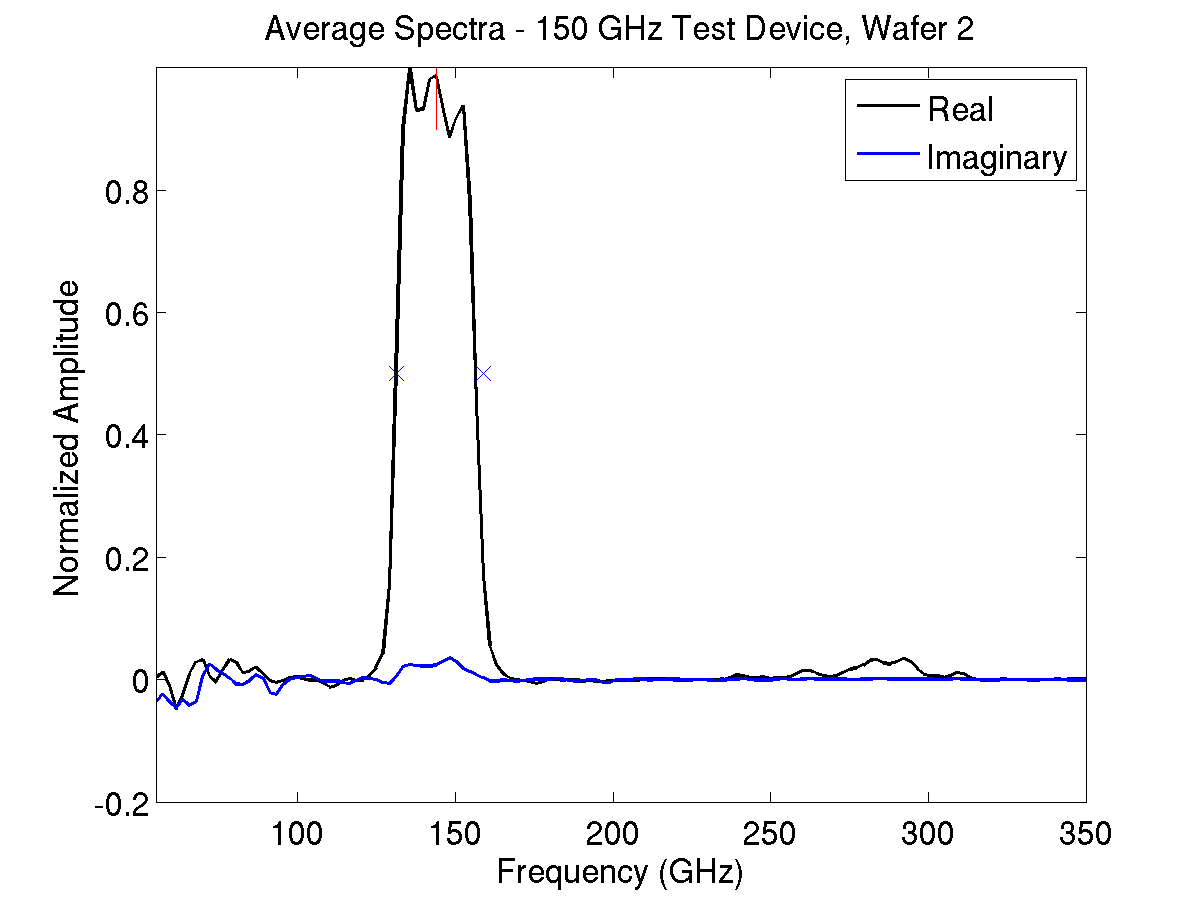}
\caption[Measured Detector Frequency Bands]{The measured frequency band for an early 150 GHz monochroic detector. The black line is the real part of the Fourier transform, while the blue line is the imaginary component. Ideally, the imaginary component should be exactly zero. The non-zero imaginary component in band is due to imperfect reconstruction of the center of the white light fringe, causing a small amount of real power to leak into the imaginary component. The real feature at \aprx300 GHz is a ``blue leak'', a harmonic of the desired 150 GHz band which was imperfectly suppressed by this early filter. Discovering such features and correcting them is one reason for conducting FTS measurements of detector bands.}
\label{fig:meas_detector_band}
\end{center}
\end{figure}

Figure \ref{fig:meas_detector_band} shows the measured frequency band for an early 150 GHz monochroic detector. The real and imaginary components of the Fourier transform are shown. All of the information about the detector band should ideally be contained in the real component of the Fourier transform. However, if the white light fringe is not perfectly centered in the timestream, then real power will leak into the imaginary component. This is the origin of the in-band feature in the imaginary component, where the power leakage is $3.5\%$ of the real in-band power.

Another salient feature of this plot is the real power at \aprx 300 GHz. This is a ``blue leak'', a harmonic of the desired 150 GHz band which was imperfectly suppressed by this early filter. This blue leak was corrected in later versions of the filter. The discovery of such features is an important reason for conducting FTS measurements of prototype detectors.

\section{Current Development Status}

As of July 2015, we are able to produce full detector wafers which include the sinuous antennas, triplexer filters, and TESs. The thermal and electrical properties of these detectors have been measured, the band properties of the individual filters have been measured in earlier prototypes. We are able to produce detectors with the desired thermal and electrical properties, and individual filters with the expected bands. The combined triplexer filters are still being tested.

We are currently in the process of measuring and iteratively improving the optical and polarization properties of the final antenna and triplexer designs. We are also working to improve issues related to the production of full wafers of detectors, such as the detector yield, and the consistency of detector properties across the wafer. 

In addition to the detector tests described above, there are several more test which we will perform before deploying the new instrument.

First, it is important to map detector response as a function of the spherical angles $\theta$ and $\phi$, which is referred to as beam mapping. To perform this measurement we will use a thermal source mounted below the test cryostat, and which is capable of rotating in $\theta$ and $\phi$ on a sphere centered on the detectors in the test cryostat.

Second, given that the temperature fluctuations in the CMB exceed the polarized fluctuations by one to two orders of magnitude, it is important that detectors aligned to one linear polarization mode have very low response to the orthogonal polarization mode. To verify this, we will conduct measurements of a chopped thermal source behind a rotating wire grid polarizer. Fieldable detectors should have better than 99\% polarization efficiency to prevent crosspolar power leakage.

Lastly, as stated in Section \ref{sec:detector_structure}, a feature of the novel log-periodic sinuous antennas used for SPT-3G is that the polarization sensitivity angle, $\phi$, can rotate by up to $5^\circ$ across the three octave bandwidth. We will measure this polarization rotation by observing a rotating Gunn oscillator, a polarized millimeter wave source with a tunable frequency. The Gunn oscillator can be tuned with a precision of $\delta f < 1 GHz$, across the entire relevant frequency range, $80GHz < f < 250GHz$.

Work on the new cryostat, improved optics, and readout electronics is also in progress, and the SPT-3G instrument is currently expected to deploy in the Austral summer of 2016.

%Cold Readout Electronics 

\doublespacing

\chapter{Digital Frequency Multiplexed Readout}
\label{ch:readout}

\section{Cryogenic Readout}

One of the many challenging aspects of cryogenic instrumentation is readout. Any data from 
the instrument, and in particular the data from the focal plane, must be amplified and 
passed on to room temperature readout and analysis systems with low noise. Superconducting Quantum Interference Devices, or SQUIDs, are low-noise cryogenic amplifiers well suited to this purpose. 

\begin{figure}
\begin{center}
\includegraphics[width=5in]{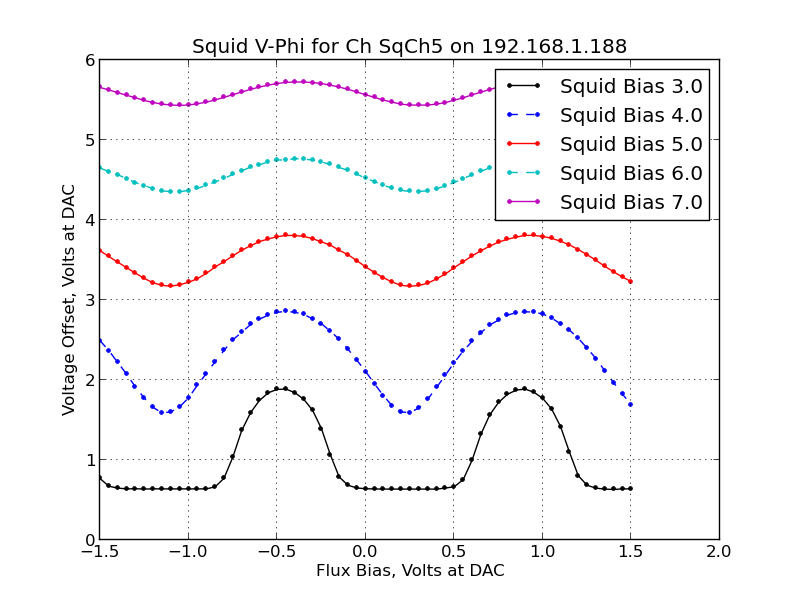}
\caption[SQUID $V-\Phi$ Curves]{When the magnetic flux, $\Phi$, through the SQUID increases, it induces a sinusoidal response in the voltage across the SQUID, referred to as a $V-\Phi$ curve. The DC voltage offset can be varied to maximize the amplitude of this response, while maintaining a smooth sinusoidal response. An operational flux bias point at which the response of the SQUID is most linear is then chosen.}
\label{fig:v-phi}
\end{center}
\end{figure}

The SQUID was invented by researchers at the Ford Motor Company in 1964 \cite{jaklevic64} as part of a blue sky research project. To amplify a signal, an inductor $L$ is coupled to a SQUID, which is biased with a constant voltage, $V_{bias}$. A change in the current passing through the inductor induces a change in the magnetic flux through the SQUID, $\Phi$, resulting in a sinusoidal voltage response (See Figure \ref{fig:v-phi}). The offset voltage can be tuned to vary the DC level of the SQUID, and maximize the amplitude of the sinusoidal voltage response. In practice a series array of SQUIDs is used to amplify the readout signal.

Furthermore, for an instrument with many channels to read out, the heat load from the temperature differential along the readout lines (which necessarily pass from the cold $<1$K stage out to the warm 300K environment) must not exceed the cooling capacity of the refrigerator maintaining the cryogenic system. The readout heat load should in fact be substantially less than the cooling capacity, because there will be other sources of heating as well, including conductance through the cryostat, radiation (from the sky, the optics of the instrument, and the cryostat itself), and ohmic heating. Convection is essentially eliminated by pumping out the cryostat to vacuum, first with a turbopump system capable of reaching pressures on the order of 1 mTorr, and subsequently by cryopumping, as the cryostat is cooled to 4K. Cryopumping is extremely efficient at condensing all remaining gasses, including nitrogen and helium. In a typical cryostat cooled to 4K the pressure will be $\ll 1 \times 10^{-5}$ Torr.
 
To reduce the readout heat load in an instrument like SPTpol, with $\aprx 1600$ detector channels, it is necessary to multiplex the readout, sending the data from multiple detectors on the same readout line. Both time-domain multiplexing (tMux) \cite{battistelli08}, and frequency-domain multiplexing (fMux) \cite{dobbs08} systems are used for this purpose. BICEP \cite{Bicep2collaboration14}, Keck \cite{Kernasovskiy12}, SPIDER \cite{runyan11}, and PIPER \cite{Lazear14}, for example use tMux systems, while EBEX \cite{reichborn10} and Polarbear \cite{Hattori13, Barron14} use fMux systems. SPTpol uses a digital frequency multiplexing, or DfMux, system \cite{george12, austermann12, story12a, sayre14}. 

\section{Digital Frequency Multiplexing}

The principle behind frequency multiplexing is that multiple amplitude modulated signals can be sent on the same wire, if each signal has a distinct frequency, $\omega_i$. In practice, the signal will not be a perfect delta function, but will have finite bandwidth, usually with low amplitude tails extending out large distances in $\omega$. It is therefore important to ensure that the spacing between frequencies is sufficient to reduce crosstalk between them to an acceptable level. This will be discussed further in Section \ref{sec:lc_design}.

\begin{figure}
\includegraphics[width=\columnwidth]{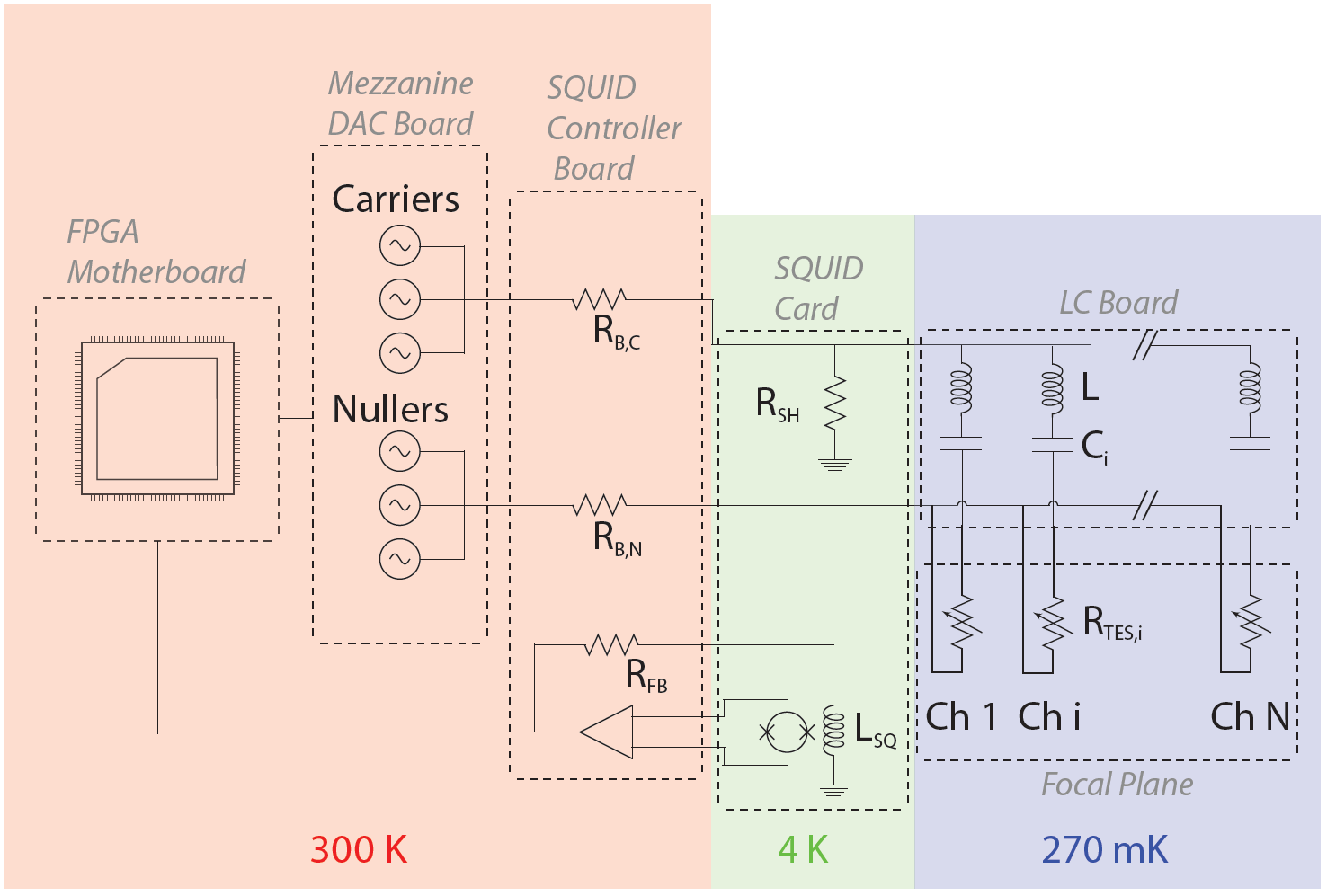}
\caption[Digital Frequency Multiplexing Circuit]{This is the basic architecture for a digital frequency multiplexing (DfMux) readout system. A carrier signal consisting of a comb of bias voltages at a series of frequencies is generated by the DfMux board and sent down a single line to the cold electronics. This voltage bias is converted to a current bias by the shunt resistor, $R_\mathrm{SH}$, on the 4K SQUID Card, before being sent to the ultracold 270mK stage. There an array of LCR resonant circuits in parallel allows a set of TES bolometers to see only the bias signal to which each LCR circuit is tuned. The bias signal at each TES is then modulated by changes in $R_\mathrm{TES}$, and the signal is summed across the array. In Digital Active Nulling (DAN) mode, the nuller signal is dynamically updated to cancel out the carrier signal, leaving only the modulated signal from the TESs. This modulated current passes through the inductor, $L_\mathrm{SQ}$, coupled to the SQUID, which converts the magnetic flux into an amplified voltage response which can be read out and recorded by the DfMux board. Figure from Sayre, 2014 \cite{sayre14}.}
\label{fig:lc_circuit}
\end{figure}

\begin{figure}
\begin{center}
\includegraphics[width=5in]{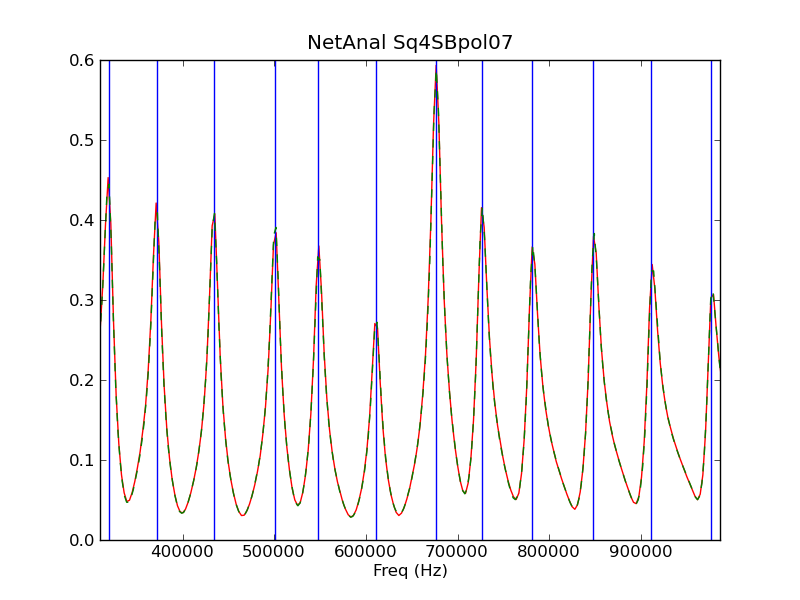}
\caption[Network Analysis for SPTpol LC Board]{A measured network analysis of an LC comb (SQUID 4 on SQUID Board 7) in the SPTpol experiment. This is a 12x multiplexing system, with $60$kHz nominal spacing between channels. The green dashed curve is the measured network analysis of this comb from the 2012 observing season, while the red curve is from the 2013 observing season, demonstrating good stability over time. The vertical blue lines are the measured locations of the resonance peaks.}
\label{fig:netanal}
\end{center}
\end{figure}

A schematic of the DfMux circuitry is shown in Figure \ref{fig:lc_circuit}. In the SPTpol system, a voltage bias is sent to 12 individual TES bolometers on a single pair of wires, so that the total bias voltage on the line is:
\begin{equation}
V_{\mathrm{Tot}} = \sum_i{V_{\mathrm{bias},i}e^{i\omega_{i}t}},
\end{equation}
where $V_{\mathrm{bias},i}$ is the bias voltage at frequency $\omega_i$. This total voltage is referred to as a frequency comb, with each frequency channel being a ``tooth'' in the comb (See Figure \ref{fig:netanal}). Each TES is the resistor, $R_{\mathrm{TES},i}$, in an LCR resonant circuit, with the twelve legs of each readout comb connected in parallel. The inductance, $L$, is the same for each element in the comb, while the capacitance, $C_i$, varies so that each element has a unique resonance frequency, $\omega_i = \sqrt{1/LC_i}$. To first order, each TES then only sees the bias frequency to which its resonant circuit is tuned. In SPTpol, $L = 20\mu$H, and $1.3 \mathrm{nF} < C_i < 16$ nF, so that $280 \mathrm{kHz} < \omega_i < 1000$kHz.

The impedance of each LCR element in the comb at a frequency $\omega$ is:
\begin{equation}
Z_i(\omega) = R_{\mathrm{TES},i} + i \omega L - \frac{i}{\omega C_i}.
\end{equation}
At the resonant frequency, $\omega_i$, this reduces to simply $R_{\mathrm{TES},i}$. The voltage bias $V_{\mathrm{bias},i}$ is sourced by the DfMux readout boards, then converted to a current bias by a resistor $R_\mathrm{bias}$, which is significantly larger than the total impedance from all the LCR circuits in the comb. To ensure that each LCR element sees a constant voltage, a shunt resistor with $R_\mathrm{SH} << R_{\mathrm{TES},i}$ is wired in parallel with the LCR comb. The majority of the current then passes through the shunt resistor, and each TES sees a voltage bias of $V_{bolo,i} = (V_{\mathrm{bias},i}/R_{\mathrm{TES},i})R_\mathrm{SH}$.

In fact, the TES is a variable resistor, $R_{\mathrm{TES},i} \aprx 0.8\Omega$, whose resistance varies with it's temperature (and therefore the optical loading). Fluctuations in $R_{\mathrm{TES},i}$ result in changes to the current at the resonant frequency, $I(\omega_i)$. 

Because of the non-linear response, and limited dynamic range of the SQUIDs used to amplify this signal, nulling circuits are used to reduce the dynamic range of the readout signal (See Figure \ref{fig:lc_circuit}). In analog nulling mode, the carrier bias waveform is inverted to produce a nuller waveform, which is introduced on the output side of the LCR circuit to cancel out the carrier, in principle leaving only the induced sky signal, $\delta I(\omega_i)$. A negative feedback flux locked loop (FLL) improves the linearity of the SQUID response by feeding the voltage output back through the flux bias resistor, $R_\mathrm{FB}$, to adjust the current at $L_\mathrm{SQ}$ toward zero. This method has the disadvantage of limiting the available bandwidth of the SQUID, because of phase shifts in the wiring of the FLL, which must run from the warm (300K) side of the SQUID amplifier chain back to the cold (4K) squid inductor. Given the line lengths involved, this results in a highest stable bias frequency of $\aprx 1.2$MHz. (The lower limit of $\aprx 100$kHz is the result of AC coupled components in the readout electronics.) 

In the improved Digital Active Nulling (DAN) mode, the signal at each channel frequency is monitored and dynamically nulled at $L_\mathrm{SQ}$, with a bandwidth of $\aprx 10$kHz. This  cancels out both the carrier waveform and induced current. The nuller current itself is then used as the measurement of the TES current, and the SQUID is only part of the DAN integrator circuit. In this design there is no need for the FLL, allowing for much larger bias frequencies, up to potentially tens of megahertz. (This advancement was crucial to extending the 12x frequency multiplexing of SPTpol to the 64x multiplexing necessary for SPT-3G.)

In practice the nulling is never perfect, but the carrier signal is greatly suppressed, and the reduced dynamic range of the readout signal means the sky signal can be amplified by a larger factor by the SQUID array. The signals from each element of the comb can then be digitized and demodulated in software by the warm electronics.

Because the frequency band of each channel does not completely cut off in finite bandwidth, there is some power from each element in the comb at the frequencies of the other elements. Each channel then hears some amount of off-resonance crosstalk from the other channels, which depends on the difference in frequency between them. The impedance in the $i^{th}$ channel at the frequency of the $j^{th}$ channel is:
\begin{equation}
Z_i(\omega_j) = R_{TES,i} + i\omega_jL - \frac{i}{\omega_j C_i}.
\end{equation}
The total current in the $i^{th}$ channel from on-resonance and the crosstalk from the $j^{th}$ channel is then:
\begin{equation}
I_i(t) = \frac{V_i e^{i \omega_i t}}{Z_i(\omega_i)} + \frac{V_j e^{i \omega_j t}}{Z_i(\omega_j)}.
\end{equation}
If we compare the magnitude of the current from the $j$th element to the on-resonance current, assuming $V_i = V_j$ we find:
\begin{equation}
\left| \frac{I_i^{\omega_j}}{I_i^{\omega_i}} \right| = \frac{R_{\mathrm{TES},i}}{\sqrt{R_{\mathrm{TES},i}^2 + \left( \omega_j L - \frac{1}{\omega_j C_i} \right)^2}}.
\end{equation}
If we make the approximation \cite{dobbs12b}: 
\begin{equation}
\sqrt{R_{\mathrm{TES},i}^2 + \left( \omega_j L - \frac{1}{\omega_j C_i} \right)^2} \approx 2 \Delta \omega L,
\end{equation}
where $\Delta \omega = | \omega_j - \omega_i  |$ is the frequency spacing between the $i^{th}$ and $j^{th}$ elements, and since $P = I^2 R$, then we get the following simplified equation for the crosstalk power from the $j^{th}$ element in the $i^{th}$ element:
\begin{equation}
\frac{| P_i^j |}{| P_i^i |} \approx \frac{R_{\mathrm{TES},i}^2}{(2 \Delta \omega L)^2}.
\end{equation}
The amount of crosstalk between channels is the first consideration in the design of a frequency schedule for a multiplexing readout system, since it leads to false correlations between map pixels, and a degradation of signal to noise. The frequency schedule for SPTpol, and how it attempts to minimize crosstalk, will be discussed in the following section.

\section{LC Board Design and Assembly}
\label{sec:lc_design}

\subsection{Design Criteria}
In SPTpol, we specified a nominal frequency spacing of 60kHz between channels in the LC comb. With $L = 20\mu$H, and $R_\mathrm{TES} \ \aprx \ 0.8 \Omega$, and assuming similar resistances and voltage biases on each TES, 60kHz spacing gives $ \aprx 0.3\%$ of the on-resonance power as crosstalk from an adjacent channel. Summing over the whole comb, the total crosstalked power should be $\lesssim 1\%$ of the on-resonance current for each TES in the comb. The challenge was then to generate the desired capacitances from an array of commercially available surface mount ceramic chip capacitors, while maintaining good frequency separation. Table \ref{tab:lc_freq_schedule} shows my final design, as it was implemented for SPTpol. Each of the capacitances listed was produced using between one and three ceramic chip capacitors, soldered in parallel by stacking them on the surface mount pads. The minimum frequency between two adjacent channels in this schedule is 49kHz, at which point the crosstalking power from the neighbor is still $\aprx 0.5\%$. 

\begin{table*}
\begin{center}
{\scriptsize
\caption{DfMux Frequency Schedule}
\begin{tabular}{l|ccc|ccc}
\hline \hline
 & & Comb A & & & Comb B & \\
 Channel & C (nF) & Freq (kHz) & $\Delta$Freq (kHz) & C (nF) & Freq (kHz) & $\Delta$Freq (kHz) \\
\hline
 1 & 15.6 & 282 & - & 12.67 & 313 & - \\
 2 & 10.4 & 346 & 64 & 9.2 & 367 & 54 \\
 3 & 7.48 & 407 & 61 & 6.8 & 427 & 60 \\
 4 & 5.87 & 460 & 53 & 5.24 & 487 & 60 \\
 5 & 4.8 & 509 & 49 & 4.3 & 537 & 50 \\
 6 & 3.77 & 574 & 65 & 3.38 & 606 & 69 \\ 
 7 & 3.17 & 626 & 52 & 2.88 & 657 & 51 \\
 8 & 2.7 & 678 & 52 & 2.47 & 709 & 52 \\
 9 & 2.27 & 740 & 62 & 2.15 & 760 & 51 \\
 10 & 1.97 & 794 & 54 & 1.83 & 824 & 64 \\
 11 & 1.68 & 860 & 66 & 1.57 & 890 & 66 \\
 12 & 1.5 & 910 & 50 & 1.36 & 956 & 66 \\     
\hline
\label{tab:lc_freq_schedule}
\end{tabular}}
\begin{tablenotes}
\item Note -- The LC frequency schedule for the SPTpol experiment. The inductance for all channels is $20 \mu$H. The warm capacitances of each channel are given, along with the expected cold frequency of the channel. The frequencies shown include an expected $4\%$ increase in the resonant frequency at the operational temperature. Also shown is the spacing between frequency channels. At the nominal frequency spacing of $60$kHz there is $\aprx0.3\%$ crosstalk between adjacent channels. At the minimum actual spacing of $49$kHz, there is only $\aprx0.5\%$ crosstalk. Two offset combs, labeled A and B, are used to minimize crosstalk between warm wiring carrying the summed signals from different combs.
\end{tablenotes}
\end{center}
\end{table*}

In addition to the off-resonance crosstalk from neighbors on the LC comb, there exist other modes of crosstalk between channels that must be taken into consideration (For a more detailed discussion of crosstalk see Dobbs et al., 2012 \cite{dobbs12b}.): 

\begin{itemize}

\item Two offset frequency schedules, labeled A and B in Tables \ref{tab:lc_freq_schedule} and \ref{tab:lc_wiring}, are used to minimize inductive crosstalk between wires carrying different combs.

\item If the $i^{th}$ element of a comb coincides with a harmonic of the $j^{th}$ element, power can leak between the two channels. In addition to maintaining good spacing between channels, I also ensured that the higher frequency channels were far ($> 10$kHz) from all harmonics of the lower frequencies. 

\item Adjacent wires in the stripline cable connecting the detector wafer to the LC board can crosstalk through capacitive coupling. To minimize this effect, adjacent frequencies in the comb are assigned to non-adjacent wires on the stripline. Table \ref{tab:lc_wiring} shows this: adjacent wires are also adjacent odd or even numbered pins on the zero insertion force (ZIF) connector on the LC board.

\item Neighboring inductors on an inductor chip can be inductively coupled. To minimize crosstalk at this point, adjacent frequencies are assigned to non-adjacent inductors (See Table \ref{tab:lc_wiring}). Each frequency comb is spread across three separate inductor chips, and no neighboring inductors are adjacent in frequency, even if they belong to separate combs. 

\item Crosstalk from one detector in a pixel to the other results in crosspolar response in that detector. To minimize crosstalk between them in the readout, they are given non-adjacent frequencies.

\item Neighboring pixels will also see correlated atmospheric noise, and sky signals. The mapping from focal plane to LC boards was designed to keep spatially close pixels proximal in frequency space. This way, any crosstalk between channels will introduce correlations between spatially close pixels, which will be filtered out by the lowpass filter applied in mapmaking to remove atmospheric noise.

\end{itemize}

\begin{table*}
\begin{center}
{\scriptsize
\caption{LC Board Wiring Layout}
\begin{tabular}{lccccccc}
\hline \hline
 ZIF & ZIF & Inductor & Inductor & Pixel & Detector & LC & LC \\
 Odd & Even & Chip & Num. & Num. & Num. & Comb & Channel \\
\hline
 1 & 2 & L1 & 8 & 1 & 1 & A & 6 \\
 3 & 4 & L2 & 8 & 1 & 2 & A & 1 \\
 5 & 6 & L3 & 8 & 2 & 1 & A & 7 \\
 7 & 8 & L1 & 7 & 2 & 2 & A & 2 \\
 9 & 10 & L2 & 7 & 3 & 1 & A & 8 \\
 11 & 12 & L3 & 7 & 3 & 2 & A & 12 \\
 13 & 14 & L1 & 6 & 4 & 1 & A & 4 \\
 15 & 16 & L2 & 6 & 4 & 2 & A & 10 \\
 17 & 18 & L3 & 6 & 5 & 1 & A & 5 \\
 19 & 20 & L1 & 5 & 5 & 2 & A & 11 \\
 21 & 22 & L2 & 5 & 6 & 1 & A & 3 \\
 23 & 24 & L3 & 5 & 6 & 2 & A & 9 \\
 25 & 26 & L1 & 4 & 7 & 1 & B & 3 \\
 27 & 28 & L2 & 4 & 7 & 2 & B & 9 \\
 29 & 30 & L3 & 4 & 8 & 1 & B & 1 \\
 31 & 32 & L1 & 3 & 8 & 2 & B & 7 \\
 33 & 34 & L2 & 3 & 9 & 1 & B & 6 \\
 35 & 36 & L3 & 3 & 9 & 2 & B & 12 \\
 37 & 38 & L1 & 2 & 10 & 1 & B & 4 \\
 39 & 40 & L2 & 2 & 10 & 2 & B & 10 \\
 41 & 42 & L3 & 2 & 11 & 1 & B & 2 \\
 43 & 44 & L1 & 1 & 11 & 2 & B & 8 \\
 45 & 46 & L2 & 1 & 12 & 1 & B & 5 \\
 47 & 48 & L3 & 1 & 12 & 2 & B & 11 \\
 49 & 50 & L6 & 1 & 13 & 1 & A & 3 \\
 51 & 52 & L5 & 1 & 13 & 2 & A & 9 \\
 53 & 54 & L4 & 1 & 14 & 1 & A & 2 \\
 55 & 56 & L6 & 2 & 14 & 2 & A & 8 \\
 57 & 58 & L5 & 2 & 15 & 1 & A & 7 \\
 59 & 60 & L4 & 2 & 15 & 2 & A & 1 \\
 61 & 62 & L6 & 3 & 16 & 1 & A & 12 \\
 63 & 64 & L5 & 3 & 16 & 2 & A & 6 \\
 65 & 66 & L4 & 3 & 17 & 1 & A & 5 \\
 67 & 68 & L6 & 4 & 17 & 2 & A & 11 \\
 69 & 70 & L5 & 4 & 18 & 1 & A & 4 \\
 71 & 72 & L4 & 4 & 18 & 2 & A & 10 \\
 73 & 74 & L6 & 5 & 19 & 1 & B & 4 \\
 75 & 76 & L5 & 5 & 19 & 2 & B & 10 \\
 77 & 78 & L4 & 5 & 20 & 1 & B & 3 \\
 79 & 80 & L6 & 6 & 20 & 2 & B & 9 \\
 81 & 82 & L5 & 6 & 21 & 1 & B & 2 \\
 83 & 84 & L4 & 6 & 21 & 2 & B & 8 \\
 85 & 86 & L6 & 7 & 22 & 1 & B & 1 \\
 87 & 88 & L5 & 7 & 22 & 2 & B & 7 \\
 89 & 90 & L4 & 7 & 23 & 1 & B & 11 \\
 - & - & L6 & 8 & R1 & - & B & 5 \\
 - & - & L5 & 8 & R2 & - & B & 6 \\
 - & - & L4 & 8 & R3 & - & B & 12 \\        
\hline
\label{tab:lc_wiring}
\end{tabular}}
\begin{tablenotes}
\item Note -- The wiring mapping for a single side of and SPTpol LC Boards. Two sided PCBs were used, with the same mapping on both sides. This mapping was designed to minimize several sources of crosstalk between channels, including between the two orthogonal polarizations in each pixel, between adjacent wires in the stripline cable connecting the focal plane and LC Board, between adjacent inductors on the eight-inductor chips, and between the warm wires for each LC comb. The last three channels are used for calibration resistors on the LC board, and do not map to detectors on the focal plane. 
\end{tablenotes}
\end{center}
\end{table*}

\subsection{Electronic Components}

\begin{figure}
\begin{center}
\includegraphics[width=5in]{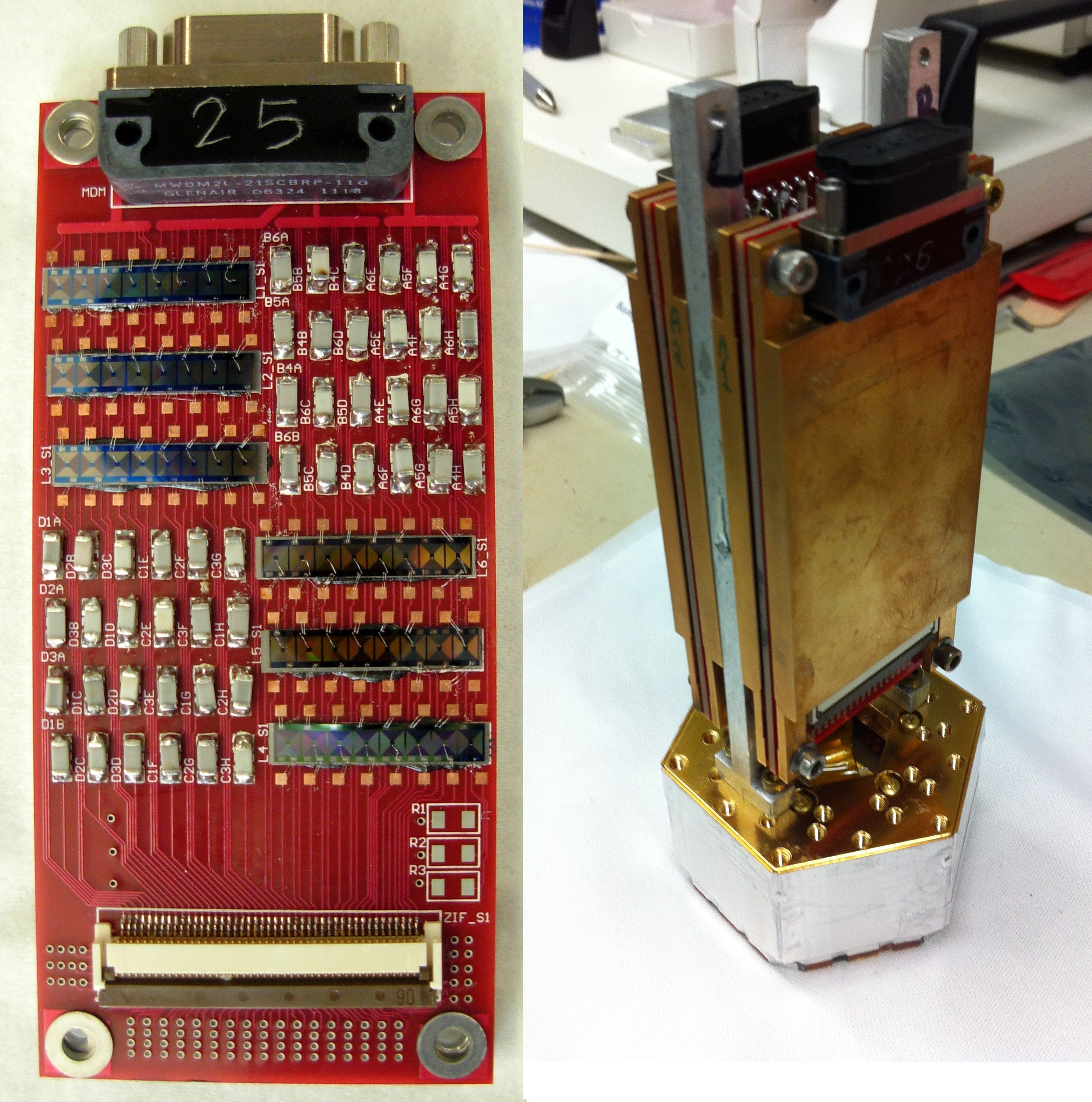}
\caption[SPTpol LC Board and Wafer]{(\textit{Left}) \ LC Board 25 after assembly. The stacked ceramic chip capacitors are visible on the lower left and upper right quadrants, while the inductor chips are visible on the upper left and lower right quadrants. The connector at bottom is the 90-pin ZIF connector which connects the LC board to the detector wafer via a stripline cable. At the top is the 21-pin micro-D-Sub connector which connects to the 4K SQUID boards. (\textit{Right}) \ Two LC Boards with gold plated aluminum shields attached, mounted on the back of a 150GHz detector module.}
\label{fig:lc_board}
\end{center}
\end{figure}

The LC circuitry is implemented on a four layer PCB (See Figure \ref{fig:lc_board}), with tinned traces, which are superconducting at the operational temperature of the board (250mK). There are four LC combs on each side of the board, for a total of 96 channels per board. Each side of the LC board is connected to the focal plane by a stripline cable, with a 90-pin ZIF connector on the LC board side, and wire bonds to the detector wafers on the focal plane side. Only 45 of the LC channels on each side the board are connected to detectors on the focal plane, the remaining three are connected to $1 \Omega \ (\aprx R_\mathrm{TES}) $ resistors on the LC board, and used for measurements of electronic noise in the readout system. 

The ceramic chip capacitors were soldered onto surface mount pads. The inductor chips, each containing eight inductors, were glued onto niobium sheets, which were epoxied to the board with Stycast 2850FT. The niobium foil has a critical temperature of $9.2K$, so at the operating temperature of the boards ($\aprx 275$mK) it is superconducting, and acts as a magnetic shield to prevent inductive coupling through the PCB to the circuits on the opposite side. Furthermore, the capacitors and inductors were arranged into quadrants on the board, with the opposite layout on the reverse side of the PCB, again to reduce inductive or capacitive coupling through the board. The inductors are electrically connected with wirebonds to pads on the PCB. Double bonds were used throughout for reliability.

Originally Panasonic capacitors were used, but these were found to have unacceptably high equivalent series resistance (ESR), which shifts the resonant frequency, and increases the total resistance of the LC comb, which must be subdominant to the bolometer resistance. To resolve this problem, capacitors from various vendors were tested, eventually resulting in capacitors from Murata and TDK being selected. The capacitance shifts down by several percent at the operating temperature, so the capacitors had to be tested cryogenically to completely evaluate them. In the process of testing the capacitors, I discovered that the capacitance of the ceramic chips shifted randomly, and permanently, after the first thermal cycle. Thereafter it became our standard practice to thermally cycle the capacitors several times by immersing them in liquid nitrogen before testing or installing them.

The yield rate of the inductor chips from fabrication was low enough that it was necessary to test the chips individually before installation. The inductors chips were each tested at 4K, by dunk testing in liquid helium on test boards consisting of eight-channel LC combs. $\aprx 70\%$ of the chips had eight good inductors out of eight, however it was also possible to use chips with seven good inductors out of eight in places. Since three of the channels on each side of the board connected to calibration resistors, some of these channels could be wired to known bad inductors without adversely affecting the readout yield on the focal plane.

The summed signal from the LC combs was read out through a 21-pin micro-D-Sub connector (Glenair part number MWDM2L-21SCBRP-.110). The boards were covered on both sides by gold plated aluminum shields, for physical protection of the inductors and wirebonds, and to prevent RF coupling to the boards. These have the added advantage of reducing the radiative heat load on the cold stage of the instrument, by covering the otherwise IR-black PCBs with a reflective surface. The LC boards were mounted on the back of the focal plane with aluminum jigs, two per hex wafer.

$50 \Omega$ termination resistors were used between the signal pins of the readout micro-D-Sub connector. The particular resistors used proved to be problematic. We used Vishay Dale thin film resistors, part numbers PAT49.9CCT-ND and PAT49.9BCT-ND. The resistive element in these products is a tantalum nitride thin film, $\aprx 1$nm thick, deposited with reactive sputtering. When the LC boards were first tested cryogenically at 275mK, strange behavior was found which was eventually traced back to the termination resistors. It was determined that they had become superconducting at some point above the operational temperature, resulting in shorts across all the readout channels. 

Upon further research, it was found that while bulk TaN has no documented superconducting transition down to 1.28K \cite{kilbane75}, TaN$_x$ thin films have superconducting transitions between 4.2K and 8.15K \cite{kilbane75,horn47,gerstenberg64,chaudhuri13}. 

These thin films are usually produced by reactive sputtering, and their ratio of Ta and N is typically not the stoichiometric ratio (1:1). The variation in $\Tc$ depends on the ratio of Ta to N in the material. Kilbane and Habig \cite{kilbane75} found that the maximum $\Tc$ occurs approximately at the stoichiometric ratio ($52\pm5\%$), with the $\Tc$ falling by $\aprx0.5$K for each percent change in the nitrogen content away from the optimum ratio.

The transition in TaN thin films is believed to be the result of their crystalline structure. TaN thin films have a face-centered cubic (FCC) lattice structure. Bulk TaN, or Ta$_2$N thin films, have hexagonal or tetragonal crystallographic phases, and exhibit no superconducting transition, to the lowest measured temperatures (1.5K for Ta$_2$N, 1.28K for bulk TaN).

The offending TaN termination resistors were replaced with Yageo $51 \Omega$ nickel chromium resistors, part number 311-51BCT-ND, which remain normal down to our operating temperature of 275mK.

A valuable lesson for all cryogenic scientists can be derived from this experience: much anxiety and frustration can be avoided by ensuring your resistors do not have superconducting transitions above your operating temperature.

\subsection{Assembly}

Seventeen boards were required to read out the SPTpol focal plane, and a total of 26 were assembled by Liz George, Nick Harrington, and I at the University of California, Berkeley. This included extra LC boards to be used in detector testing cryostats, and spare boards to remain on site at the South Pole. This Herculean labor entailed cryogenically testing over 400 inductor chips, soldering and measuring over 4,000 surface mount capacitors, soldering 5,400 ZIF connector contacts, and making over 20,000 wire bonds, using more than 100m of bonding wire. 

Several important things were learned during the course of the design, assembly, and testing of the SPTpol LC boards which informed the design of the readout system for SPT-3G, the next generation instrument on the South Pole Telescope. While the ESR of the commercially available ceramic chip capacitors was acceptable for SPTpol, it would not be sufficient for the more stringent requirements in SPT-3G. Similarly, the yield of the inductors was too low to be feasible for SPT-3G, and more robust inductors were needed. 

Individual testing of components would also not be possible, given the order of magnitude increase in the number of channels required. Assembly itself would need to be simplified for similar reasons. The testing of components and subsequent assembly of the SPTpol LC boards required four person months of effort. For SPT-3G, a similar testing and assembly pathway would require $\aprx 40$ person months.

All these factors lead to the development of custom made chips which contained all the inductors and capacitors, and which had better cryogenic electrical properties than generic commercial products. This allows chips to be tested as a whole, and removes all internal assembly and bonding of the capacitors and inductors. Furthermore, automatic wirebonding was implemented for the assembly of the LC boards, and the bonding of the stripline cables to the detector wafers, using automatic wirebonding machines at FNAL and LBNL.

\section{Network Mapping}

To read out all the data channels on an instrument, it is necessary to develop a hardware map. That is, a mapping of the connections from the detector level up through the LC channels, SQUIDS, and readout boards, which specifies all details of the readout chain for each detector. Most of this mapping is hard coded when the instrument is assembled: which detectors are wired to which LC boards, and which LC boards are connected to which DfMux readout boards is completely known. However, there is a computational problem to mapping the the expected frequencies of the LC comb to the observed resonance peaks after the LC boards are cooled to their operating temperature.

If all the projected LC channels on a comb are present, at precisely the expected frequencies, then it is trivial to match the observed resonances to the expected channels, and thus to the proper bolometers. However, in practice peaks may be missing, and there may be various shifts in the observed frequencies which make it more challenging to map the observed resonances to the proper channels.

In SPTpol, it is still marginally feasible to examine each comb by eye, and assign the channels properly. However this is time consuming, and for SPT-3G will be completely unworkable. To simplify the process of developing a network map, and to prepare for SPT-3G, I wrote software to map the observed resonances to the expected LC channels. This software can correct for uniform shifts in all the frequencies, and shifts in which the offset is a linear function of frequency. If fewer than the expected number of resonances are observed, it fits for which of the expected channels are missing, using the total distance the observed resonances must be shifted to match the expected frequencies as the fitting metric.

%Ysz and Mass Scaling Relations

\doublespacing

\chapter[\YSZ \ and Mass Scaling Relations]{\YSZ \ and Mass Scaling Relations \footnote{This is an author-created, un-copyedited version of an article accepted for publication in \textit{The Astrophysical Journal} (Saliwanchik et al., 2015 \cite{saliwanchik15}). The publisher is not responsible for any errors or omissions in this version of the manuscript or any version derived from it. The Version of Record is available online at http://dx.doi.org/10.1088/0004-637X/799/2/137.}
}
\label{ch:ysz-m}

\section{Introduction}

Galaxy clusters are the largest gravitationally collapsed systems in the observed universe, 
and their abundance as a function of mass and redshift is a sensitive probe of the growth of structure in the universe.  
The ability to accurately and precisely determine cluster masses is essential for using them to constrain cosmological parameters. 
Typically this is done through cluster observables, which do not directly measure cluster mass, but can be related to it through scaling relations \citep{vikhlinin09, rozo10, mantz10b, benson13}.
The Sunyaev-Zel'dovich (SZ) effect \citep{sunyaev72} is caused by the inverse Compton scattering of CMB photons off of hot intracluster gas. It is a measure of the line-of-sight integral of the cluster pressure and is expected to be a low-scatter proxy for cluster mass \citep{carlstrom02, kravtsov06a}.
In particular, the integrated Comptonization of a cluster, \YSZ, is expected to have a low intrinsic scatter with cluster mass and to be relatively insensitive to cluster astrophysics \citep{barbosa96, holder01a, motl05, nagai07, fabjan11}.

However, for SZ observations where the cluster size is on the order of the instrument beam size or smaller, there is typically a degeneracy in the constraints on the amplitude and shape of the assumed cluster profile \citep[e.g.,][]{benson04, planck11-5.1a, planck13-29}.
One potential way of handling this degeneracy is to employ a Bayesian analysis method. A number of experiments have used Monte Carlo methods to characterize the profiles of galaxy clusters in recent years, including Bolocam \citep{sayers13b}, the Arcminute Microkelvin Imager (AMI) \citep{ami12a, ami12b}, the Planck Collaboration \citep{planck12-5, planck12-2}, and the Atacama Cosmology Telescope (ACT) and Sunyaev-Zel'dovich Array (SZA) \citep{reese12}.
In this work, we present a Markov-Chain Monte Carlo (MCMC) analysis method for analyzing observations of the SZ effect, which determines \YSZ\ while marginalizing over other SZ model parameters. A feature of the MCMC method is that the \YSZ\ estimates it produces are well constrained even for clusters with relatively small radii on the sky. This method is related to the method presented by Montroy et al. \citep{montroy15}, which uses the same likelihood, but employs a rapid grid method to directly evaluate the likelihood throughout the parameter space.

We apply this MCMC method to simulated and real observations from the SPT.
Various experiments have examined scaling relations between SZ signal and optical or X-ray data. The Planck Collaboration has examined \YSZ-\LX\ scaling relations \citep{planck11-10} and \YSZ-$M_\mathrm{X}$ scaling relations \citep{planck13-20}. Segal et al. \citep{sehgal11}, Sifon et al. \citep{sifon13}, and Hasselfield et al. \citep{hasselfield13} investigate the scaling between central Comptonization (\yzero) and mass, or \YSZ\ and mass for ACT clusters.
Previous analyses of clusters observed in the SPT-SZ survey used the cluster detection significance, $\xi$,  as a proxy for cluster mass \citep{vanderlinde10, andersson11, benson13, reichardt13}. 
Here we show that \YSZ\ integrated over a fixed angular aperture near the SPT beam size and $\xi$ have comparable fractional scatter in their respective mass scaling relations.  
\YSZ, however, is more easily compared to cluster parameters derived from other measurements.

\section{Cluster Sample and Observations}
\label{sec:clusters_and_obs}

\subsection{SZ Observations}
\label{sec:obs}

In 2007-2011, the SPT surveyed 2500 \sqdeg \ in three frequency bands centered at 95, 150, and $220\,$ GHz. This survey is referred to as the SPT-SZ survey. The cluster sample used in this work is drawn from the two fields ($\aprx100$ \sqdeg \ each) observed with the SPT in 2008, one centered at right ascension (RA) $5^\mathrm{h}30^\mathrm{m}$, declination (Dec) $-55^\circ$ (J2000), and one at RA $23^\mathrm{h}30^\mathrm{m}$, Dec $-55^\circ$. A nearly identical cluster sample was used by Vanderlinde et al. \citep{vanderlinde10}(hereafter V10), Andersson et al. \citep{andersson11}(hereafter A11), and Benson et al. \citep{benson13}(hereafter B13).

Observing procedures, data processing, and detection algorithms for these clusters are described in detail in V10 and Staniszewski et al. \citep{staniszewski09}, and are summarized here. Details of the data processing pipeline are also described by Schaffer et al. \citep{schaffer11}.

Each field was observed by scanning the telescope back and forth in azimuth at $0.25^\circ/\mathrm{s}$, and then stepping in elevation and repeating until the entire field was covered. This process covers a 100 \sqdeg \ field in \aprx 2 hours. Field scans were repeated several hundred times until the noise in the co-added maps reached a completion depth of 18 \uk-arcmin for \onefifty. (See Staniszewski et al. \citep{staniszewski09}, V10, or Williamson et al. \citep{williamson11} for a description of field depth measurements.) The timestreams of the individual detectors were filtered to remove sky signal that was spatially correlated across the focal plane and long timescale detector drift. The combination of these filters effectively removes signals with angular scales larger than \aprx$0.5^\circ$. Data from individual detectors were combined using inverse-variance weighting, and the resulting maps were calibrated by comparison to the WMAP 5-year CMB temperature anisotropy power spectrum \citep{lueker10a}.

\subsection{Cluster Detection}
Clusters are identified in the SPT maps using a matched filter (MF) \citep{haehnelt96, herranz02a, herranz02b, melin06}.
Specifics on this procedure can be found in Staniszewski et al. \citep{staniszewski09} and V10 for single frequency cluster detection, and in Williamson et al. \citep{williamson11} and Reichardt et al. \citep{reichardt13} for multi-frequency detection.
To locate clusters, the maps are multiplied in Fourier space with a filter matched to the expected spatial signal-to-noise profile of galaxy clusters. The matched filter, $\psi$, is given by: 
\begin{equation}
\psi(k_x,k_y) = \frac{B(k_x,k_y) S(|\vec{k}|)}{B(k_x,k_y)^2 N_{\mathrm{astro}}(|\vec{k}|) + N_{\mathrm{terr}}(k_x,k_y)},
\label{eq:psi} \end{equation}
where $B$ is the instrument response after timestream filtering, $S$ is the source template, and the noise has been divided into astrophysical ($N_{\mathrm{astro}}$), and terrestrial ($N_{\mathrm{terr}}$) components. $N_{\mathrm{astro}}$ includes power from lensed primary CMB anisotropies, an SZ background from faint undetected clusters, and millimeter-wave emitting point sources. The noise power spectrum $N_{\mathrm{terr}}$ includes atmospheric and instrumental noise, estimated from jackknife maps. The source template is a two dimensional projection of an isothermal $\beta$-model, with $\beta$ set to 1 \citep{cavaliere76}:
\begin{equation} \label{eq:beta}
\Delta T = \Delta T_0 (1+\theta^2/\thcoresq)^{-1},
\end{equation}
where the central SZ temperature decrement $\Delta T_0$ and the core radius \thcore \ are free parameters.

Clusters are detected using a (negative) peak detection algorithm similar to SExtractor \citep{bertin96}. The significance of a detection, $\xi$, is defined to be the highest signal-to-noise (S/N) ratio across all \thcore.  In our analysis we use the unbiased significance,  $\zeta = \sqrt{\langle \xi \rangle^2 - 3}$, where $\langle \xi \rangle$ is the average detection significance of a cluster across many noise realizations (V10).

It is important to note here that maximizing the signal-to-noise in the MF maximizes the likelihood used in the MCMC method below (Section \ref{sec:methods}), at a fixed \thcore. 
The two methods are exploring the same likelihood, the difference is that the MF is optimized for detecting clusters not parametrizing them, since it does not simultaneously explore all dimensions in the parameter space.

\subsection{Optical and X-ray Observations}
\label{sec:Xray_obs}

The optical and X-ray observations for the clusters used in this work have previously been described in A11 and B13, which we briefly describe here. 
All eighteen clusters have redshift measurements, fifteen of which are spectroscopic, and fourteen of the clusters have X-ray measurements. 

Optical $griz$ imaging and photometric redshifts for these clusters were obtained from the Blanco Cosmology Survey \citep{desai12}, and from pointed observations using the Magellan telescopes \citep{high10}. 
Of the fifteen clusters with spectroscopic redshifts, eight were obtained by the Low Dispersion Survey Spectrograph (LDSS3) on the Magellan Clay 6.5-m telescope \citep{high10}, and one by the Inamori Magellan Areal Camera and Spectrograph (IMACS) on the Magellan Baade 6.5-m telescope \citep{brodwin10}. 
The final six cluster redshifts were measured with IMACS and GMOS on Gemini South \citep{ruel13}. 
X-ray follow-up observations were performed with \emph{Chandra} ACIS-I and \emph{XMM-Newton} EPIC (A11, B13).

\section{MCMC Analysis Methods}
\label{sec:methods}

In this work we follow a Bayesian approach to parameter estimation, using a Markov-Chain Monte Carlo method to estimate the parameters of our cluster source model.
For an review of the essential differences between the Frequentist and Bayesian statistical methods, see Monroe \cite{monroe12}.
The application of MCMC methods to the detection and characterization of compact astrophysical sources in noisy backgrounds was proposed by Hobson and McLachlan \citep{hobson03}, and several experiments have used MCMC methods for parametrizing SZ signals from galaxy clusters. Bonamente et al. \citep{bonamente04, bonamente06}, and LaRoque et al. \citep{laroque06} used MCMC methods to analyze SZ data from BIMA and OVRO, in conjunction with X-ray data from \emph{Chandra}, and fit $\beta$-model profiles to galaxy clusters. 
Muchovej et al. \citep{muchovej07}, Culverhouse et al. \citep{culverhouse10}, and Marrone et al. \citep{marrone09} parameterized SZA clusters, and Halverson et al. \citep{halverson09} parameterized the Bullet Cluster using APEX-SZ data, all using the $\beta$-model. Culverhouse et al. \citep{culverhouse10}, and Marrone et al. \citep{marrone09, marrone12} additionally estimated cluster \YSZ\ values.
In recent years there has been a surge of interest in MCMC methods for parametrizing clusters from a number of experiments, including Bolocam \citep{sayers13b}, AMI \citep{ami12a, ami12b}, the Planck Collaboration \citep{planck12-5, planck12-2}, and ACT and SZA \citep{reese12}.
Here we estimate galaxy cluster \YSZ\ values and \YSZ-$M$ scaling relations in addition to estimating $\beta$-model parameters.

\subsection{Posterior Distribution for a Compact Source}

We use a Metropolis-Hastings algorithm implementation of the MCMC method for parameter estimation \citep{metropolis53, hastings70}.
For the case of a compact object with source template $S(\mathcal{H})$ in a two dimensional astronomical dataset $D$ with Gaussian noise, the Bayesian likelihood has the form:
\begin{equation}
\mathrm{P}(D|\mathcal{H}) = \frac{\mathrm{exp}(-\frac{1}{2}[D-S(\mathcal{H})] C^{-1} [D-S(\mathcal{H})]^*)}{(2\pi)^{N_{\mathrm{pix}}/2} |C|^{1/2}},
\label{eq:Pr_1band} \end{equation}
where $C$ is the noise covariance matrix for the dataset $D$, and $N_{\mathrm{pix}}$ is the number of pixels in $D$ \citep{hobson03}. In this method, $C$ is composed of the combined $N_{\mathrm{astro}}$ and $N_{\mathrm{terr}}$ noise terms in the matched filter $\psi$ (equation \ref{eq:psi}).

We are interested in parametrizing galaxy clusters using the SZ effect, which is the spectral distortion they produce in the blackbody CMB spectrum. At two of the SPT's observing frequencies (95 and \onefifty) this distortion is manifested as a decrement in CMB power, while the net change in CMB power at $220\,$ GHz is negligible.

Equation \ref{eq:Pr_1band} is easily generalizable to the case of astronomical images in multiple frequency bands, where the unnormalized log likelihood may be calculated in the Fourier domain as \\ \\
$ \begin{array}{@{\hspace{0mm}}r@{\;}l@{\hspace{0mm}}}
\mathrm{Log} \left( \overline{\mathrm{P}}(D|\mathcal{H}) \right) =
\end{array} $
\begin{equation} -\frac{1}{2}\sum_{\bar{k},\nu_i,\nu_j}\left[\widetilde{D}_{\nu_i}(\bar{k})-\widetilde{s}^{\mathcal{H}}_{\nu_i}(\bar{k})\right]N^{-1}_{\nu_i \nu_j}(\bar{k})\left[\widetilde{D}_{\nu_j}(\bar{k})-\widetilde{s}^{\mathcal{H}}_{\nu_j}(\bar{k})\right]^*,
\end{equation}
where $\bar{k}$ is the two dimensional Fourier space vector, $\widetilde{D}_{\nu_i}(\bar{k})$ is the Fourier transform of the map for frequency $\nu_i$, $\widetilde{s}^{\mathcal{H}}_{\nu_i}(\bar{k})$ is the frequency dependent Fourier transform of the cluster model for parameter set $\mathcal{H}$, and $N_{\nu_i\nu_j}(\bar{k})$ is the frequency dependent covariance matrix for the $\nu_i$ and $\nu_j$ frequency maps. Here $N_{\nu_i\nu_j}(\bar{k})$ is simply the multiband extension of the covariance matrix $C$ in equation \ref{eq:Pr_1band}. It is convenient to perform the likelihood calculations in Fourier space rather than physical space because $N_{\nu_i\nu_j}(\bar{k})$ is diagonal in Fourier space, assuming stationary noise.

The frequency dependent covariance matrix is computed as follows. For a given $\bar{k}$, the two frequency matrix for CMB+Noise covariance is:
\begin{equation}
N_{\nu_i\nu_j}(\bar{k}) = 
\begin{bmatrix}
  C(\bar{k}) B_1(\bar{k})^2 + N_1(\bar{k}) & C(\bar{k}) B_1(\bar{k}) B_2(\bar{k}) \\
  C(\bar{k}) B_1(\bar{k}) B_2(\bar{k}) & C(\bar{k}) B_2(\bar{k})^2 + N_2(\bar{k})
\end{bmatrix}
\end{equation}
where $C(\bar{k})$ is the CMB power at $\bar{k}$, $B_1(\bar{k})$ and $N_1(\bar{k})$ are the beam and noise for the first frequency, and $B_2(\bar{k})$ and $N_2(\bar{k})$ are the beam and noise for the second frequency.

The covariance becomes more complicated when including point sources and the SZ background because the signal component has a different magnitude at different frequencies. If $Q_1(\bar{k})$ and $Q_2(\bar{k})$ are the point source or SZ covariance at two of our observing frequencies then we have:
\begin{equation}
\begin{aligned}
& N_{\nu_i\nu_j}(\bar{k}) = \\
& \begin{bmatrix}
C(\bar{k}) B_1(\bar{k})^2 + N_1(\bar{k})  + B_1(\bar{k})^2 Q_1(\bar{k}) & C(\bar{k}) B_1(\bar{k}) B_2(\bar{k}) +  B_1(\bar{k}) B_2(\bar{k}) \sqrt{Q_1(\bar{k}) Q_2(\bar{k})}\\
C(\bar{k}) B_1(\bar{k}) B_2(\bar{k})  + B_1(\bar{k}) B_2(\bar{k}) \sqrt{Q_1(\bar{k}) Q_2(\bar{k})} & C(\bar{k}) B_2(\bar{k})^2 + N_2(\bar{k}) + B_2(\bar{k})^2 Q_2(\bar{k})
\end{bmatrix}
\end{aligned}
\end{equation}

Before the log-likelihood is calculated, to account for the filtering of the dataset $\widetilde{D}_{\nu_i}(\bar{k})$ as described in Section \ref{sec:obs}, both the covariance $N_{\nu_i\nu_j}(\bar{k})$, and the source template $\widetilde{s}^{\mathcal{H}}_{\nu_i}(\bar{k})$, are multiplied by a Fourier space filter function which emulates the timestream filtering used to produce $\widetilde{D}_{\nu_i}(\bar{k})$.

\subsection{Implementation}

Our MCMC is modeled after the generic Metropolis-Hastings method described by Hobson and McLachlan \citep{hobson03}, and is implemented in MATLAB \footnote{Mathworks Inc., Natick MA, 01760}. 

In this work, we use the MCMC method for cluster parametrization, not detection. Our testing found that it was more computationally costly and not more effective at cluster detection than the MF method.  
Throughout this work, our MCMC is run over a relatively small area of sky (512 pixels $\times$ 512 pixels, or \aprx $2^\circ \times 2^\circ$) centered on a cluster which has already been identified. 

Cluster parameter recovery is tested in single and multi-frequency simulations below (\S \ref{sec:params}), but we use only \onefifty \ when investigating scaling relations (for observed and simulated clusters) to match the SPT cluster analysis in B13, from which our sample is derived. We use the $\beta$-model source template given in equation \ref{eq:beta}.
Montroy et al. \citep{montroy15} demonstrate, using simulations and methods similar to those described in \S \ref{sec:params}, that \YSZ\ is recovered accurately with a $\beta$-model for either $\beta$-model or Arnaud profile \citep{arnaud10} input clusters.  

Clusters, as described by the $\beta$-model, are characterized by four parameters: their location on the sky in RA and Dec, the magnitude of the SZ temperature decrement \tzero, and the core radius \thcore. We apply priors in the form of uniform probability distributions in each parameter. Given that we are characterizing clusters that have already been detected by the MF, our position priors can be quite tight. We impose a simple square-box prior on RA and Dec, centered at the MF cluster location and extending $\pm 1.25'$. Our \tzero \ and \thcore \ priors restrict these parameters to broadly reasonable values given the expected mass and redshift range of our cluster sample. Our SZ temperature decrement prior is $-2.5~\mathrm{mK} \le \tzero \le 0.0~\mathrm{mK}$, and our radius prior is $0.025' \le \thcore \le 2.5'$. \thcore \ is not allowed to fall to zero for numerical reasons.

For a detailed examination of the SPT beam functions and noise properties, see Schaffer et al. \citep{schaffer11}. Figure 2 of that work shows how the SPT beams scale with physical radius and $\ell$. Figure 7 shows the signal+noise and noise PSDs for an SPT map, both from the raw map, and corrected for the beam and transfer functions.

\begin{figure}
\centering
\includegraphics[height=5.75in, angle=270]{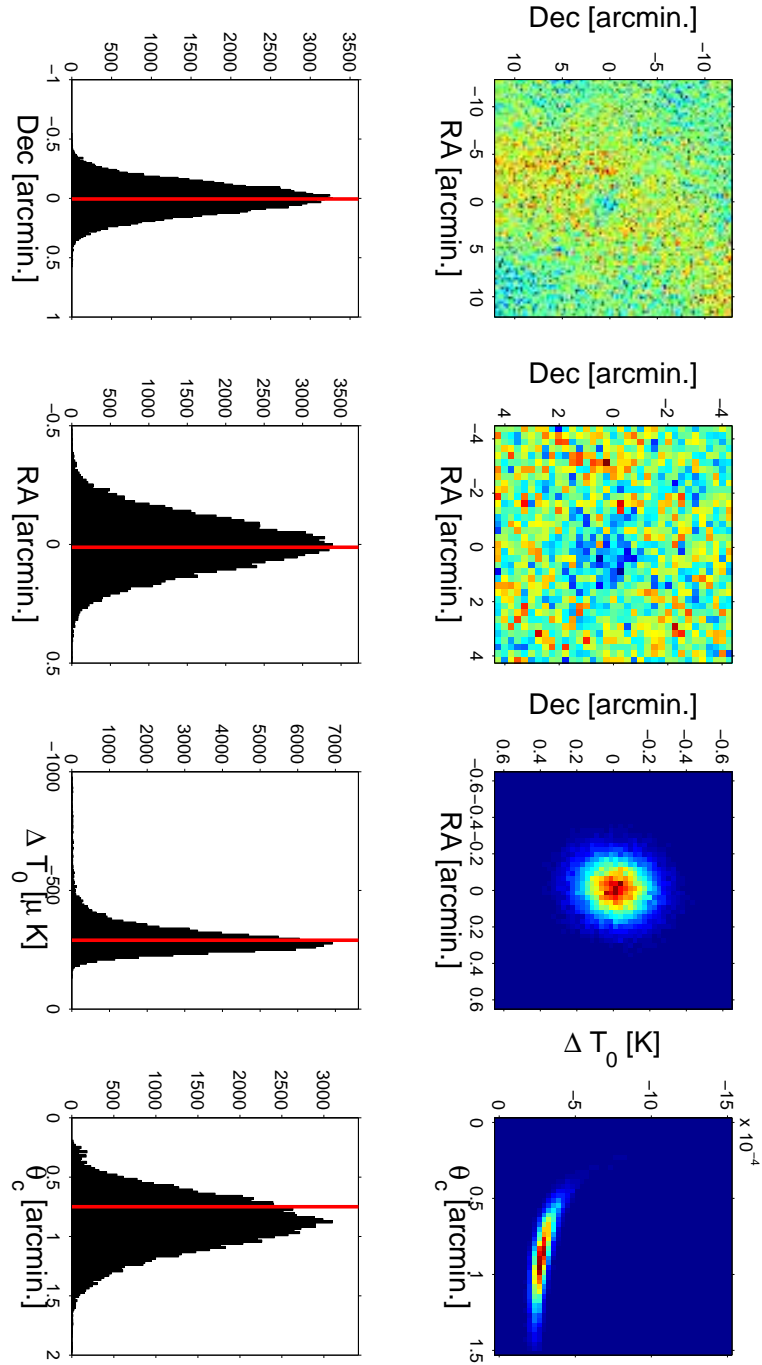}
\caption[Cluster parameter marginalized distributions]{From left to right the top row shows (1) a $25' \times 25'$ section of an SPT sky map centered on cluster SPT-CL J$2341$-$5119$ ($\xi = 9.65$, $z = 0.9983$) (2) a close up of $7.5' \times 7.5'$ centered on the cluster location, (3) the estimated posterior
distribution of the cluster position, marginalizing over \tzero \ and \thcore, and (4) the estimated posterior distribution of \tzero \ and \thcore, marginalizing over position. Likewise the bottom row shows one dimensional marginalized distributions of the parameters (5) Declination (6) Right Ascension, (7) \tzero, and (8) \thcore. 
Vertical red lines in the bottom row indicate the matched filter parameter values for this cluster.}
\label{fig:param_dist}
\end{figure}

Burn-in, as evaluated by stability of the likelihood values, is typically complete within several hundred steps. For the 12,000 simulated cluster realizations in \S \ref{sec:params} we cut the first $10^3$ steps, using the rest of the $10^4$ steps to characterize the probability surface. In the scaling relation analysis discussed in \S \ref{sec:scaling} many fewer clusters were analyzed, allowing the chain length to be extended to $10^5$ steps, from which we exclude the first  $10^4$ steps in order to ensure convergence. We define recovered parameter values to be the median of the MCMC equilibrium distribution for each parameter, marginalizing over the other parameters.  Uncertainties are given by the $68\%$ confidence interval of the marginalized distribution for each parameter, centered on the median value. Figure \ref{fig:param_dist} shows the parameter distributions for a typical cluster detected with the SPT.

\section{Simulations}
\label{sec:sims}

\subsection{Simulated Thermal SZ Cluster Maps}

We used two sets of simulations; one uses $\beta$-model clusters (defined by \tzero \ and $\thcore$) to investigate cluster parameter recovery (\S \ref{sec:params}), while the second uses cluster gas profiles inferred from dark matter light cone simulations to calibrate \YSZ-$M$ scaling relations (\S \ref{sec:scaling}). The second set of simulations is described in detail in Shaw et al. \citep{shaw10}, and will be referred to as the S10 simulations for convenience. The thermal SZ (tSZ) cluster profiles used in each set are discussed in more detail in the relevant sections below.

\subsection{Astrophysical Backgrounds}
\label{sec:astro}

We use simulated maps of astrophysical backgrounds that include contributions from the CMB and extragalactic point sources. 
Simulated CMB anisotropies were generated based on realizations of the gravitationally lensed WMAP 5-year \LCDM \ CMB power spectrum.

The extragalactic point source population at \onefifty \ consists of two classes of objects: ``dusty'' sources dominated by thermal dust emission from star formation bursts, and ``radio'' sources dominated by synchrotron emission.  We use the source count model of Negrello et al. \citep{negrello07} at 350 GHz, which is based on physical modeling by Granato et al. \citep{granato04} for dusty sources. Source counts at \onefifty \ are estimated by assuming the flux densities scale as $S_{\nu} \propto \nu^\alpha$, where $\alpha = 3$ for high-redshift protospheroidal galaxies, and $\alpha = 2$ for late-type galaxies. For radio sources we use the De Zotti et al. \citep{dezotti05} model at \onefifty, which is in agreement with observed radio source populations at $S < 100$ mJy \citep{vieira10, marriage11a, planck11-5.1a, marsden13, mocanu13}. 

Point source population realizations were generated by sampling from Poisson distributions for each population in bins with fluxes from 0.01 mJy to 1000 mJy. Sources were randomly distributed across the map. Correlations between sources or with galaxy clusters were not modeled, following V10.
These \onefifty \ simulated point source populations were used for the scaling relation simulations of \S \ref{sec:scaling}, but not for the multiband pipeline checks of \S \ref{sec:params}.

\subsection{Simulated Observations}
\label{sec:instr}

Ideally, we could emulate the SPT transfer function for the $95\,$ GHz and \onefifty\ frequency bands by producing synthetic timestreams from simulated maps convolved with the SPT beam, observing them using the SPT scan strategy, and convolving the resulting timestreams with detector time constants. We would then produce maps by processing the simulated timestreams as in \S \ref{sec:obs}. However, this full emulation of the SPT transfer function is computationally intensive. To simplify this process and produce a large number of sky maps, we model the transfer function as a two dimensional Fourier filter. V10 shows that this approximation introduces systematic errors in the recovered cluster $\xi$ values of less than 1\%.

The instrumental and atmospheric noise in the SPT maps were estimated by creating difference maps, which were constructed to have no astrophysical signal.
Each field consists of several hundred individual observations.
We randomly multiply half of the observations by -1, and then coadd the full set of observations.
We repeat this several hundred times, each time calculating the two-dimensional spatial power spectrum, which we average to estimate the instrumental and atmospheric noise in the coadded SPT map.
This averaged noise spectrum is used to generate random map realizations of the SPT noise, which are added to the simulated maps.

\section{Pipeline Checks}
\label{sec:params}

\subsection{Cluster Model}
\label{sec:TM_sims}

We use mock observations of clusters in simulated sky maps to evaluate the accuracy and bias of the recovered cluster parameters.
We begin with simulated maps that contain the astrophysical signals described in Section \ref{sec:astro}.
To this we add mock clusters with an assumed $\beta$-model profile, with known SZ decrements and radii, at specified locations.
Simulated SPT observations are then performed on these maps.  
Three different cluster core radii ($0.25'$, $0.5'$, and $1.0'$) are used, combined with eight values for peak Comptonization between 175 \uk \ and 2 mK, spanning the range of values typically found for SPT-detected clusters with $\xi > 5$. 
These cluster profiles are convolved with the SPT transfer function, and then placed in the simulated maps.
For each combination of $\beta$-model cluster parameters we create five pairs of simulated maps (\onefifty \ and $95\,$ GHz) by placing 100 copies of the cluster at random locations in five unique noise maps. 
This results in 500 noise realizations for each combination of cluster parameters, or 12,000 clusters total. 
As usual, \aprx $2^\circ \times 2^\circ$ cutouts are made around each cluster, and the MCMC is run on each patch. 

In \S \ref{sec:fund_parms} and \S \ref{sec:YSZ} we test parameter recovery in the single-band (150 GHz) and multiband ($95\,$ GHz and \onefifty) cases.
The $220\,$ GHz data contain no SZ information, but could in principle be used to remove primordial CMB anisotropy. However, the noise level of the $220\,$ GHz maps, dominated by residual atmospheric emission, is larger than the intrinsic astrophysical confusion caused by CMB anisotropy.  Therefore, we do not use the $220\,$ GHz SPT measurements to fit the cluster model.

\subsection{Position, Radius, and Amplitude}
\label{sec:fund_parms}

\begin{figure}
\includegraphics[bb = 40 264 547 519, clip=true, angle=0, width=\columnwidth]{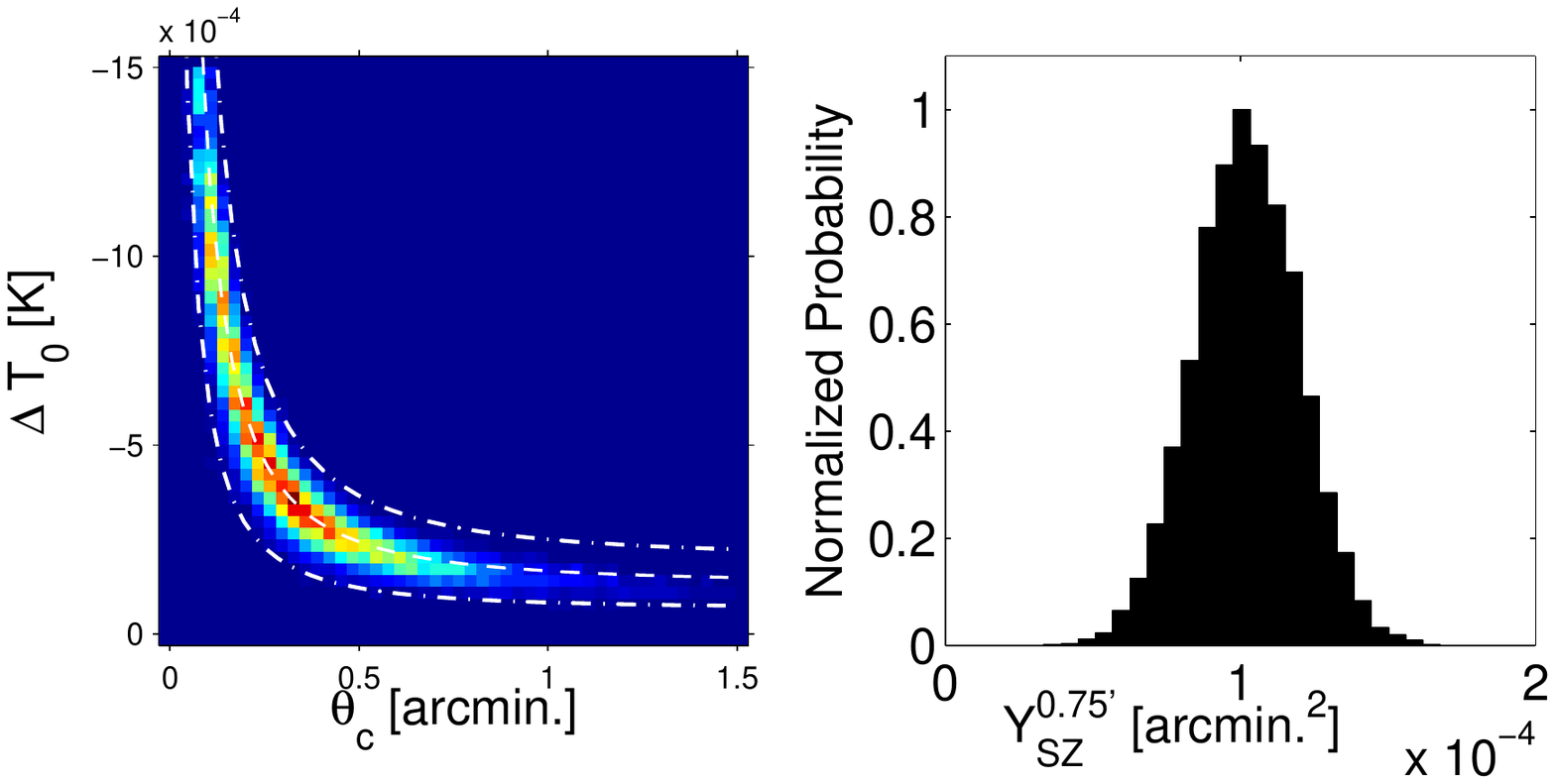}
\caption[\tzero-\thcore\ degeneracy and \YSZ\ marginalized distribution]{SPT-CL J$0533$-$5005$ ($\xi = 5.59$, $z = 0.8810$), an SPT observed cluster with core radius $\thcore < 1'$. The left figure shows the estimated posterior distribution of \tzero\ and \thcore, and the right figure shows the posterior distribution of \YSZ\ calculated from the $\tzero + \thcore$ distribution. For clusters near the SPT beam size ($\aprx1'$ FWHM at \onefifty) and selection threshold, the position is well constrained, however the radius and amplitude are degenerate.  Despite this, the integrated Comptonization, \YSZ, is well constrained. The over-plotted curves in the left figure are \YSZ\ iso-curves. The dashed line is the recovered \YSZ\ for this cluster, while the dot-dashed lines are $\pm50\%$ \YSZ.}
\label{fig:degen}
\end{figure}

We first examine the recovered values of the four baseline cluster parameters: the right ascension (RA) and declination (Dec) position, \tzero, and \thcore.  
The cluster positions are measured accurately (See Figure \ref{fig:param_dist}), and we find no bias in either position parameter, for both the single-band and multiband cases.
For clusters near the SPT beam size ($\aprx1'$ FWHM at \onefifty) and selection threshold, the amplitude and shape of the cluster will not be well constrained, however the integrated signal within the SPT beam will be.
A similar degeneracy has previously been noted in other cluster analyses \citep[e.g.,][]{benson04, planck11-5.1a, planck13-29}.
In Figure \ref{fig:degen}, we show the recovered \tzero\ and \thcore\ distributions for a typical cluster in the SPT catalog (SPT-CL J$0533$-$5005$, $\xi = 5.59$, $z = 0.8810$, $\thcore < 1.0'$).
While the position is well-constrained, there is a significant degeneracy between the constraints on \thcore\ and \tzero.

\subsection{Integrated Comptonization}
\label{sec:YSZ}

In general, the cylindrically projected integrated Comptonization of a cluster is calculated by integrating the source function, $S(\theta)$, out to a given angular aperture \thint:
\begin{equation}
\YSZ = 2 \pi \int_0^{\thint}S(\theta) \ \theta \ \mathrm{d}\theta.
\label{eq:ysz_theta}
\end{equation}
For much of this work, \thint \ will be a constant angular aperture. We distinguish this estimator of \YSZ\ from others by referring to it as \Ytheta \ hereafter.

In the case of a two dimensional projection of a spherical $\beta$-model with $\beta=1$ (equation \ref{eq:beta}), this integral can be solved analytically:
\begin{equation}
\Ytheta = \frac{\pi \tzero \thcoresq}{\fx \tcmb} \ \mathrm{Log} \left[1+\left(\frac{\thint}{\thcore}\right)^2\right],
\label{eq:y_theta}
\end{equation}
where \thcore\ is the core radius in arcminutes, \tzero\ is the central temperature decrement in units of $K_\mathrm{CMB}$, the equivalent CMB temperature fluctuation required to produce the observed power fluctuation, \tcmb\ is the CMB blackbody temperture of 2.725~K, and \fx\ is given by:
\begin{equation}
\fx = \left( x \frac{e^x+1}{e^x-1} - 4 \right) \left[ 1+\delta(x,T_e) \right],
\end{equation} 
where $x = h \nu / k \tcmb$, and $\delta(x,T_e)$ accounts for relativistic corrections to the SZ spectrum \citep{itoh98,nozawa00}. For the details of the calculation of \fx\ for the SPT see A11.

We use this equation to calculate \Ytheta \ for every step in the MCMC chain, and thus to produce a marginalized distribution of \Ytheta \ values.
In these simulations, integration to a radius approximately corresponding to the \onefifty\ SPT beam diameter (roughly the range $0.75' < \thint < 1.25'$) produces \Ytheta \ distributions that are well constrained despite the degeneracy of \thcore \ and \tzero, with minimal error in recovered cluster \Ytheta \ values.  
Integration in this section is performed to \thint \ $= 0.75'$, though other values are explored for scaling relations in \S \ref{sec:scaling} below. 
Note that we calculate \Ytheta\ from the marginalized distribution of the source model parameters, not by integrating the flux on the sky.
It is also important to note that the likelihood is calculated (in frequency space) over the full $2^\circ \times 2^\circ$ patch of sky, \thint\ is simply the radius out to which the best fit $\beta$-model is integrated.

If the redshift of a cluster is known it is also possible to integrate \YSZ\ within an angular aperture corresponding to a specific physical radius, $\rho$:
\begin{equation}
\thint = \rho \ \dainv, \nonumber
\end{equation} 
where \da\ is the angular diameter distance to the redshift $z$. In Sections \ref{sec:scaling} and  \ref{sec:spt_clusters}, we examine \YSZ\ integrated within a constant physical radius, $\rho$, for all clusters in a sample. We will refer to this quantity as \Yrho.

\begin{figure}
\includegraphics[width=\columnwidth]{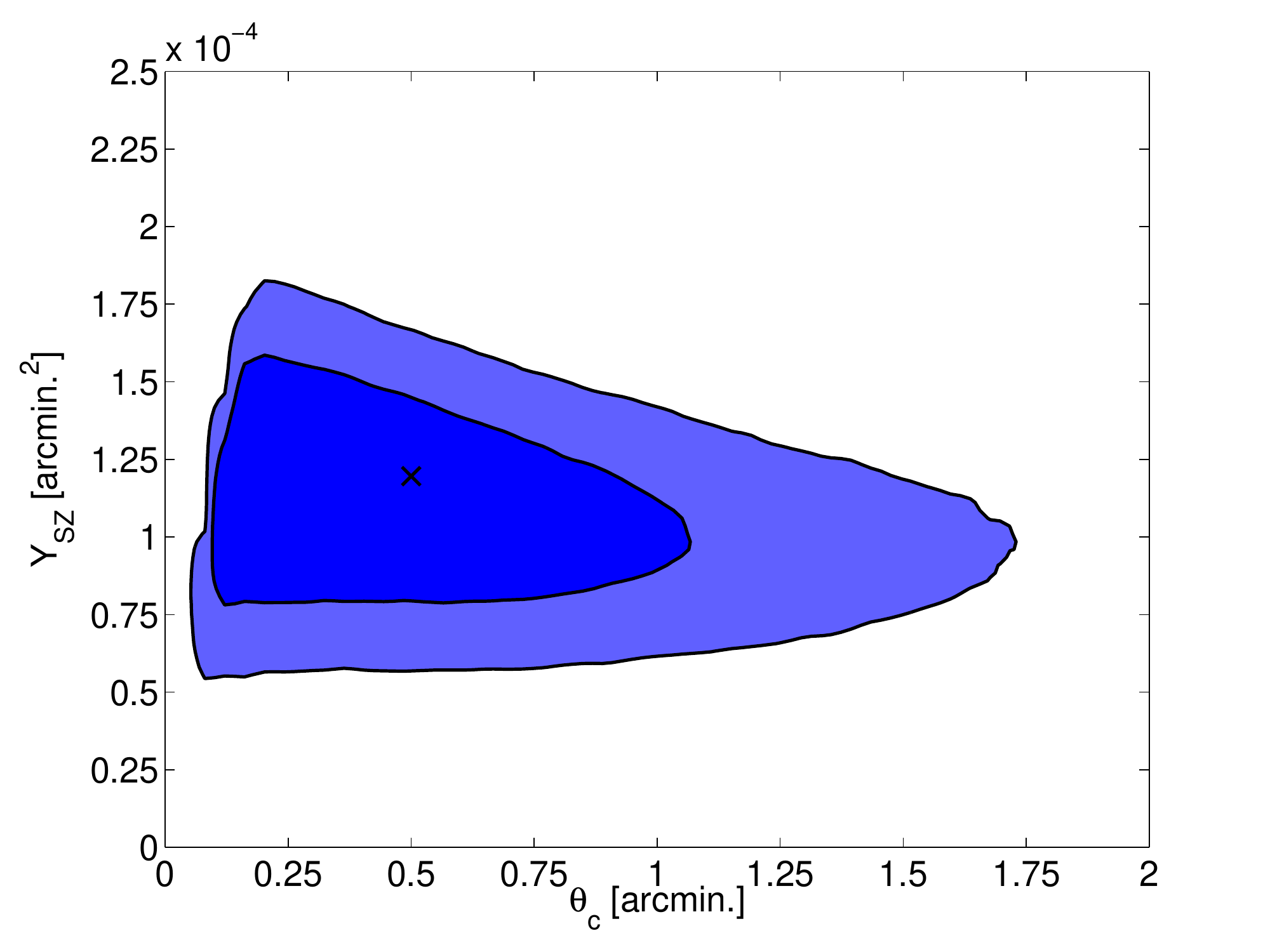}
\caption[\YSZ-\thcore\ constraints]{The marginalized constraints on \YSZ\ and \thcore\ from 500 noise realizations of a typical simulated cluster with radius smaller than the SPT \onefifty \ beam ($\thcore = 0.5'$, $\tzero = 300 \mu K$, $\xi = 6.2$). The contours show the $68\%$ and $95\%$ confidence regions. The `X' marks the input \thcore\ and \YSZ\ values ($\Ysf = 1.20 \times 10^{-4}$ arcmin.$^2$). Despite only having an upper bound on \thcore, \YSZ\ is well constrained.}
\label{fig:ysz_vs_theta}
\end{figure}

In Figure \ref{fig:degen} we show a typical SPT cluster in which \Ytheta\ is well constrained despite the degeneracy between \tzero\ and \thcore.
Figure \ref{fig:ysz_vs_theta} shows \YSZ\ and \thcore\ parameter distributions for 500 runs of a typical simulated cluster with a radius smaller than the SPT \onefifty\ beam size ($\thcore = 0.5'$, $\tzero = 300 \mu K$, $\xi = 6.2$). 
The $68\%$ confidence interval for \YSZ\ in these simulations is typically $\aprx 14\%$ of the central value.
The cutoff at low \thcore\ is due to the small, but non-zero, minimum priors on \thcore\ and \tzero, this is not a feature of the data likelihood.  
Despite only having an upper bound on \thcore, \YSZ\ is still well constrained.
We find that \YSZ\ is well constrained and consistent regardless of whether the analysis is single-band (\onefifty) or multiband ($95\,$ GHz and \onefifty).

\begin{figure}
\includegraphics[width=\columnwidth]{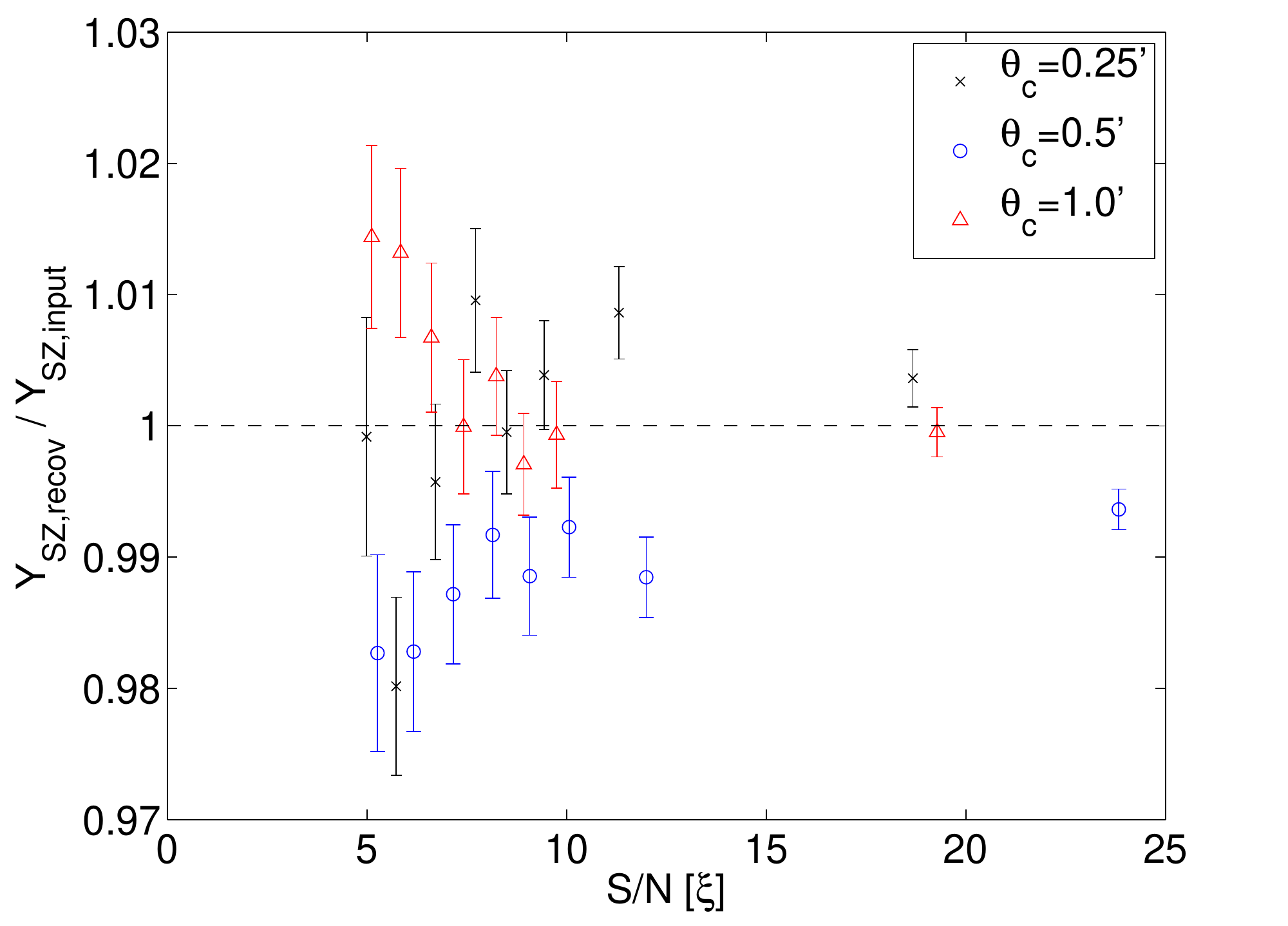}
\caption[\YSZ\ recovery in simulations]{Average ratios of recovered to input \YSZ\ for 24 $\beta$-model source profiles, generated from different combinations of \thcore \ and \tzero. Each point is the mean recovered \YSZ\ for a simulated cluster with 500 independent noise realizations. The errorbars represent the error on the mean of these recovered \YSZ\ values.}
\label{fig:Y_ratio}
\end{figure}

In Figure \ref{fig:Y_ratio}, we show the ratio of the recovered to input \YSZ\ as a function of core radius and cluster detection significance, $\xi$, for 24 different combinations of \thcore \ and \tzero, each with 500 independent noise realizations.
Despite a slight apparent bias for some \thcore\ values, we find no significant bias as a function of the detection significance, and recover Ysz accurately to $<2\%$ in all cases.
On average recovered \Ytheta \ values are $0.27\%$ lower than input values, which is below the $0.49\%$ error in the mean.

\section{Scaling Relations from Simulated Clusters}
\label{sec:scaling}

In this section, we compare \YSZ\ and $\zeta$ as SZ observables for the SPT-SZ survey, focusing on their scatter with cluster mass.
To do this, we use maps derived from the S10 simulations, which are intended to provide more realistic cluster profiles than the $\beta$-model clusters used in \S \ref{sec:params}.

The steepness of the galaxy cluster mass function will introduce bias in a scaling relation fitted in the presence of noise or intrinsic scatter in the population. Therefore, in \S \ref{sec:tsz_sims} we fit \Ytheta-$M$ scaling relations for clusters in simulated tSZ-only maps, to minimize the selection bias. These maps contain none of the celestial or instrumental noise spectra described in \S \ref{sec:sims} \ (CMB, point sources, atmospheric noise, and instrumental noise), only tSZ signal. 

In \S \ref{sec:sky_sims} we fit for a \Ytheta-$M$ scaling relation using clusters in S10 simulation maps containing the full astrophysical and instrumental noise terms to evaluate the performance of the MCMC in the presence of noise. The main reason to consider intrinsic and measurement error is that both are important in terms of the cosmological analysis. For example, both intrinsic and measurement uncertainty affect the selection of the cluster sample: a larger measurement uncertainty would cause lower mass clusters to scatter into the cluster sample, decreasing the purity.  Therefore we consider the total scatter, which includes both sources, to evaluate the performance of our method.

\subsection{Simulated Clusters}
\label{sec:shaw_sims}
 
The S10 simulations are based on a dark matter lightcone simulation, with cosmological parameters consistent with the WMAP 5-year data and large-scale structure measurements \citep{dunkley09}. To include baryons in the simulations, Shaw et al. \citep{shaw10} apply the semi-analytic gas model of Bode et al. \citep{bode07}, specifically their fiducial model, to the dark matter halos identified in the output of the lightcone simulation. From the simulations, we construct two dimensional SZ intensity maps at \onefifty \ of clusters with virial mass (\Mvir) greater than $5 \times 10^{13} \msun h^{-1}$ by summing the electron pressure density along the line of sight. The resulting maps are projections of all the clusters in the lightcone simulation onto a simulated sky. Forty $10^\circ \times 10^\circ$ maps were produced by this procedure, together with catalogs of cluster masses, redshifts, and positions.  

\subsection{\YSZ-$M$ Scaling Relation Fitting Methods}
\label{sec:var_theta}

We assume a scaling between \YSZ\ and $M$ of the form:
\begin{equation}
\YSZ = \asz \left(\frac{\Mvir}{3 \times 10^{14} \, \msun h^{-1}}\right)^{\bsz} \left(\frac{E(z)}{E(0.6)}\right)^{\csz},
\label{eq:scaling_rel} \end{equation}
parametrized by the normalization \asz, the mass scaling \bsz, and the redshift evolution \csz, and where $E(z) \equiv H(z)/H_0$.
For self-similar evolution, $\bsz = 5/3$ and $\csz = 2/3$ (e.g., Kravtsov et al. \citep{kravtsov06a}). 
The pivot points of the scaling relation were defined to match the approximate mean mass and redshift for the SPT cluster sample.  

We fit the \Ytheta-\Mvir\ scaling relation by minimizing the fractional scatter, $\mathcal{S}$, in \Ytheta, defined as:
\begin{equation}
\mathcal{S} = \sqrt{\frac{1}{N} \displaystyle\sum_{n=1}^N \left( \frac{Y^{\mathrm{recov}}_{n}-Y^{\mathrm{input}}_{n}}{Y^{\mathrm{input}}_{n}} \right)^2},
\label{eq:frac_scatter} \end{equation}
where $Y^{\mathrm{recov}}_{n}$ is the integrated Comptonization recovered by the MCMC for the $n^{th}$ cluster, $Y^{\mathrm{input}}_{n}$ is the corresponding Comptonization calculated from the input catalog mass and the assumed scaling relation (equation \ref{eq:scaling_rel}), and we sum over $N$ simulated clusters. The scaling relation parameters \asz, \bsz, and \csz \ are varied using a grid search method, and the scatter $\mathcal{S}$ is calculated at each point in the parameter space. The combination of parameters that minimizes $\mathcal{S}$ is taken to be the best-fit set of parameters. This definition of fractional scatter is used to fit \Ytheta-\Mvir\ scaling relations in \S \ref{sec:tsz_sims} and in \S \ref{sec:sky_sims}.

\subsection{Results for Simulated Thermal-SZ-Only Maps}
\label{sec:tsz_sims}

We run both the MCMC and MF methods on tSZ-only maps from the S10 dark matter lightcone simulations described in \S \ref{sec:shaw_sims}. 
These simulated tSZ maps contain only thermal SZ signal, and no CMB, point sources, atmospheric noise, or instrumental noise.

We measure the SZ signal in these maps using the methods described in Sections \ref{sec:TM_sims} and \ref{sec:YSZ} for clusters with $\Mvir > 4 \times 10^{14} \msun h^{-1}$, and redshift $0.3 < z < 1.2$. 
We then use the cluster virial masses and equation \ref{eq:scaling_rel} to find the best fit scaling relation parameters by minimizing the fractional scatter in equation \ref{eq:frac_scatter}. We do this for both the \YSZ-$\Mvir$ and $\zeta$-$\Mvir$ scaling relations, which allows for direct comparison of these analysis methods. 
The redshift range corresponds to the redshift range of observed SPT clusters, and the mass criteria corresponds to the mass of clusters at the lower SPT significance limit of the Reichardt et al. \citep{reichardt13} cluster catalog ($\xi = 4.5$), at the survey median redshift of $z = 0.6$. 

As a baseline for the scatter in the measured \YSZ-\Mvir\ scaling relations for these simulations, we examine the intrinsic scatter between \Mvir\ and \Yvir, the contribution to the SZ flux from within the spherical virial radius for each cluster.
We fit the \Yvir-\Mvir\ scaling relation parameters using the same method as for measured \YSZ\ values, and find the fractional scatter in the best-fit scaling relation to be $16\%$. 

\begin{figure*}
\clearpage
\includegraphics[angle=0, width=\textwidth]{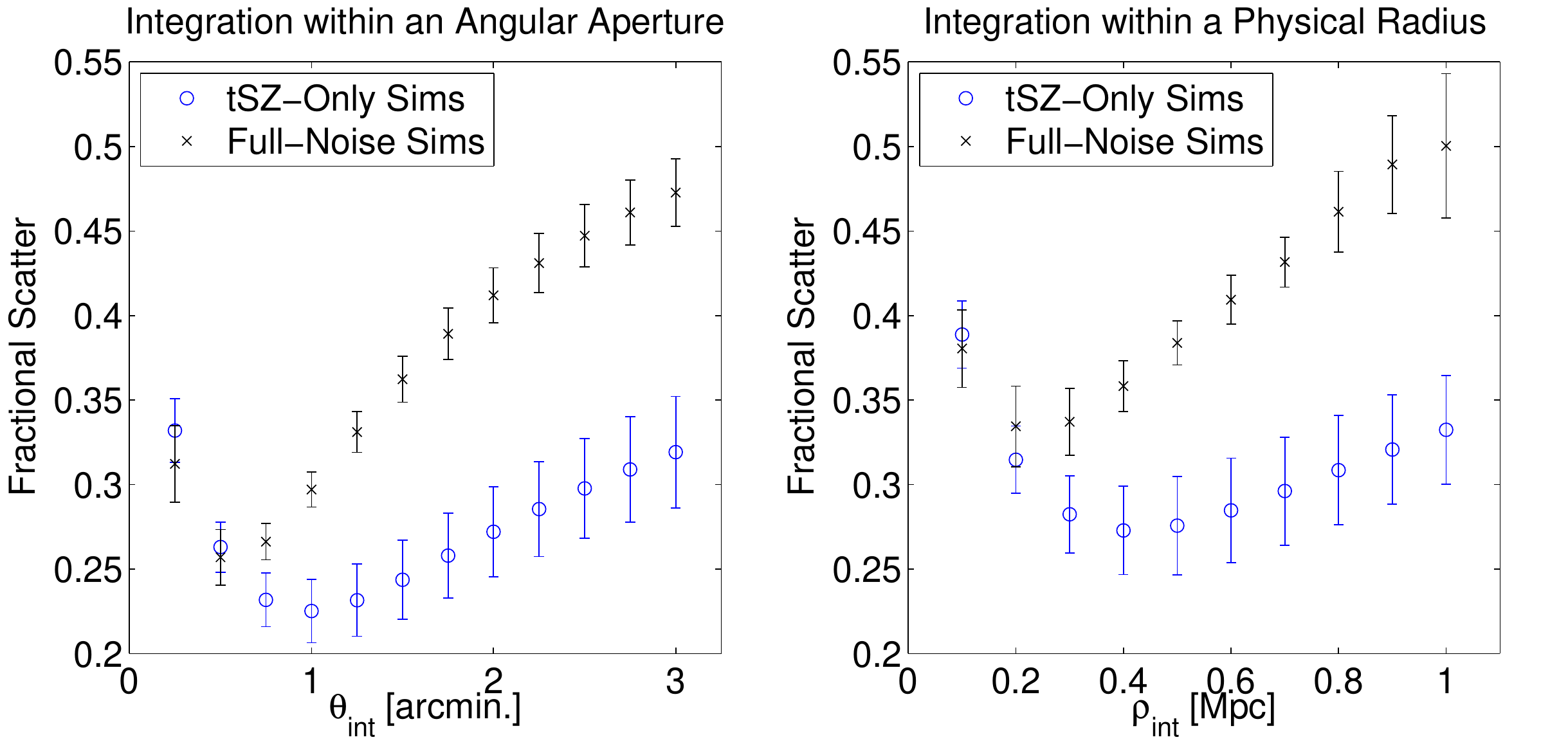}
\caption[Scatter in \YSZ-\Mvir\ scaling relations]{Fractional scatter vs. integration radius for tSZ-only and full noise simulations. Left panel: fractional scatter vs. integration angle in arcminutes. Right panel: fractional scatter vs. integration radius in megaparsecs. The scatter in the tSZ-only simulations is essentially the intrinsic scatter in the population, since only tSZ fluctuations are present. Adding the other noise terms shifts the scatter up, and the minimum down in angular or physical scale because those noise terms dominate at large angles. The optimal angular apertures correspond roughly to the optimal physical radii at the median redshift of the cluster sample, $z = 0.6$.}
\label{fig:frac_scatter}
\end{figure*}

We fit \Ytheta-$\Mvir$ relations for a range of angular apertures, $\theta$, with \Ytheta\  defined in equation \ref{eq:y_theta}. 
Figure \ref{fig:frac_scatter} shows the fractional scatter as a function of the integration angle for angles ranging from $0.25'$ to $3.0'$. 
We find that the fractional scatter in \Ytheta \ does not vary significantly with angular aperture, with a broad minimum in the scatter at $\aprx 0.75'$ - $1.0'$ (\Ysf). 
The exact location of the minimum scatter shifts between the tSZ-only maps and the full-noise S10 maps, but is near $0.75'$ in both cases (see Figure \ref{fig:frac_scatter}).
For simplicity, and for comparison between the different simulated maps and observed clusters, we use the \Ysf-\Mvir\ scaling relation as our nominal scaling relation.
The \Ysf-$\Mvir$ scaling relation has $23\pm2\%$ fractional scatter in \YSZ, which is slightly less than the $27\pm2\%$ scatter in the $\zeta$-$\Mvir$ scaling relation for these clusters.  
The scatter in the tSZ-only simulations is primarily due to the intrinsic scatter in the mass to SZ observable scaling, scatter from the tSZ background is sub-dominant.
(Note, the scatter here is fractional scatter, whereas previous SPT analyses in V10 and B13 quoted a log-normal scatter, at a level consistent with the values found in this work.)
 
\begin{figure}
\includegraphics[width=\columnwidth]{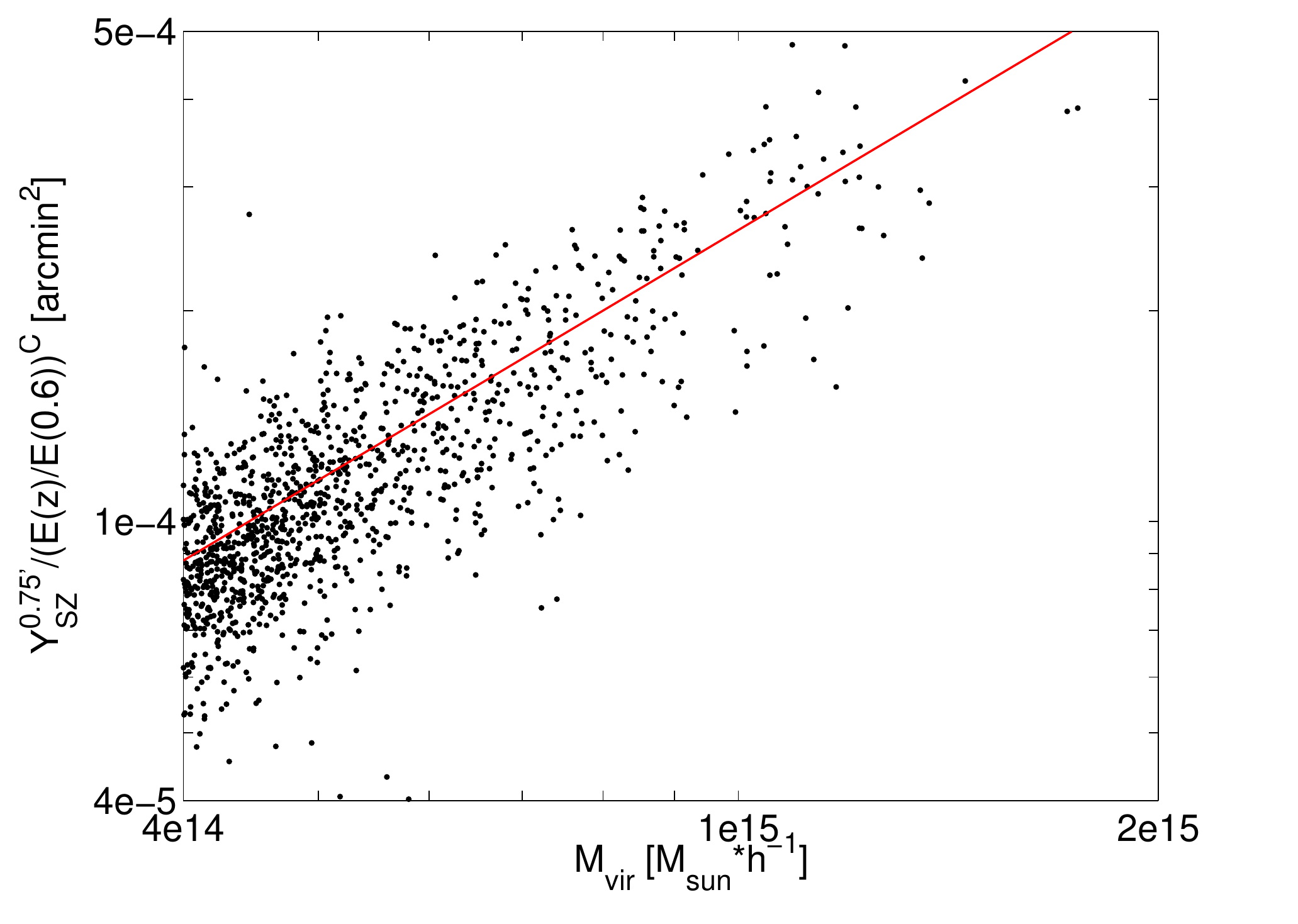}
\caption[\Ysf-\Mvir\ scaling relation for tSZ-only simulations]{\Ysf \ versus $\Mvir$ \ for 1187 mass-selected clusters in the S10 simulated tSZ-only maps, where we only include clusters with $\Mvir > 4 \times 10^{14} \msun h^{-1}$ in the redshift range $0.3 < z < 1.2$. Fractional scatter in \YSZ\ is $23\pm2\%$. The solid line is the best-fit \Ysf-$\Mvir$ scaling relation found for this cluster sample.}
\label{fig:shaw_sz}
\end{figure}

Figure \ref{fig:shaw_sz} shows \Ysf \ versus $\Mvir$ for the 1187 clusters examined from this simulation. The solid line is the best-fit \Ysf-$\Mvir$ scaling relation found for these clusters.  The scaling relation parameters (\asz, \bsz, \csz, and $\mathcal{S}$) for the \Ysf \ scaling relation are given in Table \ref{tab:scaling_rels}.

We also calculate \YSZ\ within a constant physical radius, $\rho$, (\Yrho) for all the clusters in the catalog. 
The angular size of a cluster is a function of its redshift, therefore, it is interesting to measure \YSZ\ within a fixed physical radius.
In Figure \ref{fig:frac_scatter}, we plot the best-fit scatter for a range of integration radii between 0.1 to 1.0 Mpc. 
We find that the minimum fractional scatter in \YSZ\ within a fixed physical radius is higher than the minimum fractional scatter within a fixed angular aperture.
For \Yrho, the scatter is increased by the varying angular size of the chosen physical radius at different redshifts. 
The optimal physical radius corresponds roughly to the optimal angular aperture, at the median redshift of the cluster sample, $z = 0.6$. 
Clusters farther from the median redshift will have integration angles farther from the optimal angle, resulting in relatively higher scatter in \Yrho\ than in \Ytheta.

As can be seen in Figure \ref{fig:frac_scatter}, we find a broad minimum in scatter at $\aprx 0.3$ - $0.4$ Mpc, with a minimum scatter of $27\pm3\%$. This is comparable to the $\zeta$-$\Mvir$ relation for these clusters, and slightly higher than the scatter in the \Ysf-$\Mvir$ scaling relation. The scaling relation parameters for \YSZ\ within $0.3$ Mpc (\YthM), ($0.3$ Mpc being equivalent to $0.75'$ at $z = 0.6$) are given in Table~\ref{tab:scaling_rels}.

\subsection{Results for Full-Noise Simulated Maps}
\label{sec:sky_sims}

We also fit \Ytheta-$\Mvir$ scaling relations for the simulated clusters in the presence of other astrophysical and instrumental noise components (see \S \ref{sec:astro} and \S \ref{sec:instr}). The same cluster sample ($\Mvir > 4 \times 10^{14} \msun h^{-1}$, and $0.3 < z < 1.2$) was analyzed in this set of simulations as in the simulated tSZ-only maps. We will refer to this set of simulations as the full-noise S10 simulated maps. These simulations are important for understanding the cluster sample selection, which will be affected by the total noise.

The scaling relation fitting for the clusters from this set of simulations was performed as in \S \ref{sec:tsz_sims}. As in \S \ref{sec:tsz_sims}, the scatter is a weak function of angular aperture, with the minimum shifted to $\aprx 0.5'$ - $0.75'$. Figure \ref{fig:frac_scatter} shows the fractional scatter as a function of angular aperture of integration. 

For the \Ysf-$\Mvir$ scaling relation we find a fractional scatter in \YSZ\ of $27\pm1\%$. Since the scatter here includes both intrinsic scatter and the measurement uncertainty, we expect it to be larger than the scatter in \Ysf \ in \S \ref{sec:tsz_sims}. This level of scatter is comparable to the $27\pm2\%$ scatter in $\zeta$ found in the $\zeta$-$\Mvir$ scaling relation for these simulations. 

\begin{figure}
\includegraphics[width=\columnwidth]{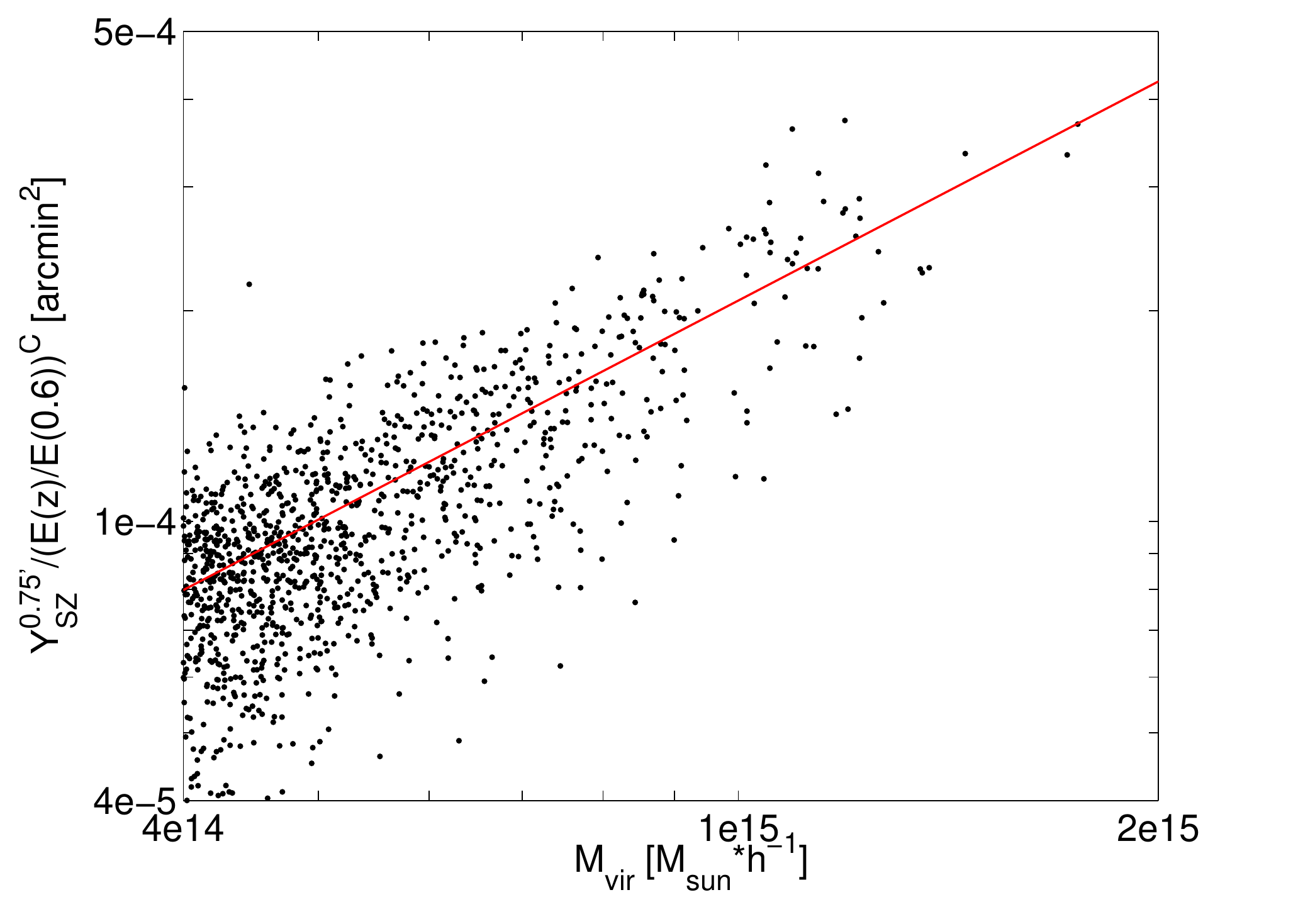}
\caption[\Ysf-\Mvir\ scaling relation for full-noise simulations]{\Ysf \ versus $\Mvir$ for 1187 mass-selected clusters in the full-noise S10 simulations, which include CMB, point sources, astrophysical noise, and realistic SPT instrument noise. We include only clusters with $\Mvir > 4 \times 10^{14} \msun h^{-1}$ in the redshift range $0.3 < z < 1.2$. Fractional scatter in \YSZ\ is $27\pm1\%$. The solid line is the best-fit \Ysf-$\Mvir$ scaling relation found for this cluster sample.}
\label{fig:shaw_sky}
\end{figure}

Figure \ref{fig:shaw_sky} shows \Ysf \ versus $\Mvir$ for the 1187 clusters analyzed from the full-noise S10 simulated maps. The solid line is the best-fit \Ysf-$\Mvir$ scaling relation found for this cluster sample. The mass scaling relation parameters for \Ysf\ in this set of simulations are given in Table \ref{tab:scaling_rels}. It will be noted that some of the scaling relation parameters here differ significantly from those in the tSZ-only simulations in Section \ref{sec:tsz_sims}. This is expected, because intrinsic and measurement scatter are not distinguished here. The full cosmoMC treatment of the data in Section \ref{sec:spt_clusters} deals with these issues. The scaling relation values from that section are the most accurate. The results in this section are meant to be illustrative only.

Using these simulations we also calculate \Yrho\ for a range of $\rho$ values, as in \S \ref{sec:tsz_sims}, and fit \Yrho-$\Mvir$ scaling relations for each $\rho$. 
Figure \ref{fig:frac_scatter} shows the fractional scatter as a function of the integration radius for a range of physical radii. 
We find a broad minimum in scatter at $\aprx0.2$ - $0.3$ Mpc, with a minimum scatter of $33\pm2\%$.
The optimal integration radius shifts down here relative to the simulated tSZ-only maps because of the scale dependence of the noise sources added in the full-noise S10 maps, which dominate the scatter in these simulations. 
In particular, the noise induced by the simulated CMB increases with angular scale, leading to a preference for smaller integration radii. 
The scatter in \Yrho\ for these simulations is slightly higher than the scatter in both the $\zeta$ and the \Ysf\ mass scaling relations. 
The scaling relation parameters for the nominal \YthM\ mass scaling relation are given in Table \ref{tab:scaling_rels}. 
The optimal physical radius again corresponds roughly to the optimal angular aperture, at the median redshift of the cluster sample.

\section{\YSZ\ for SPT Observed Clusters}
\label{sec:spt_clusters}

\subsection{\YSZ-\Mfh\ Scaling Relation Fitting Methods}
\label{sec:spt_clusters_scaling}

In this section we perform \Ytheta-$M$ scaling relation fitting for a sample of SPT observed clusters, using the same scaling relation as in the simulations (equation \ref{eq:scaling_rel}) and the X-ray determined cluster masses. 
We use only \onefifty\ data in this section to be directly comparable to the results in Benson et al. \citep{benson13}, which does not include $95\,$ GHz data.
In this section we define cluster mass as \Mfh, the mass inside a spherical radius \rfh, within which the average density is 500 times the critical density of the universe at the cluster's redshift.
To fit this scaling relation with clusters selected in the SPT-SZ survey, we have to account for the shape of the cluster mass function and the SPT survey selection, which was based on the SPT significance, $\xi$. 
This is similar to the procedure followed in previous SPT analyses \citep[B13, V10]{reichardt13}, with the added complication that in this work we must express the SPT selection function in \YSZ\ instead of $\xi$.

The (unnormalized) probability of a mass $M$ given an integrated Comptonization \YSZ\ is given by:
\begin{equation}
\overline{\mathrm{P}}(M|\YSZ) = \mathrm{P}(\YSZ|M)\mathrm{P}(M), \nonumber
\end{equation}
where $\mathrm{P}(\YSZ|M)$ is the Gaussian probability distribution with which we have been working previously, and $\mathrm{P}(M)$ is the mass function. The number of clusters is a steep function of cluster mass, which (combined with the measurement uncertainty in \YSZ) results in relatively more low-mass than high-mass clusters at a given \YSZ, an effect commonly referred to as Eddington bias.

For our cluster sample we use the eighteen clusters from B13, fourteen of which have X-ray derived masses (see \S \ref{sec:Xray_obs}), and all of which have $\xi > 5$. For this analysis we use only the \onefifty\ data, the SPT band with the highest SZ sensitivity. For a list of cluster names, $\xi$ values, and redshifts for this sample, see Table \ref{tab:cluster_table}.

To fit for scaling relations we use a method similar to the one described in B13, which we modify to account for the cluster selection based on \YSZ\ instead of $\zeta$. In B13, we used a version of the CosmoMC \citep{lewis02b} analysis package, modified to include the cluster abundance likelihood in the CosmoMC likelihood calculation. All fitting is performed assuming a standard flat \LCDM\ cosmology, parametrized with the standard six-parameters ($\Omega_c h^2$, $\Omega_b h^2$, $\Theta_s$, $n_s$, $\Delta_R^2$, and $\tau$), and using the WMAP 7-year data set. The cosmological parameters are held constant throughout, we marginalize over only the cluster scaling relation parameters. At each step in the chain, a point in the joint cosmological and scaling relation parameter space is selected. The Code for Anisotropies in the Microwave Background (CAMB) \citep{lewis00} is used to compute the matter power spectrum at twenty redshift bins between $0 < z < 2.5$, spaced logarithmically in $1+z$. The matter power spectrum, cosmological parameters, and \YSZ-\Mfh\ and \YX-\Mfh\ scaling relation parameters are then input to the cluster likelihood function. \YX\ is defined as $\YX \equiv M_\mathrm{g} T_\mathrm{X}$, where $M_\mathrm{g}$ is the cluster gas mass within \rfh, and $T_\mathrm{X}$ is the core-excised X-ray spectroscopic temperature in an annulus between $0.15$ and $1.0 \times \rfh$.

To calculate the cluster likelihood, first the matter power spectrum and cosmological parameters are used to calculate the cluster mass function, based on Tinker et al. \citep{tinker08}. Next, the mass function is converted to the predicted cluster abundance in our observable space, $N(\YSZ,\YX,z)$. This conversion is accomplished using our standard \Ytheta-\Mfh\ scaling relation (equation \ref{eq:scaling_rel}), with flat unbounded priors on all parameters, and the \YX-\Mfh\ scaling relation from B13:
\begin{equation}
\frac{\MX}{ 10^{14} \msun h^{-1} } = \left(\ax h^{3/2}\right) \left( \frac{\YX}{3 \times 10^{14} \msun \, {\rm keV}} \right)^{\bx} E(z)^{\cx},
\end{equation}
parametrized by the normalization factor \ax, the mass scaling \bx, the redshift evolution \cx, and the log-normal intrinsic scatter \dx. This scaling relation is based on the relation used in Vikhlinin et al. \citep{vikhlinin09b}. 

The priors on the \YX-\Mfh\ scaling relation parameters were: $\ax = 5.77 \pm 0.56$, $\bx = 0.57 \pm 0.03$, $\cx = -0.40 \pm 0.20$, and $\dx = 0.12 \pm 0.08$. All these priors are Gaussian. The slope and normalization priors were motivated by Vikhlinin et al. \citep{vikhlinin09b}, and the priors on the redshift evolution and scatter were motivated by the range observed in several different sets of simulations which included varying astrophysics \citep{kravtsov06a, fabjan11}. These priors are identical to those used in Benson et al. \citep{benson13}, and a more detailed description motivating them is given there.

The predicted cluster density as a function of \YSZ, \YX, and $z$ can be written as follows:
\begin{equation}
  \frac{dN(\YSZ,\YX,z | \vec{p})}{d\YSZ \ d\YX \ dz} = \nonumber
\end{equation}
\begin{equation}
  \int P(\YSZ, \YX | M, z, \vec{p}) \ P(M, z | \vec{p}) \ \Phi(\YSZ) \ dM,
  \label{eq:grid}
\end{equation}
where $\vec{p}$ is the set of cosmological and scaling relation parameters, and $\Phi(\YSZ)$ is the selection function in \YSZ. This predicted cluster density function differs from B13 in that the selection function must be transformed from a Heaviside step function at $\xi=5$ into a function of \YSZ. We assume that \YSZ\ and $\xi$ can be related with a log-normal distributed scaling relation, and that the selection in B13 can therefore be well-approximated by an error-function in \YSZ. We then define our selection function as:
\begin{equation}
\Phi(\YSZ) = \frac{1}{2} \mathrm{erf} \left( \frac{ \YSZ - \Ythresh} {\sqrt{2} \ \Ythresh \ D} \right) + \frac{1}{2},
\end{equation}
where the selection threshold, \Ythresh \ is defined as the \YSZ\ value corresponding to $\xi = 5$ at the redshift $z$. We estimate \Ythresh\ by fitting a \YSZ-$\xi$ scaling relation of the form:
\begin{equation}
\YSZ = A \xi^B E(z)^C,
\end{equation}
using the catalog of SPT observed clusters given in R13. The width of the selection error-function is given by the scatter in the \YSZ-$\xi$ scaling relation, D.

We evaluate equation \ref{eq:grid} on a $200 \times 200 \times 30$ grid in (\YSZ, \YX, $z$) space, and convolve with a Gaussian error term in \YSZ\ to account for the measurement noise. The width of the Gaussian is given by the uncertainty in \YSZ\ as a function of \YSZ, $\delta \YSZ(\YSZ)$, as determined by the cluster parametrization MCMC (see \S \ref{sec:YSZ}).

The likelihood function of the observed cluster sample is defined by the Poisson probability:
\begin{eqnarray}
  \mathrm{Log} \left( \mathcal{L}(\vec{p}) \right) = \sum_i \mathrm{Log} \left( \frac{dN(\YSZ_i,\YX_i,z_i, | \vec{p})}{d\YSZ \ d\YX \ dz} \right) - \nonumber \\
 \int \frac{dN(\YSZ,\YX,z, | \vec{p})}{d\YSZ \ d\YX \ dz} \ d\YSZ \ d\YX \ dz,
  \label{eq:lnlike}
\end{eqnarray}
where the summation is over the SPT clusters in our catalog. Note also that this is the unnormalized log-likelihood. 

There is a complication, in that \YX\ is dependent on the cosmological parameters. $\YX \equiv M_\mathrm{g} \ T_\mathrm{X}$, where $M_\mathrm{g}$ is the gas mass within \rfh, and $T_\mathrm{X}$ is the core-excised X-ray temperature in an annulus between $0.15 \times \rfh$ and $1.0 \times \rfh$. To maintain consistency with the cosmological parameters, we recalculate \YX\ for each cluster at every step in CosmoMC, given the current \YX-\Mfh\ relation and \rfh.  In the likelihood, we add $\sum_i \mathrm{Log}(\YX_i)$ to the right hand side of equation \ref{eq:lnlike} to account for the recalculation of \YX. For a detailed explanation of this correction term, see Appendix B of B13.

To account for measurement error in \YX\ and $z$ for each cluster, we marginalize over the relevant parameter, weighted by a Gaussian likelihood determined by its uncertainty. For the few clusters without observed \YX\ data, we instead weight the marginalized parameter by a uniform distribution over the allowed parameter range.

The likelihood of this set of cosmological and scaling relation parameters is then used by CosmoMC in the acceptance/rejection computation.  
Only the \Ytheta-\Mfh\ scaling relation parameters are of interest to us in this analysis. The cosmological and \YX-\Mfh\ scaling relation parameters were used as a crosscheck to verify that the results were in agreement with the analysis performed on these clusters in B13, but will not be presented here. All parameters differed from the values presented in B13 by $<< 1\sigma$.

\subsection{\YSZ-\Mfh\ Scaling Relation Results}
\label{sec:B13_scaling_rel_results}

We use CosmoMC to fit \Ytheta-\Mfh\ scaling relations for a range of angular apertures, and find a broad minimum in scatter in the range $0.5'$ - $0.75'$, with a minimum intrinsic log-normal scatter of $21\pm11\%$. 
For these scaling relation parameters we apply flat, unbounded priors.
The scatter in the $\zeta$-\Mfh\ scaling relation for these clusters is comparable, at $21\pm9\%$. 
The scaling relation parameters for the \Ysf-\Mfh\ scaling relation are given in Table \ref{tab:scaling_rels}. 

We also fit mass scaling relations for \Yrho\ integrated within a range of physical radii, $\rho$, from $0.1$ Mpc to $0.5$ Mpc.  
We find a broad minimum in scatter in the range $0.2$ - $0.3$ Mpc, with a minimum intrinsic log-normal scatter of $23\pm5\%$. 
This is comparable to the scatter in both the $\zeta$ and \Ysf\ mass scaling relations.
The parameters for the nominal \YthM mass scaling relation ($0.3$ Mpc corresponds to $0.75'$ at the survey median redshift of $z = 0.6$) are listed in Table \ref{tab:scaling_rels}.

\subsection{Cluster Masses}

To calculate the masses of the clusters, the \Ysf \ CosmoMC chains were used. The probability density function for the mass was computed on an evenly spaced mass grid for each step in the CosmoMC chains. These probability density functions were then summed to obtain a mass estimate fully marginalized over all scaling relation and cosmological parameters. This was done for CosmoMC chains containing only \Ysf \ data, and no \YX\ data, and vice versa, to obtain mass estimates based on only the SZ and X-ray data respectively. The cluster \Mfh \ masses derived from the \Ysf \ and \YX\ data (\MSZ \ and \MX \ respectively) can be found in Table \ref{tab:cluster_table}, along with the corresponding \Ysf \ and \YX\ values. \Ysf values are given in \msun keV for ease of comparison with \YX.

\begin{figure}
\includegraphics[width=\columnwidth]{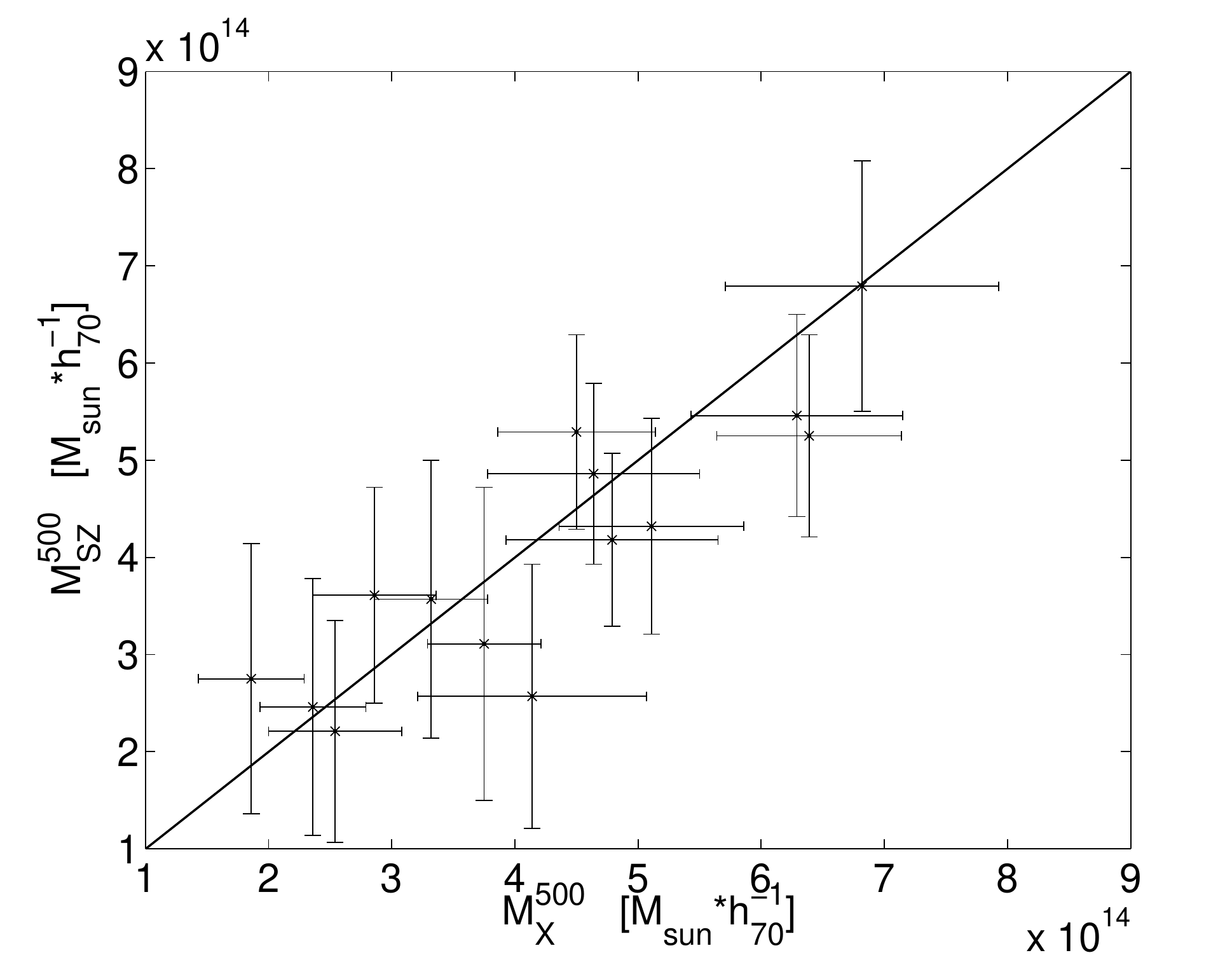}
\caption[\MSZ \ vs. \MX]{Masses computed from \Ysf \ for the 14 SPT observed clusters in Table \ref{tab:cluster_table} with follow-up X-ray observations versus corresponding \MX \ values. For reference we overplot the relation $\MSZ = \MX$.}
\label{fig:MSZvsMX}
\end{figure}

Figure \ref{fig:MSZvsMX} shows the cluster masses calculated from the \YSZ-\Mfh\ scaling relation versus the masses calculated from the \YX-\Mfh\ scaling relation for the B13 cluster sample. The solid line is the reference line $\MSZ = \MX$.

\begin{table*}
\begin{center}
{\scriptsize
\caption{\YSZ-$M$ Scaling Relation Parameters}
\begin{tabular}{l|ccccc|c}
  &  &  & MCMC & & & MF \\
\hline \hline
 Data Set & Integration & \asz & \bsz & \csz & Scatter    & Scatter \\
          & Radius      & ($\times 10^{-4}$) &      &      &    &  \\
\hline
\multirow{2}{*}{tSZ-Only S10 Sims} & $0.75'$ & $1.44\pm0.11$ & $1.20\pm0.11$ & $1.63\pm0.24$ & $23\pm2\%$ & \multirow{2}{*}{$27\pm2\%$} \\
& $0.3$ Mpc & $1.53\pm0.16$ & $1.26\pm0.17$ & $1.13\pm0.13$ & $28\pm2\%$ \\ \hline
\multirow{2}{*}{Full-Noise S10 Sims} & $0.75'$ & $1.37\pm0.10$ & $1.04\pm0.11$ & $1.02\pm0.20$ & $27\pm1\%$ & \multirow{2}{*}{$27\pm2\%$}\\
& $0.3$ Mpc & $1.49\pm0.18$ & $1.12\pm0.22$ & $0.53\pm0.25$ & $34\pm2\%$ \\ \hline
\multirow{2}{*}{B13 SPT Observed Clusters} & $0.75'$ & $1.85\pm0.36$ & $1.77\pm0.35$ & $0.96\pm0.50$ & $\ \ \ 21\pm11\%^a$  & \multirow{2}{*}{$\ \thinspace  21\pm9\%^a$} \\
& $0.3$ Mpc & $2.09\pm0.35$ & $1.43\pm0.20$ & $0.35\pm0.28$ & $\ \thinspace  26\pm9\%^a$ \\ \hline
\end{tabular}}
\begin{tablenotes}
Note -- The tSZ-only maps contain only thermal SZ signal, while the full-noise S10 maps include tSZ, CMB, point sources, atmospheric noise, and realistic SPT instrumental noise.
The values of scatter reported for the simulations are fractional scatter, while the values reported for the B13 clusters are intrinsic log-normal scatter.
In the S10 simulations virial masses are used to fit the scaling relations, while for the B13 cluster sample the masses are \Mfh.
For comparison with the scatter in each \YSZ-$M$ scaling relation we list the scatter in the corresponding MF derived $\zeta$-$M$ scaling relation for the same data set. 
\\ (a) These values are intrinsic log-normal scatter.
\end{tablenotes}
\label{tab:scaling_rels}
\end{center}
\end{table*}

\subsection{$\YSZ(\rfh$)}
\label{sec:y_five_hundred}

The self-similar model of cluster formation assumes that clusters scale in well-defined ways based on their mass, typically defined within physical radii proportional to the critical density of the universe at the cluster's redshift (e.g., Kravtsov et al. \citep{kravtsov12} and Kaiser et al. \citep{kaiser84}). For this reason, studies of the scaling relations of clusters typically measure physical observables defined by this physical radius, usually \rfh. In this section, we will calculate $\YSZ(\rfh)$, denoted \Yfh, for comparison with other published parameters for the clusters in B13.

We investigated a method for estimating \rfh\ from SZ data, as a way to measure \Yfh\ solely from SZ data. This method proved to be problematic however, because it required estimating \Mfh\ from a fixed angular aperture, and calculating \rfh\ from that estimate. This results in the scatter in the \Ysf-\Mfh\ scaling relation feeding back into the calculation of \Yfh. Instead, we use the X-ray determined \rfh\ in our calculations of \Yfh.

In Table \ref{tab:cluster_table}, we give the measured \Yfh\ values for our cluster sample.
We note that, as defined in equation \ref{eq:ysz_theta}, the MCMC fits for a cylindrically projected measure of \Yfh\, rather than the spherical de-projected value often used in other \YSZ-$M$ scaling relation results (e.g., A11, Arnaud et al. \citep{arnaud10}). \Yfh\ values are given in \msun keV here, for comparison with A11.

A11 describes a template fitting method of estimating \Yfh, which uses an SZ source template motivated from X-ray measurements of each cluster. The profile is assumed to match the product of the best-fit gas density profile to the X-ray measurements of each cluster, and the universal temperature profile of Vikhlinin et al. \citep{vikhlinin06}. These profiles are multiplied together to produce the radial pressure profile, and projected onto the sky using a line-of-sight integral through the cluster. A11 then constructs a spatial filter using equation \ref{eq:psi}, and this X-ray derived source model. The X-ray determined cluster position is used to place priors on the cluster location to prevent maximization bias in the recovered \Yfh \ values. \Yfh \ is calculated by integrating the source model over a solid angle corresponding to \rfh, as in equation \ref{eq:ysz_theta}.

\begin{figure}
\includegraphics[width=\columnwidth]{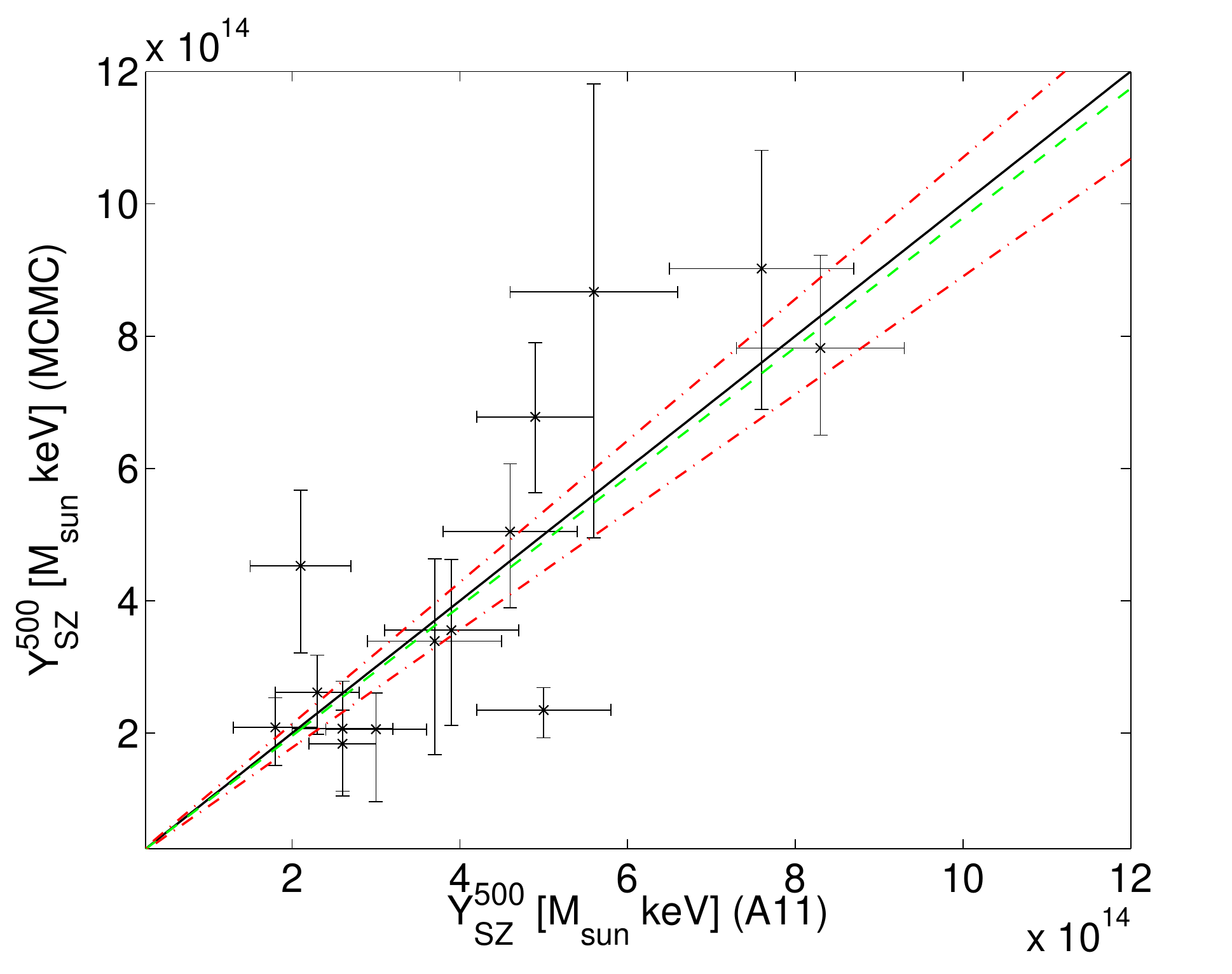}
\caption[$\Yfh \mathrm{(MCMC)}$ vs. $\Yfh \mathrm{(A11)}$]{\Yfh \ for the 14 SPT clusters from Table \ref{tab:cluster_table} with follow-up X-ray observations, calculated by the MCMC method described here, and by the MF method of Anderson et al. \citep{andersson11}. We also show the reference line $\Yfh \mathrm{(MCMC)} = \Yfh \mathrm{(A11)}$ (solid), the best-fit line (green dashed), and the uncertainty in the fit defined as the range for which $\Delta \chi^2 < 1$ compared to the best-fit (red dot-dashed). The best-fit normalization is $A = 0.98 \pm 0.09$, demonstrating that the scaling relation is consistent with equality between $\Yfh \mathrm{(MCMC)}$ and $\Yfh \mathrm{(A11)}$.}
\label{fig:mcmc_vs_a11}
\end{figure}

In Figure \ref{fig:mcmc_vs_a11}, we plot the \Yfh \ estimated by the MCMC method against the \Yfh \ estimated by the template fitting method in A11. The best-fit relation between the two is $\Yfh(\mathrm{MCMC}) = (0.98\pm0.09) \ \Yfh(\mathrm{A11})$, where the uncertainty is the range for which $\Delta \chi^2 < 1$ ($68\%$ confidence limit) compared to the best-fit. (For a treatment of the calculation of $\chi^2$ with uncertainty in both variables see, for example, Numerical Recipes in C++, Section 15.3 \citep{press02}.) We see that these two methods of calculating \Yfh \ are consistent, that is, the best-fit scaling relation is consistent with equality between $\Yfh(\mathrm{MCMC})$ and $\Yfh(\mathrm{A11})$. 
The scatter about the expected one-to-one line here is dominated by differences in cluster model shape between the two methods (X-ray derived SZ profile versus $\beta$-model).

\begin{figure}
\includegraphics[width=\columnwidth]{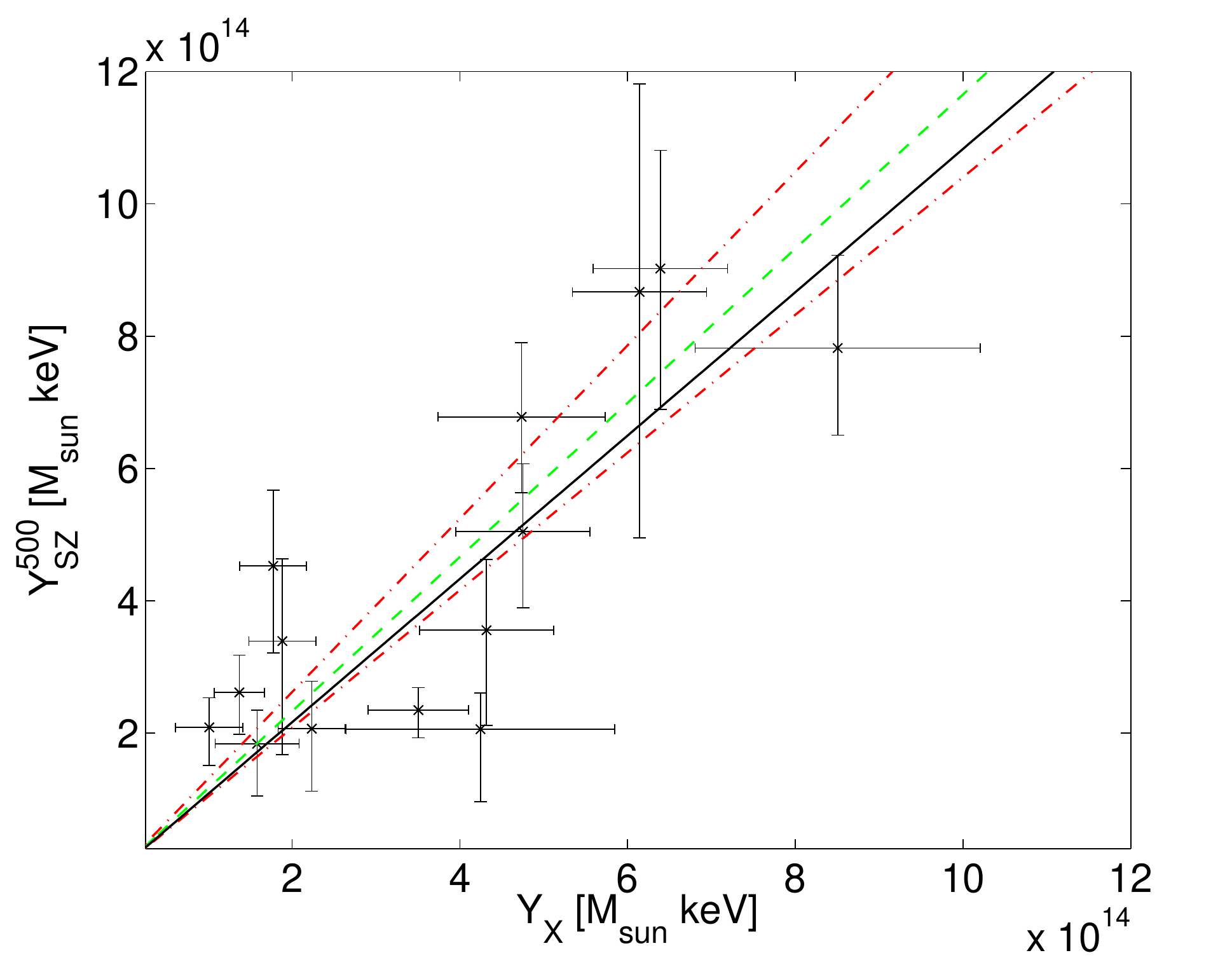}
\caption[\Yfh \ vs. \YX]{\Yfh(MCMC) \ versus \YX\ for the 14 SPT clusters from Table \ref{tab:cluster_table} with follow-up X-ray observations. We also show the expected scaling relation from Arnaud et al. \citep{arnaud10}: $\Yfh = 1.08 \ \YX$ (solid), the best-fit line (green dashed), and the uncertainty in the fit defined as the range for which $\Delta \chi^2 < 1$ compared to the best-fit (red dot-dashed). The best-fit normalization is $A = 1.17 \pm 0.12$, consistent with the expected scaling between \Yfh \ and \YX.}
\label{fig:ysz_vs_yx}
\end{figure}

We also verify that our \Yfh \ values for these clusters are in agreement with the \YX\ values presented in B13, given the expected \YSZ-\YX\ scaling. Figure \ref{fig:ysz_vs_yx} shows the \Yfh \ values of our catalog of SPT observed clusters plotted against their \YX\ values from B13. 

We can make a prediction of the relationship between \YSZ\ and \YX\ based on the universal pressure profile from Arnaud et al. \citep{arnaud10}, based on X-ray measurements of a representative sample of local, massive clusters.  Even though \YSZ\ and \YX\ are effectively measures of the cluster pressure, they depend on the details of the shape of the profile differently, which can still vary somewhat between clusters.  Assuming the Arnaud et al. \citep{arnaud10} pressure profile, we predict a relationship of $\Yfh = 1.08 \ \YX$, where \Yfh \ is integrated within a fixed angular aperture corresponding to \rfh \ (often called a cylindrical projection). In Figure \ref{fig:ysz_vs_yx}, we plot the \YSZ\ estimated by the MCMC method against the \YX\ measured in B13. We fit a scaling relation of the form $\Yfh = A \ \YX$, and find that the best-fit normalization is $A = 1.17\pm0.12$, consistent with the expected normalization. This fit has a total $\chi^2$ of $19.46$ for 14 degrees of freedom, with a probability to exceed of $P = 0.15$. The uncertainty in the normalization is the range for which $\Delta \chi^2 < 1$ compared to the best-fit.

\begin{table*}
\begin{center}
{\scriptsize
\caption{SPT Cluster Fluxes and Masses}
\begin{tabular}{lcccccc}
\hline \hline
Object Name & $z$ & \Ysf & \Yfh & \YX & \MSZ & $\MX$ \\
 & & ($10^{14} \msun \mathrm{keV}$) & ($10^{14} \msun \mathrm{keV}$) & ($10^{14} \msun \mathrm{keV}$) & ($10^{14} \msun h_{70}^{-1}$) & ($10^{14} \msun h_{70}^{-1}$) \\
\hline
SPT-CL J0509-5342 & $0.463$ & $0.9\pm0.1$ & $3.6^{+1.4}_{-1.1}$ & $4.3\pm0.8$ & $4.32\pm1.11$ & $5.11\pm0.75$  \\
SPT-CL J0511-5154$^a$ & $0.74$ & $1.2\pm0.2$ & $ - $ & $ - $ & $2.79\pm1.43$ & $ - $ \\
SPT-CL J0521-5104$^a$ & $0.72$ & $1.1\pm0.2$ & $ - $ & $ - $ & $2.46\pm1.32$ & $ - $ \\
SPT-CL J0528-5259 & $0.765$ & $1.1\pm0.2$ & $1.8^{+0.8}_{-0.5}$ & $1.6\pm0.5$ & $2.21\pm1.14$ & $2.54\pm0.54$  \\
SPT-CL J0533-5005 & $0.881$ & $1.4\pm0.2$ & $2.1^{+0.6}_{-0.4}$ & $1.0\pm0.4$ & $2.75\pm1.39$ & $1.86\pm0.43$  \\
SPT-CL J0539-5744$^a$ & $0.77$ & $1.0\pm0.2$ & $ - $ & $ - $ & $1.93\pm0.93$ & $ - $ \\
SPT-CL J0546-5345 & $1.067$ & $2.0\pm0.3$ & $5.0^{+1.1}_{-1.0}$ & $4.8\pm0.8$ & $4.18\pm0.89$ & $4.79\pm0.86$  \\
SPT-CL J0551-5709 & $0.423$ & $0.7\pm0.1$ & $3.4^{+1.7}_{-1.2}$ & $1.9\pm0.4$ & $3.57\pm1.43$ & $3.32\pm0.46$  \\
SPT-CL J0559-5249 & $0.611$ & $1.6\pm0.2$ & $9.0^{+2.1}_{-1.8}$ & $6.4\pm0.8$ & $5.46\pm1.04$ & $6.29\pm0.86$  \\
SPT-CL J2301-5546$^a$ & $0.748$ & $1.0\pm0.2$ & $ - $ & $- $ & $1.89\pm0.89$ & $ - $ \\
SPT-CL J2331-5051 & $0.571$ & $1.4\pm0.2$ & $2.3^{+0.4}_{-0.3}$ & $3.5\pm0.6$ & $5.29\pm1.00$ & $4.50\pm0.64$  \\
SPT-CL J2332-5358 & $0.403$ & $0.9^{+0.2}_{-0.1}$ & $8.7^{+3.7}_{-3.1}$ & $6.1\pm0.8$ & $5.25\pm1.04$ & $6.39\pm0.75$  \\
SPT-CL J2337-5942 & $0.781$ & $3.1\pm0.2$ & $7.8^{+1.3}_{-1.4}$ & $8.5\pm1.7$ & $6.67\pm1.29$ & $6.82\pm1.11$  \\
SPT-CL J2341-5119 & $0.998$ & $2.3\pm0.2$ & $6.8\pm1.1$ & $4.7\pm1.0$ & $4.86\pm0.93$ & $4.64\pm0.86$  \\
SPT-CL J2342-5411 & $1.074$ & $1.5\pm0.3$ & $2.6\pm0.6$ & $1.4\pm0.3$ & $2.46\pm1.32$ & $2.36\pm0.43$  \\
SPT-CL J2355-5056 & $0.320$ & $0.4\pm0.1$ & $2.1^{+0.9}_{-0.7}$ & $2.2\pm0.4$ & $3.11\pm1.61$ & $3.75\pm0.46$  \\
SPT-CL J2359-5009 & $0.774$ & $1.4\pm0.2$ & $4.5^{+1.3}_{-1.1}$ & $1.8\pm0.4$ & $3.61\pm1.11$ & $2.86\pm0.50$  \\
SPT-CL J0000-5748 & $0.701$ & $1.1\pm0.2$ & $2.1^{+1.1}_{-0.6}$ & $4.2\pm1.6$ & $2.57\pm1.36$ & $4.14\pm0.93$  \\
\hline
\end{tabular}}
\begin{tablenotes}
Note -- Cluster redshifts and X-ray fluxes are quoted from Benson et al. \citep{benson13}. 
\Ysf \ is the integrated Comptonization within $0.75'$, calculated with our \YSZ\ MCMC method. 
\Yfh \ is the integrated Comptonization within \rfh. 
\Ysf\ and \Yfh\ values are given in \msun keV for comparison to \YX\ and the \YSZ\ values from A11. 
\Ysf\ and \Yfh\ are cylindrically projected. 
\MSZ \ and \MX \ are estimates of \Mfh \ calculated from the same CosmoMC chains, using only the \Ysf \ and \YX\ data respectively.
\\ (a) These clusters have only SZ data, and no X-ray observations.
\end{tablenotes}
\label{tab:cluster_table}
\end{center}
\end{table*}

\section{\YSZ\ and Mass Scaling Relation Conclusions}
\label{sec:conc}

We describe and implement a method of constraining \YSZ\ generalizable to any cluster profile, and we show that this method accurately recovers \YSZ\ in simulations. 
We compare \YSZ\ to SPT cluster detection significance, focusing on scatter with mass.  
Finally, we apply this method to clusters detected in the SPT-SZ survey, and compare the estimated \YSZ\ values to \YSZ\ estimated by a template fitting method, and to \YX.

We apply our method to clusters in simulated tSZ-only maps and measure \Ytheta, the integrated Comptonization within a constant angular aperture.  
We find that \YSZ\ is measured with the lowest fractional scatter in an aperture comparable to the SPT beam size ($\aprx1'$ FWHM at \onefifty).
We fit \Ytheta-\Mvir\ scaling relations for a range of angular apertures and find a minimum fractional scatter of $23\pm2\%$ in \YSZ, at a fixed mass, with the minimum occurring for an angular aperture of $0.75'$.
We also calculate \YSZ\ within a range of physical radii, $\rho$, and find a minimum scatter in \Yrho\ at an integration radius of $0.3$ Mpc, which corresponds roughly to $0.75'$ at the survey median redshift ($z = 0.6$), with a fractional scatter of $28\pm2\%$ at a fixed mass.
Using the same simulated clusters, we also fit a $\zeta$-\Mvir\ relation, where $\zeta$ is the matched filter SZ detection significance measured by SPT, and find a fractional scatter of $27\pm2\%$.

We also analyze clusters in simulations including tSZ, CMB, point sources, atmospheric noise, and realistic SPT instrumental noise. 
In these full-noise simulations, the \Ysf-\Mvir\ scaling relation has $27\pm1\%$ scatter, the \YthM-\Mvir\ scaling relation has $34\pm2\%$ scatter, and $\zeta$-\Mvir\ scaling relation has $27\pm2\%$ scatter. 
These simulations demonstrate that scatter in \Ytheta\ is comparable to the scatter in $\zeta$.

To investigate \YSZ\ scaling relations in SPT observed clusters, we fit \Ytheta-\Mfh\ and \Yrho-\Mfh\ scaling relations to the sample of eighteen SPT clusters described and examined in Benson et al. \citep{benson13}. 
Of these, fourteen have X-ray observations and measured \YX\ values, which we use to estimate the cluster \Mfh\ masses.
We fit the scaling relations using a version of CosmoMC, similar to the one described in Benson et al. \citep{benson13}, modified to account for the cluster selection based on \YSZ\ instead of SPT significance. 
For these clusters, the \Ysf-\Mfh\ scaling relation is found to have $21\pm11\%$ intrinsic log-normal scatter in \YSZ\ at a fixed mass, the \YthM-\Mfh\ scaling relation has $26\pm9\%$ scatter, and the $\zeta$-\Mfh\ relation has $21\pm9\%$ scatter.
The \YSZ-\Mfh\ scaling relations have the advantage of being more easily comparable to scaling relations produced by other experiments, since they are based on a physical parameter instead of the SPT detection significance.

We also calculate a cylindrically projected \Yfh, the integrated Comptonization within \rfh, for the clusters in the Benson et al. \citep{benson13} sample. 
We compare the \Yfh \ values recovered by our Markov-Chain Monte Carlo method to those calculated for the same clusters by the template fitting method described in A11 and find the two methods to be consistent. 
The advantage of the MCMC based \YSZ\ estimator over the A11 method is that it does not require follow-up X-ray data to establish cluster profiles.
We further compare the MCMC derived \Yfh \ values to the \YX\ values for these clusters from Benson et al. \citep{benson13} and find that they are consistent with the expected scaling between \YSZ\ and \YX, based on the universal pressure profile of Arnaud et al. \citep{arnaud10}.

We have demonstrated, with both simulations with realistic SPT noise and SPT observed clusters, that \YSZ\ is most accurately determined in an aperture comparable to the SPT beam size. 
We have used this information in measuring \YSZ\ for the catalog of clusters observed with the SPT in the 2008 and 2009 seasons \citep{reichardt13}, and in the full SPT-SZ survey catalog \cite{bleem15}.
The SPT-SZ survey of $2500 \ \sqdeg$ was completed in November 2011, and has detected $\aprx 500$ clusters with a median redshift of $\aprx 0.5$ and a median mass of $\Mfh \ \aprx \ 2.3 \times 10^{14} \msun h^{-1}$.
The methods and results presented here will continue to inform the measurement and use of \YSZ\ for the clusters detected in the SPTpol and SPT-3G experiments.

%Conclusion

\doublespacing

\chapter{Conclusion}

The SPT-3G instrument is a significant improvement over the previous instruments on the South Pole Telescope, both in terms of the number of detectors which will be fielded, and the noise performance of the instrument optics. The final focal plane will consist of 2,690 dual polarization multichroic sinuous antenna pixels, and a total of 16,140 TES bolometers. At the time of writing, the iterative process of detector fabrication and testing is still underway. We are able to produce wafers with the full fieldable antenna, filter, and TES design. The electrical and thermal properties of the TES, and the band properties of the filters have been tested in earlier prototype detectors. Work still needs to be done to validate the functioning and optical properties of the fully integrated pixel design. 

The detectors I have helped to develop for SPT-3G will allow us to improve our understanding of cosmology, from the dark energy equation of state to constraining inflationary models through the tensor-to-scalar ratio. We may even be able to resolve the neutrino mass hierarchy problem and measure the sum of the neutrino masses. 

The SPTpol instrument is currently operating in its fourth observing season at the South Pole. It has already produced valuable results, most notably the first detection of lensing B-modes in the CMB. The cryogenic electronics I designed to read out the 1,536 TES bolometers in the SPTpol experiment have been performing within specifications for four seasons. Accessing the cryogenic portions of the readout system is a significant undertaking, since the cryostat must be warmed, unshipped from the receiver cabin, and disassembled. Since these systems cannot be repaired or replaced without a significant loss of observing time, it is essential that their performance be consistent and their failure rate low. In the four years of observing with the SPTpol instrument, none of the cryogenic readout boards have failed. 

Design and assembly insights gained from the development of the SPTpol readout electronics have informed the development of the SPT-3G cryogenic readout systems. In particular the move from commercial surface mount capacitors and small inductor chips to custom fabricated integrated LC wafers was motivated by the need for lower ESR components, and for production and testing scalability to a readout system with $\aprx10 \times$ the number of channels. Tantalum nitride resistors are now avoided, following the re-discovery of the superconducting transition in thin film TaN.

I developed a Bayesian likelihood based method for measuring the integrated Comptonization, \YSZ, of galaxy clusters, and a variant of cosmoMC for fitting \YSZ-$M$ scaling relations. These methods have been extensively verified using simulated data, and observed clusters from the SPT-SZ survey. I have demonstrated that for accurate recovery of \YSZ\, the optimal angular or physical scale for integration corresponds to the beam scale of the telescope. The likelihood method presented here has been used to calculate \YSZ\ for subsequent SPT-SZ cluster catalogs, and will continue to serve for the future SPTpol and SPT-3G galaxy cluster catalogs. Likewise the \YSZ-$M$ scaling relation fitting methods developed here will continue to inform future SPT studies of galaxy clusters, and aid in the improvement of both our understanding of the process of structure formation, and our constraints on global cosmological parameters such as the equation of state of dark energy.

\bibliographystyle{unsrt}
\bibliography{thesis}

\end{document}